\documentclass[12pt]{article}

\pdfoutput = 1

\usepackage{times}
\usepackage{authblk}
\usepackage{amsmath}
\usepackage{amssymb}
\usepackage{amsthm}
\usepackage{bm}
\usepackage{braket}
\usepackage{slashed}
\usepackage[svgnames,psnames]{xcolor}
\usepackage[nottoc]{tocbibind}
\usepackage{cite}
\usepackage{here}
\usepackage[height = 8.8 in, width = 6.45 in]{geometry}
\usepackage{graphicx}
\usepackage{adjustbox}
\usepackage{tikz-cd}
\usepackage[titletoc]{appendix}
\usepackage{comment}
\usepackage[colorlinks, linktocpage, hypertexnames = false]{hyperref}

\hypersetup{linkcolor = Crimson, citecolor = MediumBlue, urlcolor = MediumBlue}

\numberwithin{equation}{section}

\theoremstyle{definition}

\begin{document}

~\vspace{4cm}~
\begin{center}{\fontsize{15}{0}
\textbf{Topological field theories and symmetry protected topological phases with fusion category symmetries}}
\end{center}

\begin{center}
Kansei Inamura
\end{center}

\begin{center}
{\small Institue for Solid State Physics, University of Tokyo, Kashiwa, Chiba 277-8581, Japan}
\end{center}
~

\begin{abstract}
Fusion category symmetries are finite symmetries in 1+1 dimensions described by unitary fusion categories.
We classify 1+1d time-reversal invariant bosonic symmetry protected topological (SPT) phases with fusion category symmetry by using topological field theories.
We first formulate two-dimensional unoriented topological field theories whose symmetry splits into time-reversal symmetry and fusion category symmetry.
We then solve them to show that SPT phases are classified by equivalence classes of quintuples $(Z, M, i, s, \phi)$ where $(Z, M, i)$ is a fiber functor, $s$ is a sign, and $\phi$ is the action of orientation-reversing symmetry that is compatible with the fiber functor $(Z, M, i)$.
We apply this classification to SPT phases with Kramers-Wannier-like self-duality.
\end{abstract}

\setcounter{page}{0}

\thispagestyle{empty}

\newpage

\tableofcontents

\flushbottom

\section{Introduction}
Symmetry protected topological (SPT) phases are symmetric gapped phases whose ground states are unique on any closed manifold.
In the low energy limit, SPT phases are expected to be described by invertible quantum field theories, which are classified by using bordism groups \cite{Kap2014, KTTW2015, FH2016, Yon2019}.
In particular, in 1+1 dimensions, the low energy limit of SPT phases is believed to become topological quantum field theories (TQFTs).
Therefore, the classification of 1+1d SPT phases with symmetry $G$ reduces to the classification of invertible $G$-equivariant TQFTs.
When $G$ is an internal symmetry, the low energy TQFTs are defined on oriented manifolds.
Such TQFTs are called oriented TQFTs.
The algebraic descriptions of oriented $G$-equivariant TQFTs enable us to show that 1+1d bosonic SPT phases with finite internal symmetry $G$ are classified by group cohomology $H^2(G, \mathrm{U}(1))$ \cite{Tur1999, MS2006, KT2017, CGLW2013}.
On the other hand, when $G$ involves time-reversal symmetry, the low energy limit is described by unoriented TQFTs, which can be defined on unoriented manifolds.
The algebraic descriptions of unoriented $G$-equivariant TQFTs for finite groups $G$ are given in \cite{KT2017, Swe2013}.
In particular, it is shown that invertible unoriented $G$-equivariant TQFTs are classified by twisted group cohomology $H^2(G, \mathrm{U}(1)_{\rho})$ where the homomorphism $\rho: G \rightarrow \mathbb{Z}_2$ determines which elements reverse the orientation of time \cite{KT2017, CGLW2013}.

We can extend the classification of SPT phases along with the generalizations of the notion of symmetry.
In general, symmetries are characterized by the algebraic relations of topological defects.
For example, ordinary group symmetries are associated with invertible topological defects with codimension 1.
One possible generalization of the ordinary group symmetries is symmetries generated by topological defects with higher codimensions.
Such generalized symmetries are called higher form symmetries \cite{GKSW2015}.
Another generalization is to consider non-invertible topological defects with codimension 1, which do not form a group.
In particular, in 1+1 dimensions, the algebraic relations of finite numbers of topological defect lines including non-invertible ones are described by unitary fusion categories \cite{BT2018, CLSWY2019}.
The corresponding symmetries are called fusion category symmetries.
As well as ordinary group symmetries, we can discuss anomalies and gauging of them \cite{FRS2002, FFRS2010, BCP2014a, BCP2014b, BCP2015, CR2016, BT2018, TW2019, CLSWY2019}.

Fusion category symmetries have a long history in the study of two-dimensional rational conformal field theories \cite{Ver1988, PZ2001, FFRS2004, FFRS2007, CLSWY2019}.
We can also find fusion category symmetries in lattice models such as anyonic chains \cite{BG2017, FTLTKWF2007, TTWL2008, TAFHLT2008, GATLTW2009, GATHLTW2013, PBTLTV2012}.
The thermodynamic limit of anyonic chains is often described by conformal field theories, whose energy spectra cannot be gapped without breaking fusion category symmetries.
There is also a way to construct two-dimensional statistical mechanical models with general fusion category symmetries \cite{AMF2016, AFM2020}.
These models are naturally described in terms of three-dimensional topological field theories on a manifold with boundaries.
Similarly, fusion category symmetries on the boundary of 2+1d topologically ordered states are investigated in \cite{LTLSB2020}.
Recently, fusion category symmetries are also studied in the context of two-dimensional topological field theories \cite{BT2018, TW2019, KORS2020, HL2021}.
Remarkably, it is shown in \cite{TW2019, KORS2020} that oriented bosonic TQFTs with fusion category symmetry are classified by the module categories of the fusion category.
This includes the classification of bosonic fusion category SPT phases, which are gapped phases with fusion category symmetry whose ground states are unique on a circle.

In this paper, we generalize the classification of bosonic fusion category SPT phases to the case with time-reversal symmetry.
Our approach is to axiomatize two-dimensional unoriented TQFTs with fusion category symmetry and solve them under the condition that the Hilbert space on a circle without topological defect is one-dimensional.
Along the way, we reproduce the classification of bosonic fusion category SPT phases without time-reversal symmetry.

The rest of the paper is organized as follows.
In section \ref{sec: Preliminary: fusion category symmetries}, we review fusion category symmetries in two-dimensional quantum field theories.
The contents of this section are not restricted to TQFTs.
In section \ref{sec: Oriented TQFTs with fusion category symmetry}, we classify bosonic fusion category SPT phases without time-reversal symmetry.
Section \ref{sec: Consistency conditions of oriented TQFTs} is devoted to the review of oriented bosonic TQFTs with fusion category symmetry.
By solving the consistency conditions of oriented TQFTs explicitly in section \ref{sec: Bosonic fusion category SPT phases without time-reversal symmetry}, we show that fusion category SPT phases without time-reversal symmetry are classified by isomorphism classes of fiber functors.\footnote{A fiber functor is a tensor functor from a fusion category to the category of vector spaces.}
This agrees with the result in previous papers \cite{TW2019, KORS2020}, although the approach is slightly different.
Our derivation clarifies the physical interpretation of fiber functors.
In section \ref{sec: Unoriented TQFTs with fusion category symmetry}, we classify bosonic fusion category SPT phases with time-reversal symmetry.
We first formulate unoriented TQFTs with fusion category symmetry in section \ref{sec: Consistency conditions of unoriented TQFTs} and then classify SPT phases by solving the consistency conditions of unoriented TQFTs in section \ref{sec: Bosonic fusion category SPT phases with time-reversal symmetry}.
We will see that bosonic fusion category SPT phases with time-reversal symmetry are classified by the algebraic data $(Z, M, i, s, \phi)$ where $(Z, M, i)$ represents bosonic fusion category SPT phases without time-reversal symmetry, $s$ represents bosonic SPT phases only with time-reversal symmetry, and $\phi$ represents the action of orientation-reversing symmetry.
Finally, in section \ref{sec: Examples (unoriented)}, we discuss examples including the classification of SPT phases with duality symmetry.
In some cases, duality symmetries do not admit time-reversal invariant SPT phases, although they admit SPT phases without time-reversal symmetry.
This may be thought of as mixed anomalies between time-reversal symmetry and the duality symmetries.
Throughout the paper, we assume that the total symmetry splits into time-reversal symmetry and finite internal symmetry.

\section{Preliminary: fusion category symmetries}
\label{sec: Preliminary: fusion category symmetries}
In this section, we briefly review fusion category symmetries to fix the notation by following \cite{BT2018}.
For details about fusion categories, see e.g. \cite{EGNO2015}.
The basic ingredients of fusion category symmetry are topological defect lines and topological point operators.
A topological defect line of a theory with fusion category symmetry $\mathcal{C}$ is labeled by an object $x$ of a unitary fusion category $\mathcal{C}$.
In particular, the trivial defect line corresponds to the unit object $1$, which is simple.
A topological point operator that changes a topological defect $x$ to another topological defect $y$ is labeled by a morphism $f \in \mathop{\mathrm{Hom}}(x, y)$, where $\mathop{\mathrm{Hom}}(x, y)$ is a $\mathbb{C}$-vector space of morphisms from $x$ to $y$.\footnote{We note that $\mathop{\mathrm{Hom}}(x, y)$ is a subspace of topological point operators between $x$ and $y$. For example, topological point operators on which a simple topological line ends are excluded because $\mathop{\mathrm{Hom}(1, x)}$ is empty when $x$ is simple.}
Every topological point operator $f \in \mathop{\mathrm{Hom}}(x, y)$ has the adjoint $f^{\dagger} \in \mathop{\mathrm{Hom}}(y, x)$.

When we have two topological defects $x$ and $y$ that run parallel with each other, we can regard them as a single topological defect, which is labeled by the tensor product $x \otimes y$.
We can also think of the tensor product $x \otimes y$ as the fusion of $x$ and $y$.
The trivial defect $1$ acts as the unit of the fusion:
\begin{equation}
1 \otimes x \cong x \otimes 1 \cong x.
\end{equation}
The isomorphisms $l_x: 1 \otimes x \rightarrow x$ and $r_x: x \otimes 1 \rightarrow x$ are called the left and right units respectively, which are taken to be the identity morphism $\mathrm{id}_x$ by identifying $1 \otimes x$ and $x \otimes 1$ with $x$.
The order of the fusion of three topological defects $x$, $y$, and $z$ is changed by using the isomorphism $\alpha_{xyz}: (x \otimes y) \otimes z \rightarrow x \otimes (y \otimes z)$ that is called the associator.
The associators satisfy the following commutative diagram called the pentagon equation:
\begin{equation}
\begin{tikzcd}
 & (x \otimes y) \otimes (z \otimes w) \arrow[dr, "\alpha_{x, y, z \otimes w}"] & \\
 ((x \otimes y) \otimes z) \otimes w \arrow[ru, "\alpha_{x \otimes y, z, w}"] \arrow[d, "\alpha_{xyz} \otimes \mathrm{id}_w"'] & & x \otimes (y \otimes (z \otimes w))\\
 (x \otimes (y \otimes z)) \otimes w \arrow[rr, "\alpha_{x, y \otimes z, w}"'] & & x \otimes ((y \otimes z) \otimes w) \arrow[u, "\mathrm{id}_x \otimes \alpha_{yzw}"']
\end{tikzcd}
\label{eq: pentagon}
\end{equation}

\begin{figure}
\begin{minipage}{0.25 \hsize}
\begin{center}
\includegraphics[width = 3.5cm]{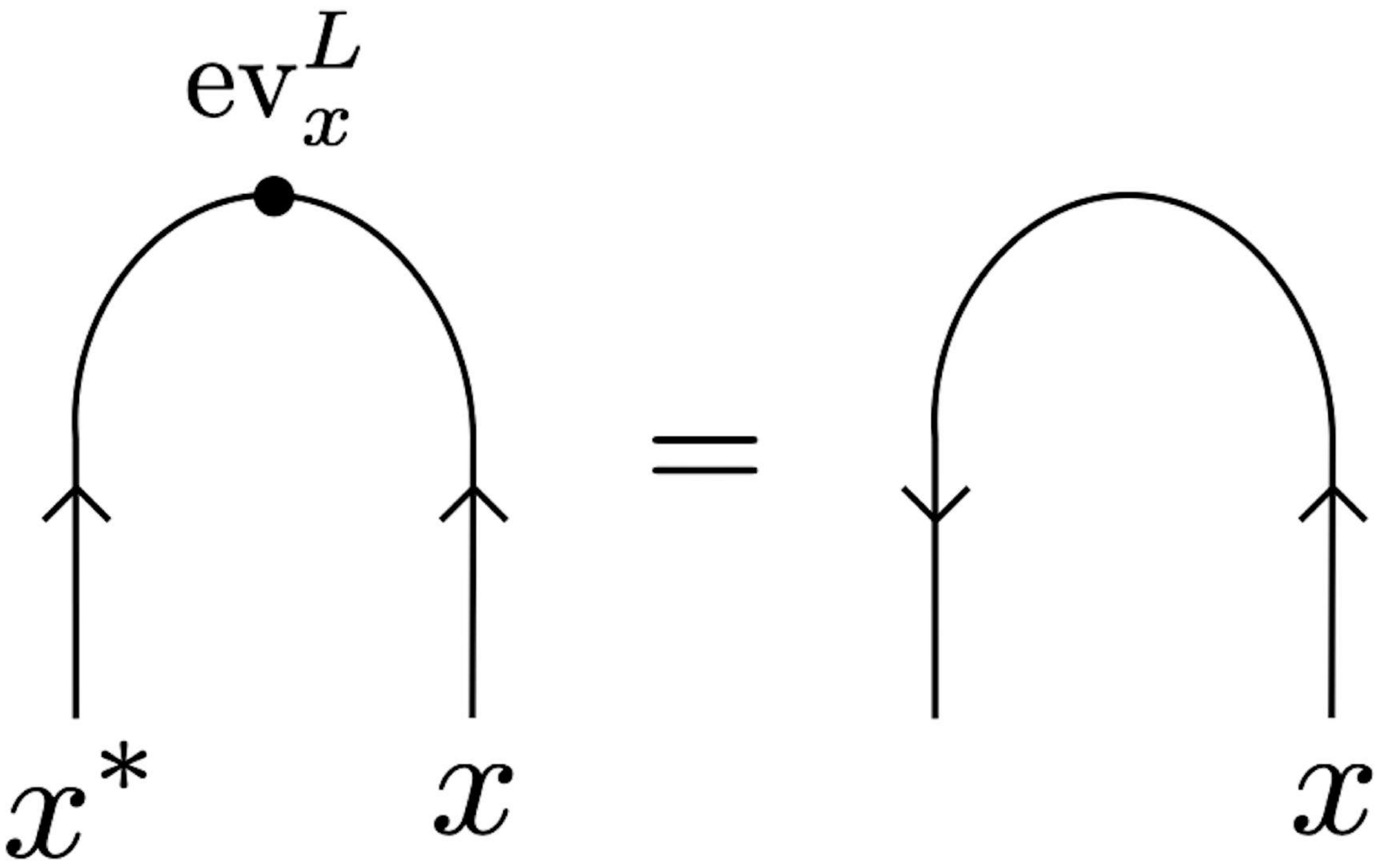}
\end{center}
\end{minipage}%
\begin{minipage}{0.25 \hsize}
\begin{center}
\includegraphics[width = 3.5cm]{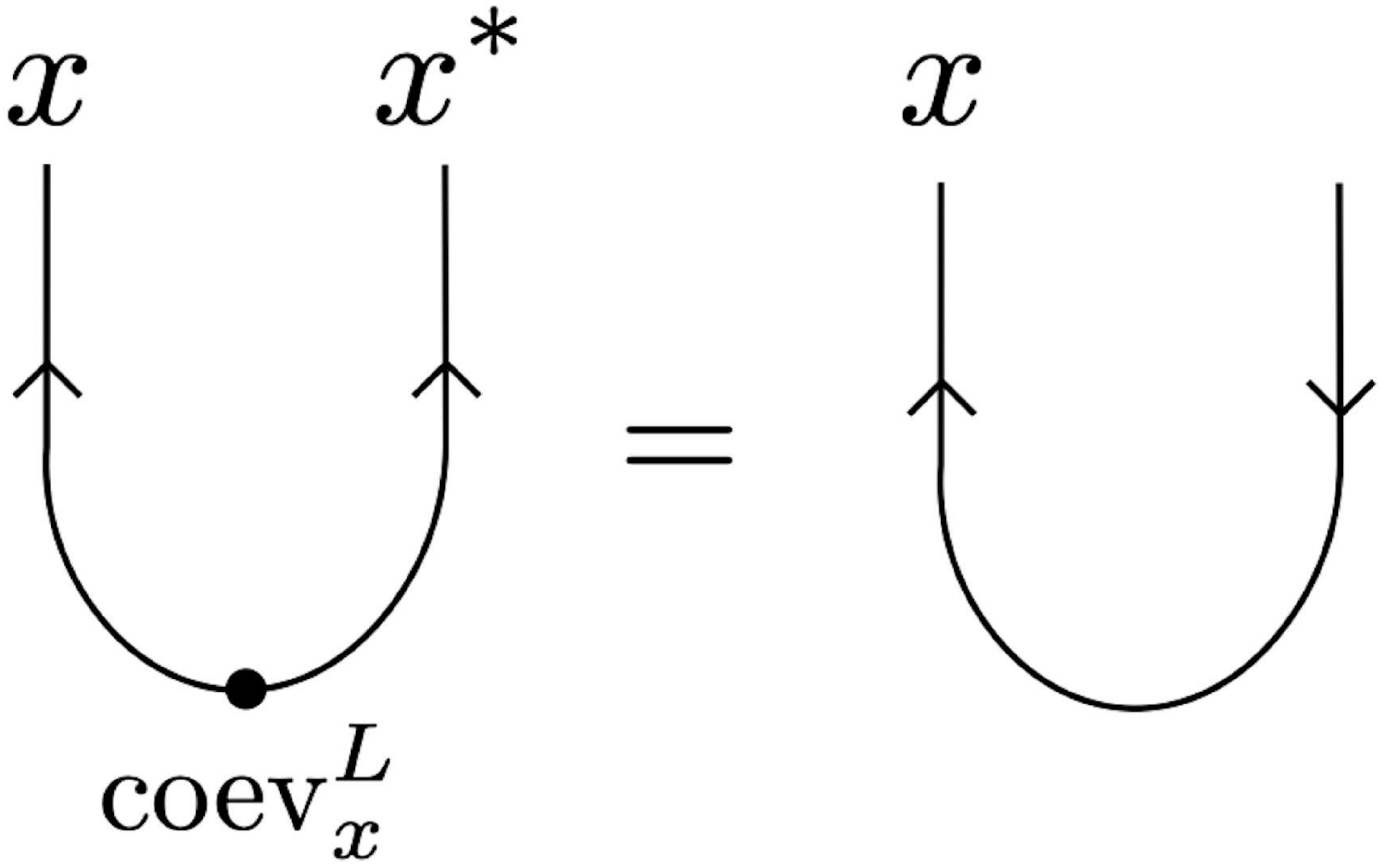}
\end{center}
\end{minipage}%
\begin{minipage}{0.25 \hsize}
\begin{center}
\includegraphics[width = 3.5cm]{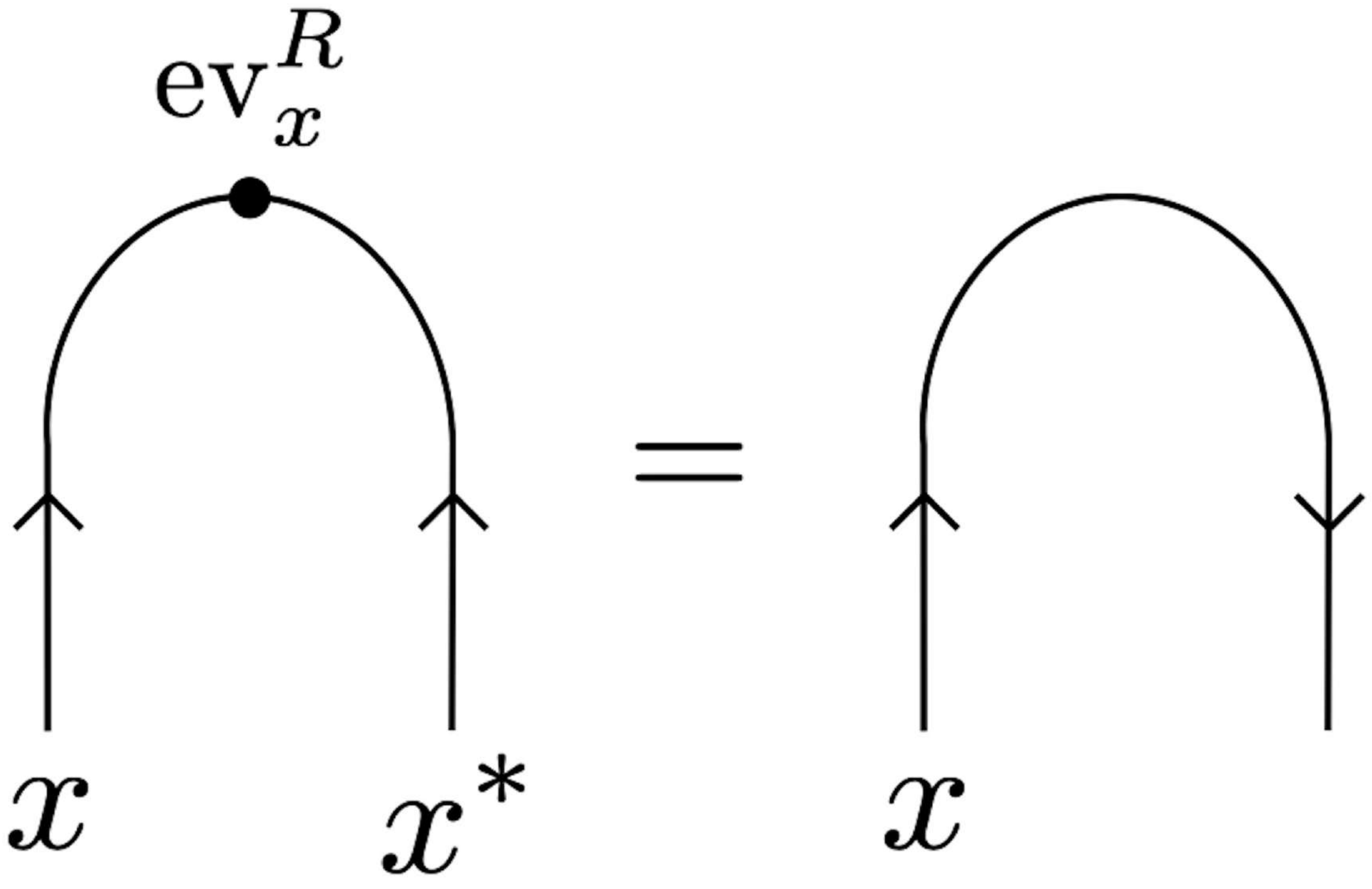}
\end{center}
\end{minipage}%
\begin{minipage}{0.25 \hsize}
\begin{center}
\includegraphics[width = 3.5cm]{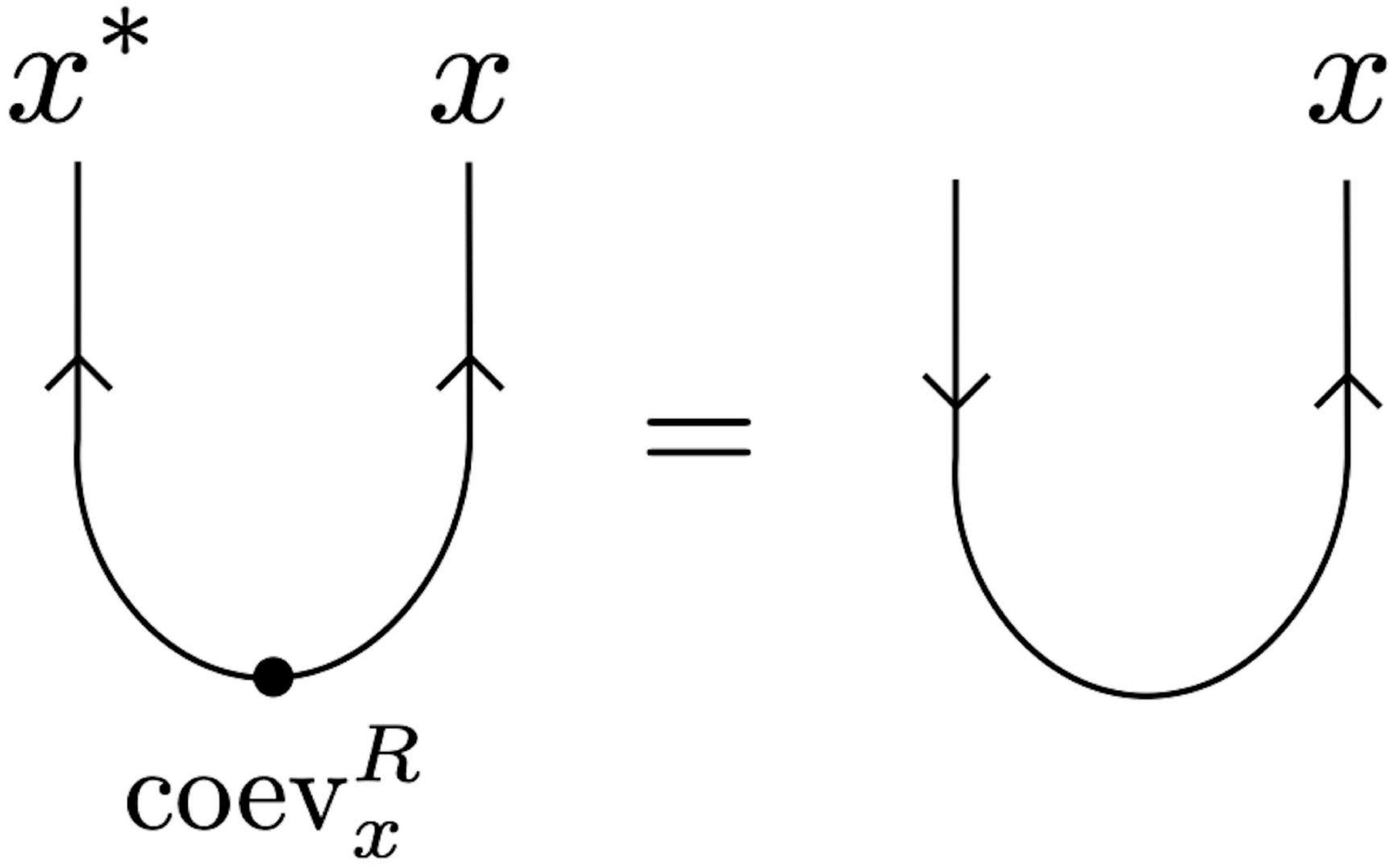}
\end{center}
\end{minipage}
\caption{The left and right (co)evaluation morphisms describe the operation of folding topological defect lines to the left and right respectively. In the folded diagrams, topological point operators corresponding to (co)evaluation morphisms are often implicit.}
\label{fig: folding}
\end{figure}

The orientation reversal of a topological defect $x$ is labeled by the dual object $x^{*}$.
In particular, the topological defect $x^{**}$ whose orientation is reversed twice is equal to the original topological defect $x$ up to natural isomorphism $a_x: x \rightarrow x^{**}$ that is called the pivotal structure.
Folding $x$ to the left is described by morphisms $\mathrm{ev}_x^L: x^* \otimes x \rightarrow 1$ and $\mathrm{coev}_x^L: 1 \rightarrow x \otimes x^*$, which are called the left evaluation morphism and the left coevaluation morphism respectively.
Similarly, folding $x$ to the right is described by the right evaluation morphism $\mathrm{ev}_x^R: x \otimes x^* \rightarrow 1$ and the right coevaluation morphism $\mathrm{coev}_x^R: 1 \rightarrow x^* \otimes x$, which are given by the adjoints of the left evaluation and coevaluation morphisms
\begin{equation}
\mathrm{ev}_x^R = (\mathrm{coev}_x^L)^{\dagger}, \quad \mathrm{coev}_x^R = (\mathrm{ev}_x^L)^{\dagger}.
\label{eq: adjoint}
\end{equation}
For a unitary fusion category, there is a canonical pivotal structure that is given by $a_x = (\mathrm{id}_{x^{**}} \otimes \mathrm{ev}_x^L) \circ ((\mathrm{ev}_{x^*}^L)^{\dagger} \otimes \mathrm{id}_x)$, see e.g. \cite{HP2017}.\footnote{The associator is omitted in this expression. In the subsequent sections, we will sometimes omit associators to simplify the notation. One can restore them wherever needed.}
We note that the left and right (co)evaluation morphisms are related by the canonical pivotal structure as follows:
\begin{equation}
\mathrm{ev}_x^L = \mathrm{ev}_{x^*}^R \circ (\mathrm{id}_{x^*} \otimes a_x), \quad (a_x \otimes \mathrm{id}_{x^*}) \circ \mathrm{coev}_x^L = \mathrm{coev}_{x^*}^R.
\label{eq: canonical}
\end{equation}
These evaluation and coevaluation morphisms are represented diagrammatically as shown in figure \ref{fig: folding}.
In particular, the zigzag-shaped defect constructed from these folding operations cannot be distinguished from the straight defect:
\begin{equation}
\adjincludegraphics[valign = c, width = 1.6cm]{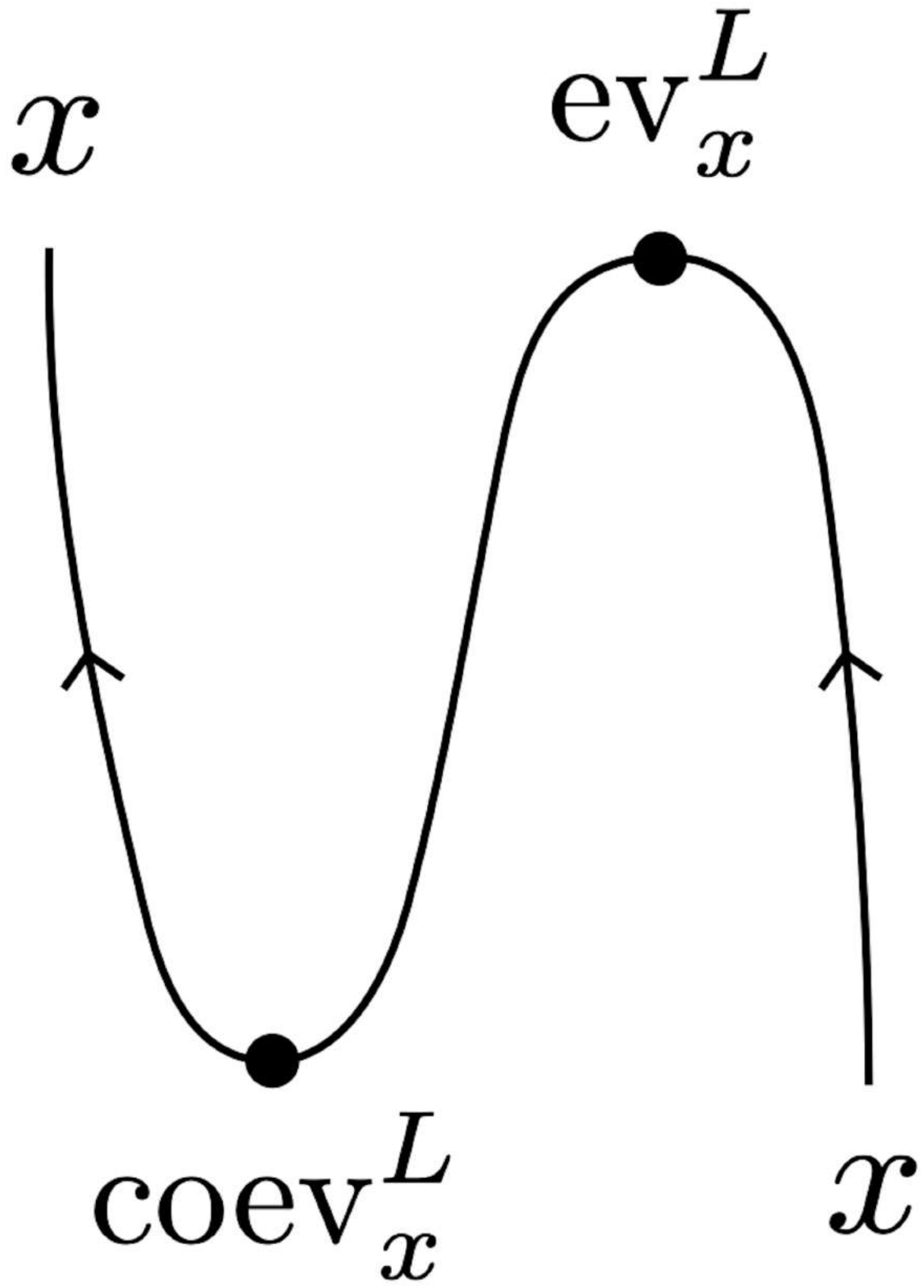} ~ = ~ 
\adjincludegraphics[valign = c, width = 1.6cm]{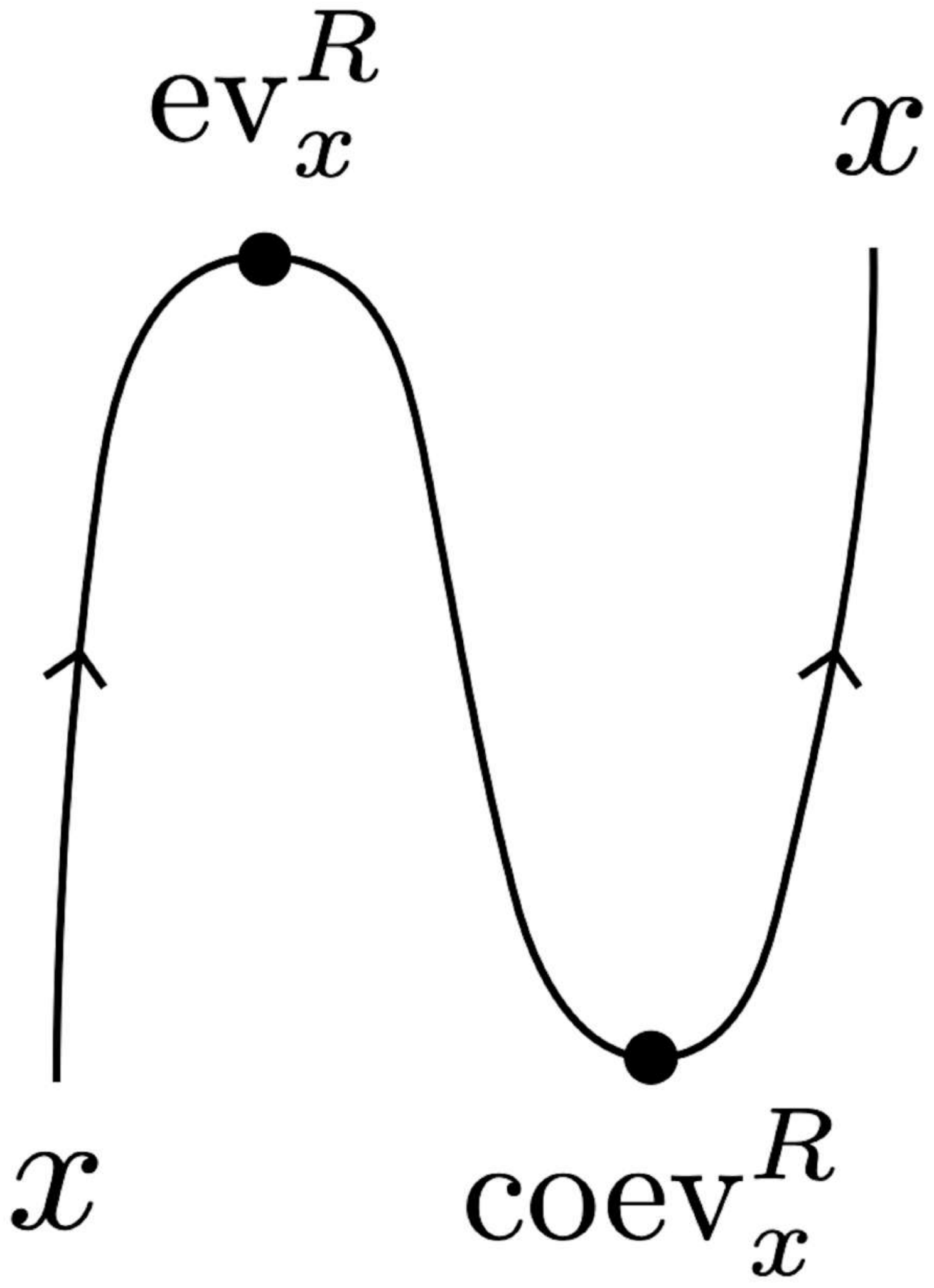} ~ = ~ 
\adjincludegraphics[valign = c, width = 0.2cm]{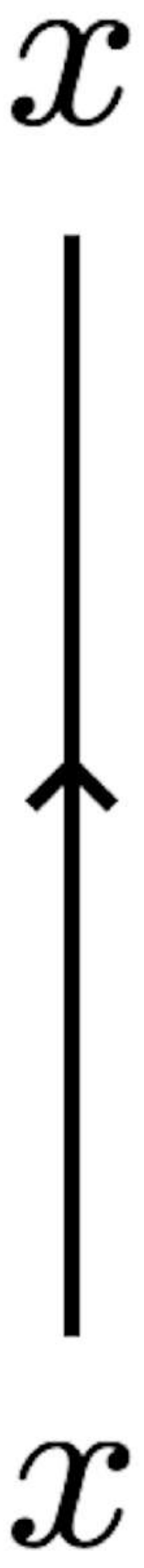}.
\label{eq: zigzag axiom}
\end{equation}
The loop diagram evaluates the quantum dimension
\begin{equation}
\mathop{\mathrm{dim}}x = \adjincludegraphics[valign = c, width = 1.8cm]{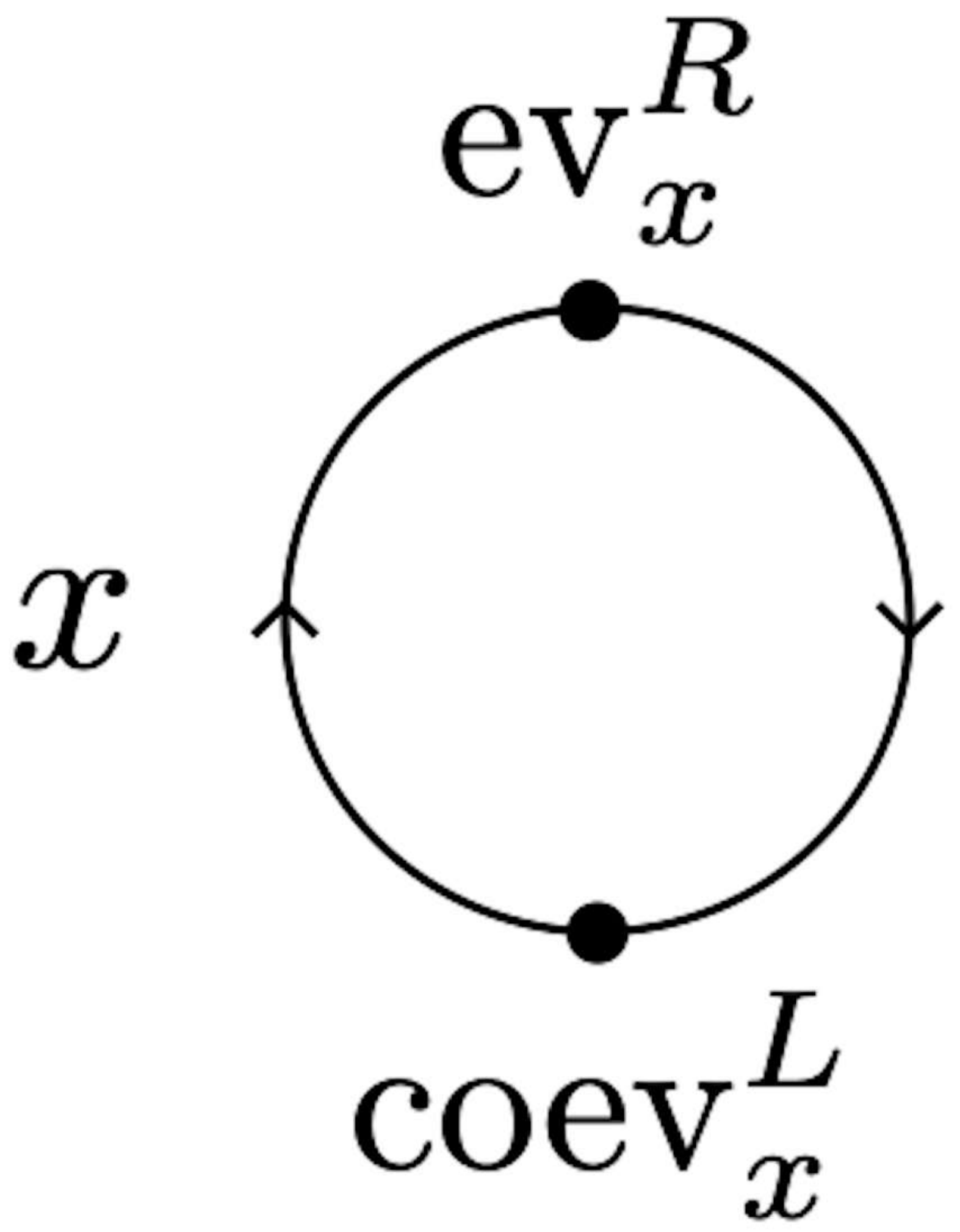},
\end{equation}
which does not depend on whether the loop is clockwise or counterclockwise, i.e. $\mathop{\mathrm{dim}} x = \mathop{\mathrm{dim}} x^*$.

We can also consider the sum $x \oplus y$ of topological defects $x$ and $y$.
When the topological defect $x \oplus y$ is put on a path $P$, the correlation function is calculated as the sum of the correlation functions in the presence of topological defects $x$ and $y$ on $P$:
\begin{equation}
\braket{\cdots (x \oplus y)(P) \cdots} = \braket{\cdots x(P) \cdots} + \braket{\cdots y(P) \cdots}.
\end{equation}

\section{Oriented TQFTs with fusion category symmetry}
\label{sec: Oriented TQFTs with fusion category symmetry}

\subsection{Consistency conditions of oriented TQFTs}
\label{sec: Consistency conditions of oriented TQFTs}
\begin{figure}
\begin{center}
\includegraphics[width = 6cm]{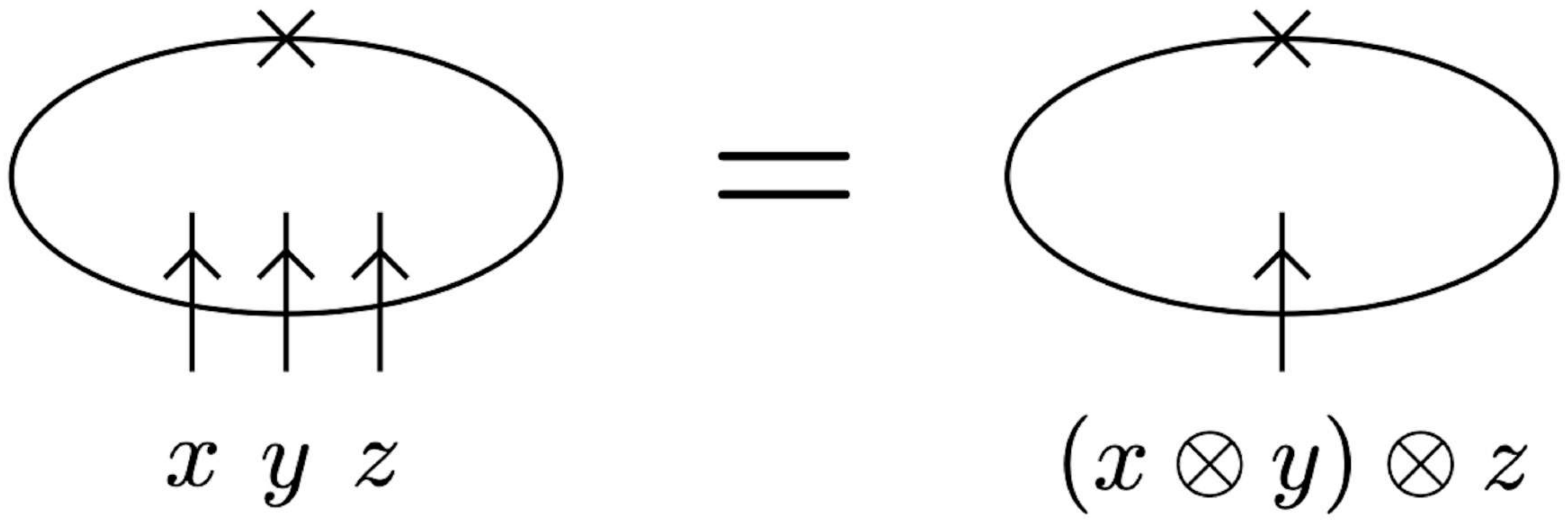}
\caption{The vector space on a circle with multiple topological defects is the same as the vector space on a circle with a single topological defect labeled by the tensor product of the topological defects. To determine the order of the tensor product, we need to specify the base point on a circle, which is represented by the cross mark in the above figure.}
\label{fig: vector space on a circle}
\end{center}
\end{figure}

In this section, we review two-dimensional bosonic oriented TQFTs with fusion category symmetry $\mathcal{C}$.
We will enumerate the consistency conditions on transition amplitudes by following \cite{BT2018}.
We first assign a vector space $V_x$ to a circle with a topological defect $x$.
When a topological defect is a direct sum $x \oplus y$, the vector space $V_{x \oplus y}$ is given by the direct sum of two vector spaces $V_x$ and $V_y$.
To define the vector space on a circle with multiple topological defects, we need to specify the base point on the circle.
Once we fix the base point, the vector space on a circle with topological defects $x, y, z, \cdots$ is defined as the vector space on a circle with a single topological defect labeled by the tensor product $((x \otimes y) \otimes z) \otimes \cdots$, see figure \ref{fig: vector space on a circle}.
The vector space on a disjoint union of circles is given by the tensor product of the vector spaces on each circle.

\begin{figure}
\begin{center}
\includegraphics[width = 6cm]{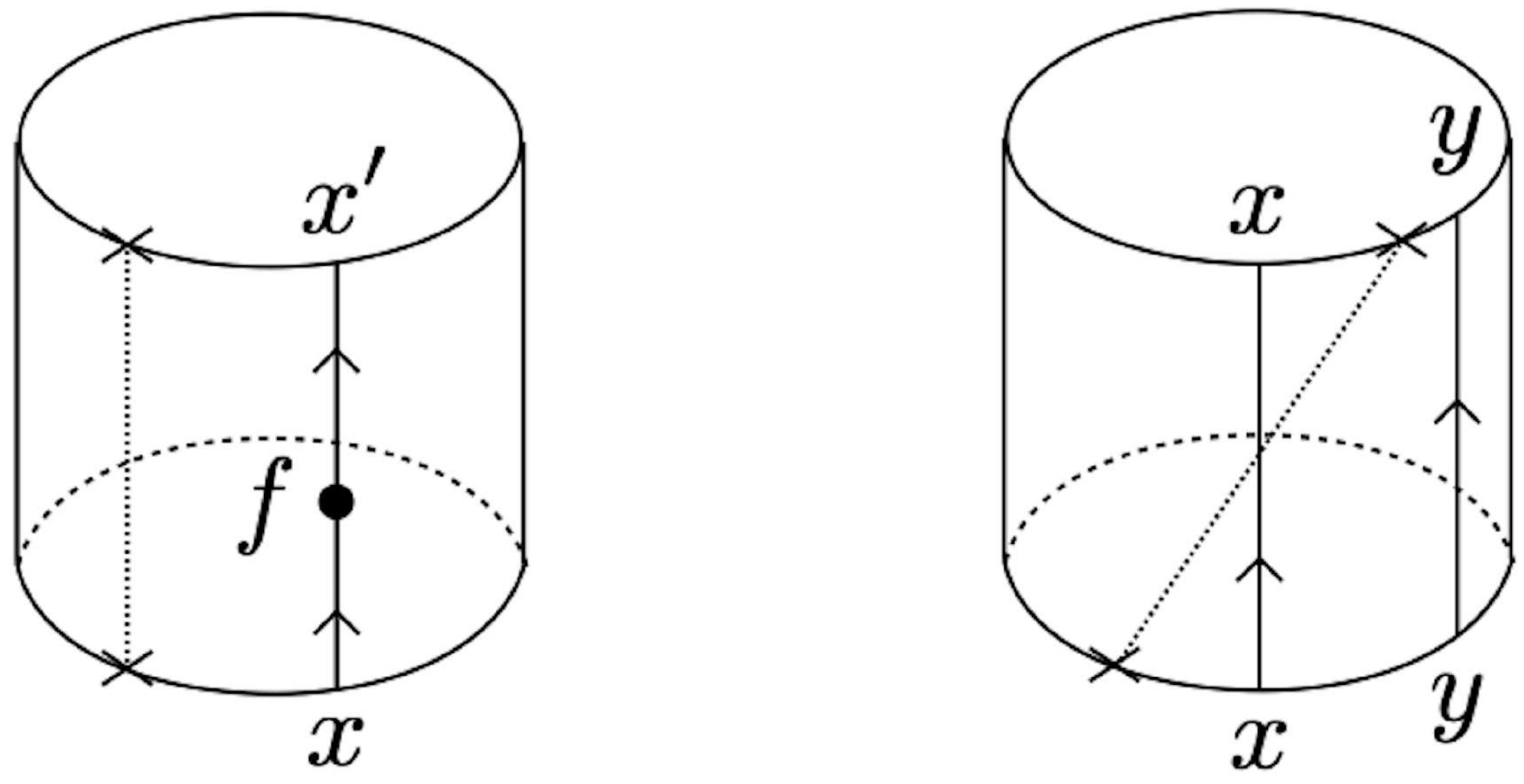}
\caption{There are two types of transition amplitudes for a cylinder. The left figure represents the transition amplitude $Z(f): V_x \rightarrow V_{x^{\prime}}$ corresponding to a topological point operator $f \in \mathop{\mathrm{Hom}}(x, x^{\prime})$. The right figure represents the transition amplitude $X_{xy}: V_{x \otimes y} \rightarrow V_{y \otimes x}$ corresponding to a change of the base point.}
\label{fig: cylinder amplitude}
\end{center}
\end{figure}

Next, we consider transition amplitudes between the vector spaces.
Let us begin with the transition amplitude for a cylinder.
There are two types of cylinder amplitudes as shown in figure \ref{fig: cylinder amplitude}: one is the transition amplitude corresponding to a topological point operator $f \in \mathop{\mathrm{Hom}}(x, x^{\prime})$, and the other is the transition amplitude corresponding to a change of the base point.
The former is denoted by $Z(f): V_x \rightarrow V_{x^{\prime}}$.
In particular, the identity morphism $\mathrm{id}_x$ corresponds to the identity map $\mathrm{id}_{V_x}$.
The latter is denoted by $X_{xy}: V_{x \otimes y} \rightarrow V_{y \otimes x}$ when the base point on a circle with two topological defects $x$ and $y$ is moved from the left of $x$ to the right of $x$.
We call the trajectory of the base point the auxiliary line.

\begin{figure}
\begin{center}
\includegraphics[width = 6cm]{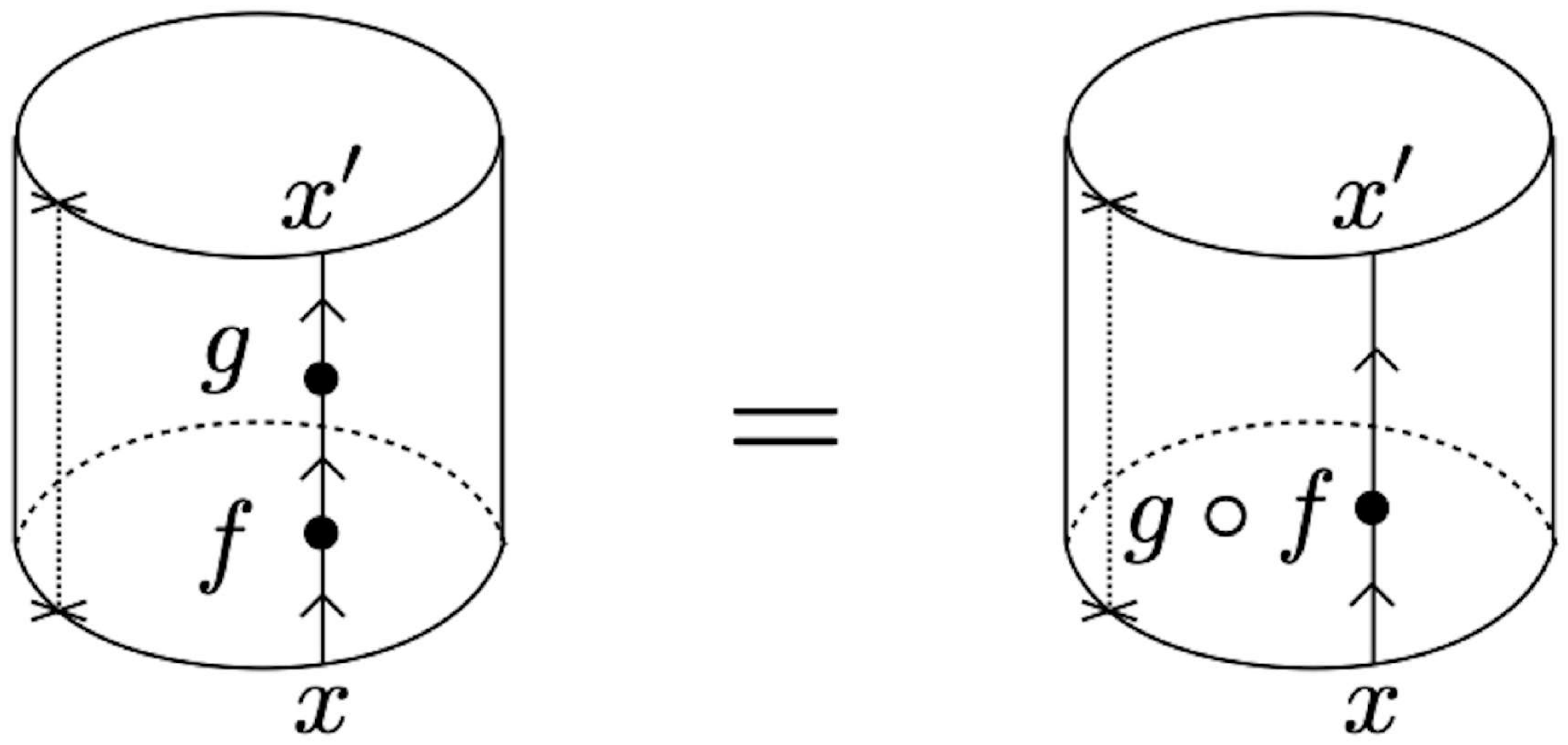}
\caption{The transition amplitude $Z(g \circ f)$ corresponding to the composition of topological point operators agrees with the composition $Z(g) \circ Z(f)$ of the transition amplitudes corresponding to each topological point operator. This is because putting a topological point operator $g \circ f$ can be regarded as putting two topological point operators $f$ and $g$ successively.}
\label{fig: composition on a cylinder}
\end{center}
\end{figure}

\begin{figure}
\begin{center}
\includegraphics[width = 9cm]{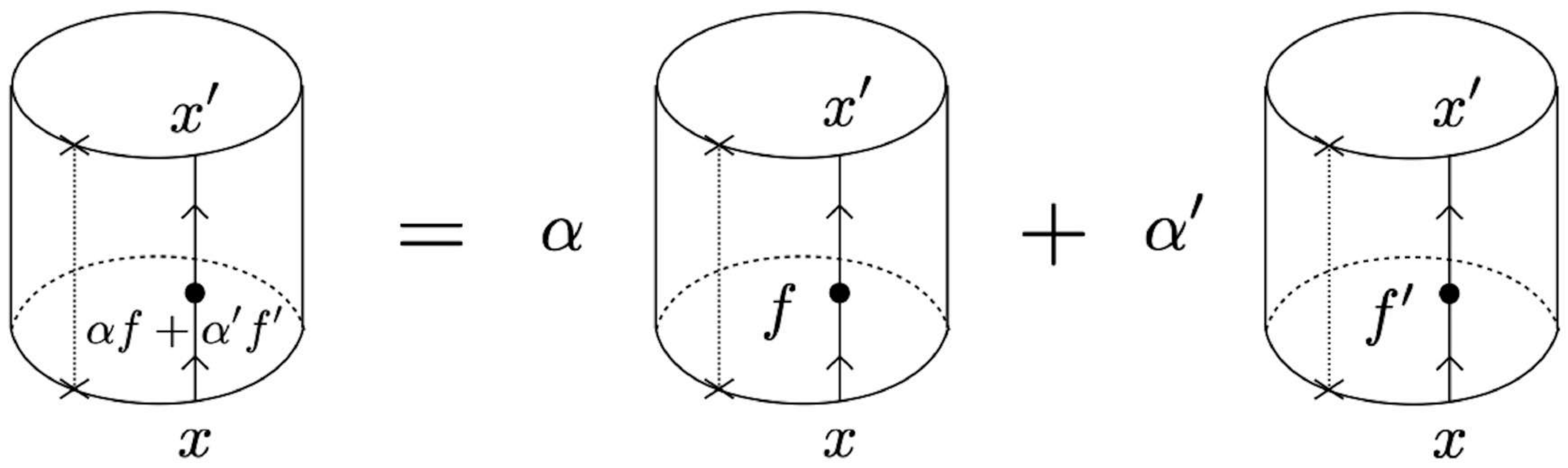}
\caption{The functor $Z: \mathcal{C} \rightarrow \mathrm{Vec}$ is $\mathbb{C}$-linear in morphisms.}
\label{fig: C-linear}
\end{center}
\end{figure}

\begin{figure}[t]
\begin{center}
\includegraphics[width = 8cm]{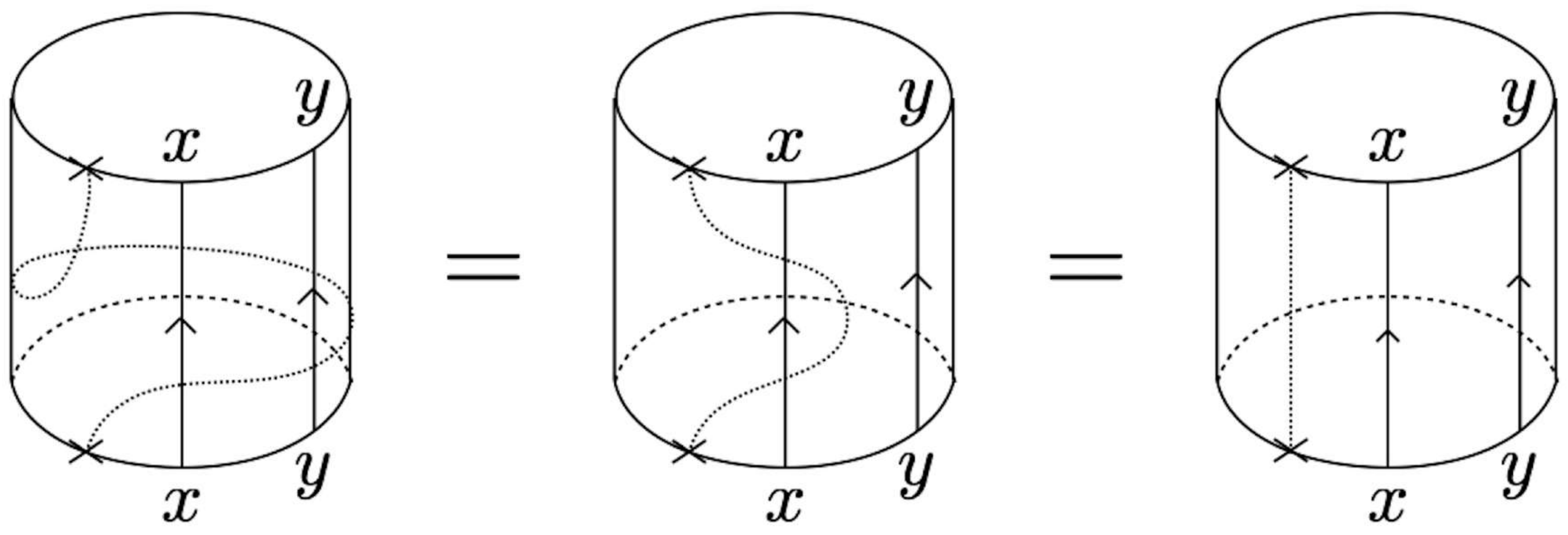}
\caption{Winding the auxiliary line around a circle is equivalent to moving the auxiliary line back and forth, both of which are equivalent to doing nothing. This implies that $X_{yx}$ is the inverse of $X_{xy}$. In particular, $X_{xy}: V_{x \otimes y} \rightarrow V_{y \otimes x}$ is an isomorphism.}
\label{fig: isomorphism X}
\end{center}
\end{figure}

\begin{figure}
\begin{center}
\includegraphics[width = 8cm]{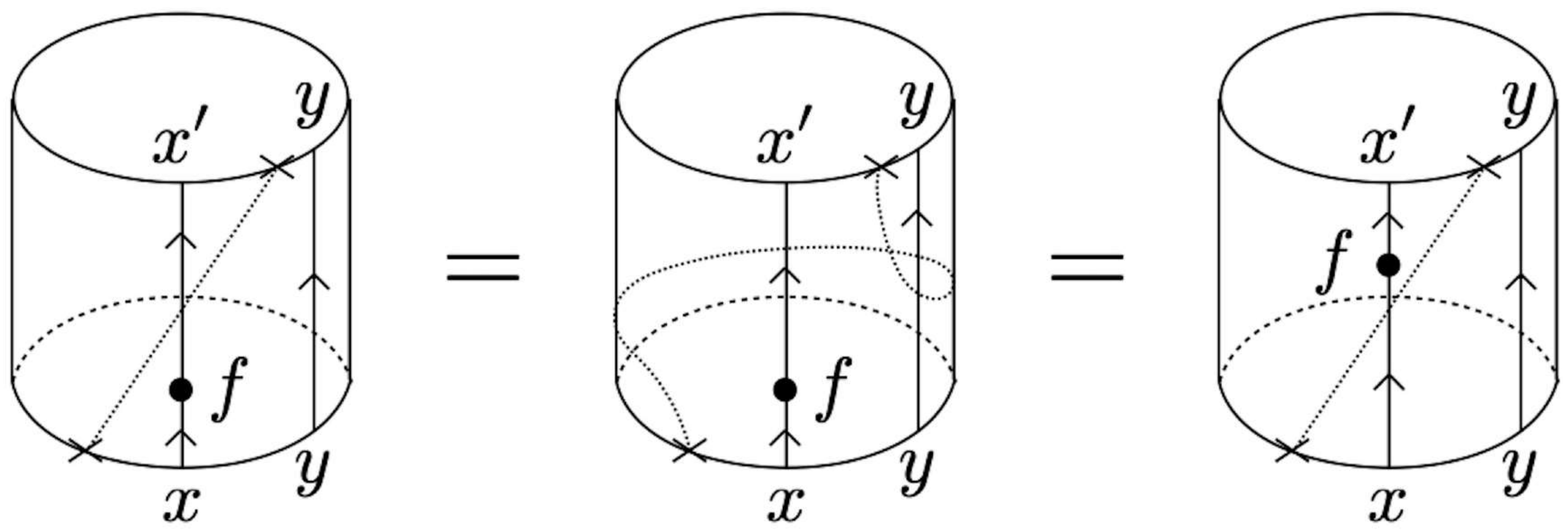}
\caption{The change of the base point $X_{xy}$ commutes with a topological point operator on $x$ because we can prevent the auxiliary line from intersecting with the topological defect $x$ by changing the path of the auxiliary line as shown in the middle figure. This means that the auxiliary line is transparent to a topological point operator on $x$. The auxiliary line is also transparent to a topological point operator on $y$ for the same reason.}
\label{fig: Z(f)X = XZ(f)}
\end{center}
\end{figure}

To make the cylinder amplitude well-defined, we require that the composition of morphisms $f$ and $g$ gives rise to the composition of the transition amplitudes $Z(f)$ and $Z(g)$, see figure \ref{fig: composition on a cylinder}:
\begin{equation}
Z(g \circ f) = Z(g) \circ Z(f).
\tag{O1}
\label{eq: composition on a cylinder}
\end{equation}
This makes $Z$ a functor from $\mathcal{C}$ to $\mathrm{Vec}$ where $Z(x)$ is defined as the vector space $V_x$ for every object $x$ of $\mathcal{C}$.
We demand that this functor is $\mathbb{C}$-linear in morphisms because a linear combination of topological point operators $f, f^{\prime} \in \mathop{\mathrm{Hom}}(x, x^{\prime})$ results in a linear combination of transition amplitudes, see figure \ref{fig: C-linear}:
\begin{equation}
Z(\alpha f + \alpha^{\prime} f^{\prime}) = \alpha Z(f) + \alpha^{\prime} Z(f^{\prime}), \quad \forall \alpha, \alpha^{\prime} \in \mathbb{C}.
\tag{O2}
\label{eq: C-linear}
\end{equation}
Furthermore, we require that the transition amplitudes do not depend on the shape of the auxiliary line if we fix the base points on the initial and final circles.
In particular, the transition amplitudes are invariant under winding the auxiliary line around a cylinder, see figure \ref{fig: isomorphism X}.
This implies that the linear map $X_{xy}: V_{x \otimes y} \rightarrow V_{y \otimes x}$ is an isomorphism, whose inverse is given by $X_{yx}$:
\begin{equation}
X_{xy} = X_{yx}^{-1}.
\tag{O3}
\label{eq: C2}
\end{equation}
We also require that a change of the base point commutes with morphisms $f \in \mathop{\mathrm{Hom}}(x, x^{\prime})$ and $g \in \mathop{\mathrm{Hom}}(y, y^{\prime})$ as shown in figure \ref{fig: Z(f)X = XZ(f)}:
\begin{equation}
\begin{aligned}
Z(\mathrm{id}_y \otimes f) \circ X_{xy} & = X_{x^{\prime}y} \circ Z(f \otimes \mathrm{id}_y),\\
Z(g \otimes \mathrm{id}_x) \circ X_{xy} & = X_{x y^{\prime}} \circ Z(\mathrm{id}_x \otimes g).
\end{aligned}
\tag{O4}
\label{eq: C4}
\end{equation}
These equations indicate that the auxiliary line is transparent to topological point operators.
The last constraint on the cylinder amplitude is that moving a base point across the topological defects $x$ and $y$ at the same time is equivalent to moving a base point across $x$ and $y$ successively.
This leads to the following consistency condition:
\begin{equation}
X_{y, z \otimes x} \circ Z(\alpha_{yzx}) \circ X_{x, y \otimes z} \circ Z(\alpha_{xyz}) = Z(\alpha_{zxy}^{-1}) \circ X_{x \otimes y, z}.
\tag{O5}
\label{eq: coherence}
\end{equation}
In particular, if we choose $x = y = 1$, we obtain $X_{1, x} = \mathrm{id}_{V_x}$, which means that moving the base point across a trivial defect does not affect the transition amplitude.\footnote{We recall that the left and right units are chosen to be the identity morphism.}
This completes the consistency conditions on the cylinder amplitude.

For later convenience, we define the generalized associator $\mathcal{A}_{p \rightarrow p^{\prime}}: V_{p} \rightarrow V_{p^{\prime}}$ as a composition of isomorphisms $X$ and $Z(\alpha)$, where $p$ is a tensor product of any number of objects and $p^{\prime}$ is a cyclic permutation of $p$ with arbitrary parentheses.
For example, the generalized associator $\mathcal{A}_{(x \otimes y) \otimes z \rightarrow (z \otimes x) \otimes y}$ is given by both sides of eq. (\ref{eq: coherence}).
In general, there are many ways to construct an isomorphism from $V_{p}$ to $V_{p^{\prime}}$ only from $X$ and $Z(\alpha)$.
However, they give rise to the same isomorphism due to eq. (\ref{eq: coherence}).
This means that the generalized associator $\mathcal{A}_{p \rightarrow p^{\prime}}$ is unique, although it has many distinct-looking expressions.

\begin{figure}
\begin{center}
\includegraphics[width = 11cm]{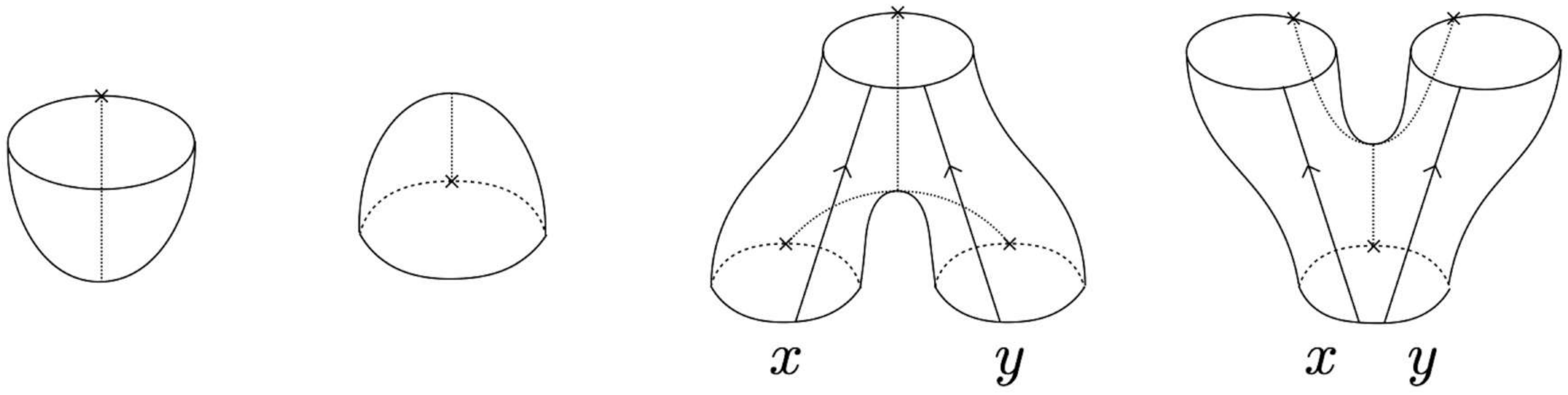}
\caption{The above figures represent the basic elements of the transition amplitude for a general surface: from left to right, the unit, the counit, the multiplication, and the comultiplication.}
\label{fig: basic elements}
\end{center}
\end{figure}

The transition amplitude for a general surface can be constructed from the cylinder amplitude and the following four basic elements: the unit $i: \mathbb{C} \rightarrow V_1$, the counit $\epsilon: V_1 \rightarrow \mathbb{C}$, the multiplication $M_{xy}: V_x \otimes V_y \rightarrow V_{x \otimes y}$, and the comultiplication $\Delta_{xy}: V_{x \otimes y} \rightarrow V_x \otimes V_y$, see also figure \ref{fig: basic elements} for diagrammatic representations.
For unitary quantum field theories, the counit $\epsilon$ and the comultiplication $\Delta_{xy}$ are the adjoints of the unit $i$ and the multiplication $M_{xy}$ respectively (with respect to the non-degenerate pairing (\ref{eq: non-degenerate})):
\begin{equation}
\epsilon = i^{\dagger}, \quad \Delta_{xy} = M_{xy}^{\dagger}.
\end{equation}
More generally, a transition amplitude changes to its adjoint when the corresponding surface is turned upside down, or equivalently when the time direction is reversed.

Any two-dimensional surface $\Sigma$ can be obtained by successively gluing cylinders and the basic elements shown in figure \ref{fig: basic elements}.
The transition amplitude for a surface $\Sigma$ is defined as the composition of the linear maps corresponding to cylinders and the basic elements that appear in the decomposition of $\Sigma$.
However, the way to decompose $\Sigma$ is not unique in general.
Therefore, we need to impose consistency conditions so that the transition amplitudes do not depend on a decomposition.
Such consistency conditions can be listed as follows \cite{BT2018}:
\begin{description}
\item [Non-degenerate pairing] (figure \ref{fig: non-degenerate})\\
\begin{figure}
\begin{center}
\includegraphics[width = 3.6cm]{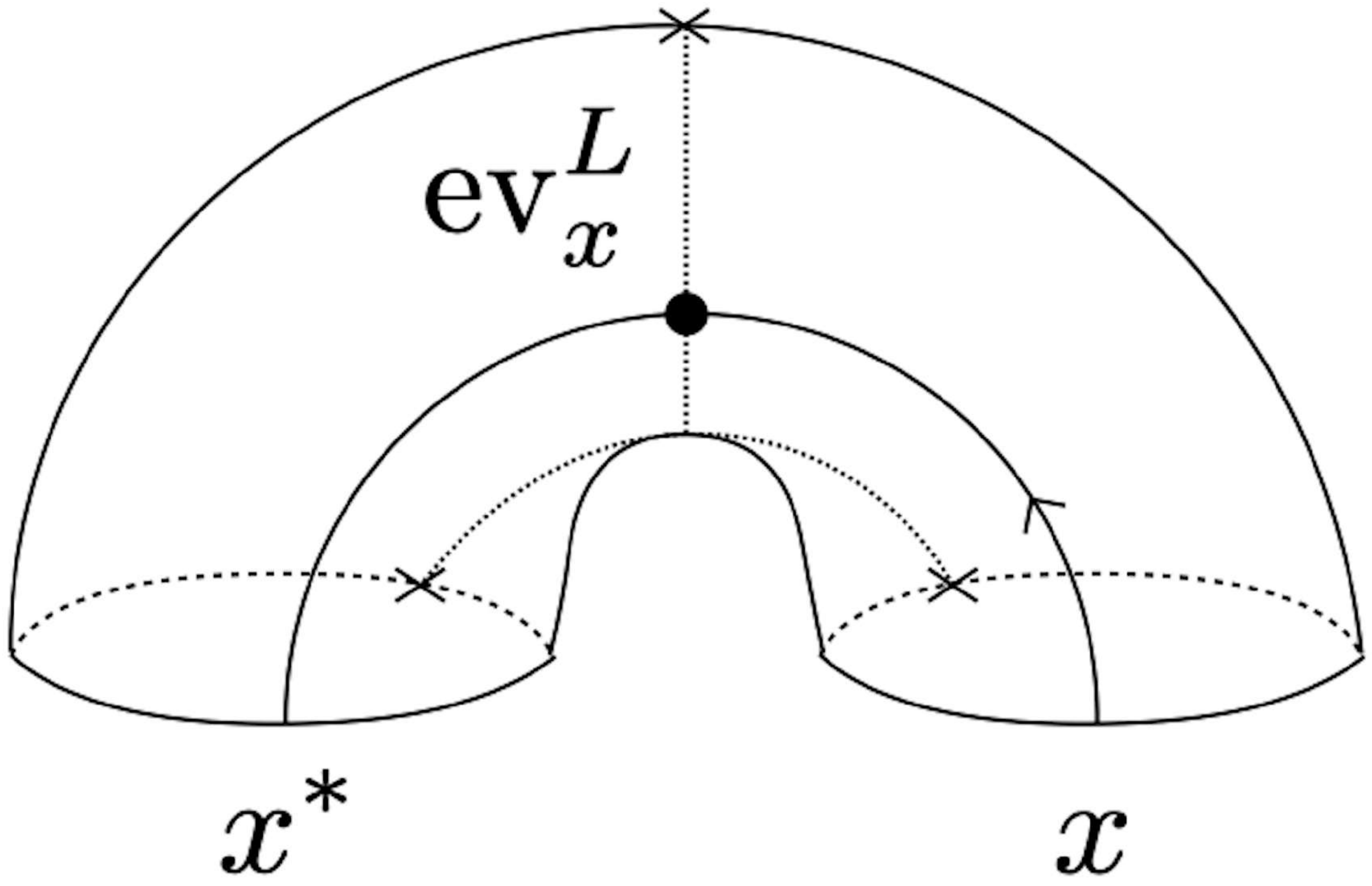}
\caption{The composition of the multiplication and the counit defines a non-degenerate pairing, which can be used to regard $V_{x^*}$ as the dual vector space of $V_x$.}
\label{fig: non-degenerate}
\end{center}
\end{figure}
\begin{equation}
\text{The pairing }\epsilon \circ Z(\mathrm{ev}_x^L) \circ M_{x^*, x}: V_{x^*} \otimes V_x \rightarrow \mathbb{C}\text{ is non-degenerate}.
\tag{O6}
\label{eq: non-degenerate}
\end{equation}
This non-degenerate pairing is related to the inner product of the Hilbert space via $CPT$ conjugation \cite{KT2017, KTY2017}.
\item [Unit constraint] (figure \ref{fig: unit constraint})\\
\begin{figure}
\begin{center}
\includegraphics[width = 8cm]{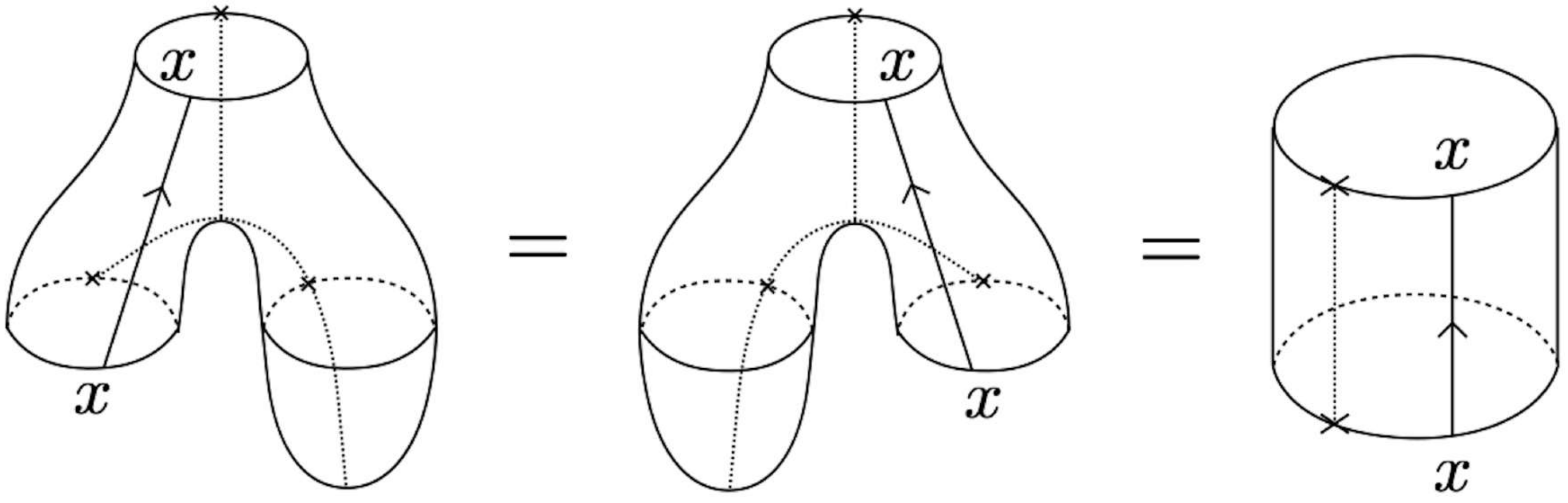}
\caption{The multiplication $M$ has the unit $i(1) \in V_1$.}
\label{fig: unit constraint}
\end{center}
\end{figure}
The unit $i$ acts as the unit of the multiplication $M$:
\begin{equation}
M_{1x}(i(1) \otimes v_x) = M_{x1}(v_x \otimes i(1)) = v_x, \quad \forall v_x \in V_x.
\tag{O7}
\label{eq: unit constraint}
\end{equation}
\item [Associativity] (figure \ref{fig: associativity constraint})\\
\begin{figure}
\begin{center}
\includegraphics[width = 8cm]{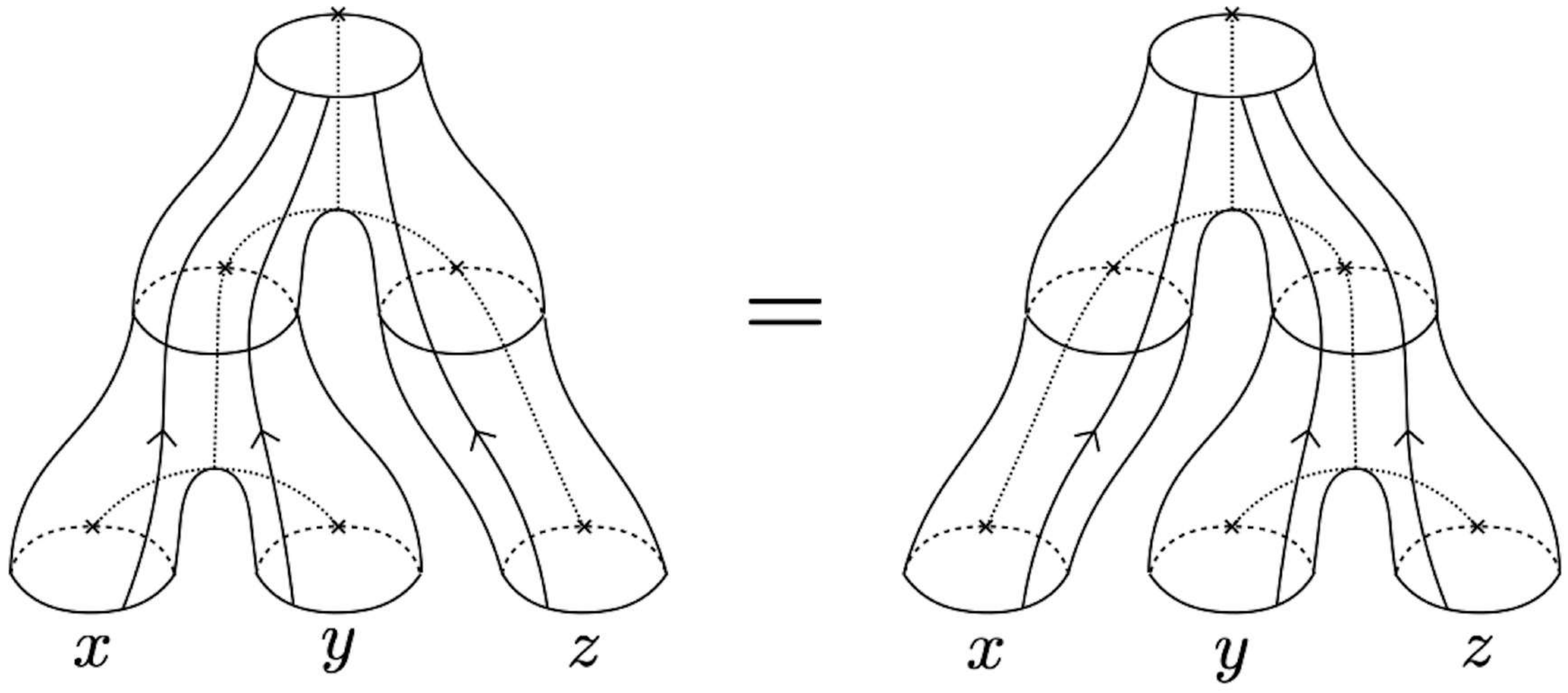}
\caption{The multiplication $M$ is associative up to associator. The topological point operators corresponding to associators are omitted in the above figure.}
\label{fig: associativity constraint}
\end{center}
\end{figure}
The multiplication $M$ is associative up to associator:
\begin{equation}
Z(\alpha_{xyz}) \circ M_{x \otimes y, z} \circ (M_{xy} \otimes \mathrm{id}_{V_z}) = M_{x, y \otimes z} \circ (\mathrm{id}_{V_x} \otimes M_{yz}) \circ \alpha_{V_x V_y V_z}^{\mathrm{Vec}},
\tag{O8}
\label{eq: associativity constraint}
\end{equation}
where $\alpha_{V_x V_y V_z}^{\mathrm{Vec}}$ is the associator of the category of vector spaces $\mathrm{Vec}$, which is chosen to be the identity map $\alpha_{V_x V_y V_z}^{\mathrm{Vec}} = \mathrm{id}_{V_x \otimes V_y \otimes V_z}$ by identifying $(V_x \otimes V_y) \otimes V_z$ with $V_x \otimes (V_y \otimes V_z)$.
\item [Twisted commutativity] (figure \ref{fig: twisted commutativity})\\
\begin{figure}[t]
\begin{center}
\includegraphics[width = 6cm]{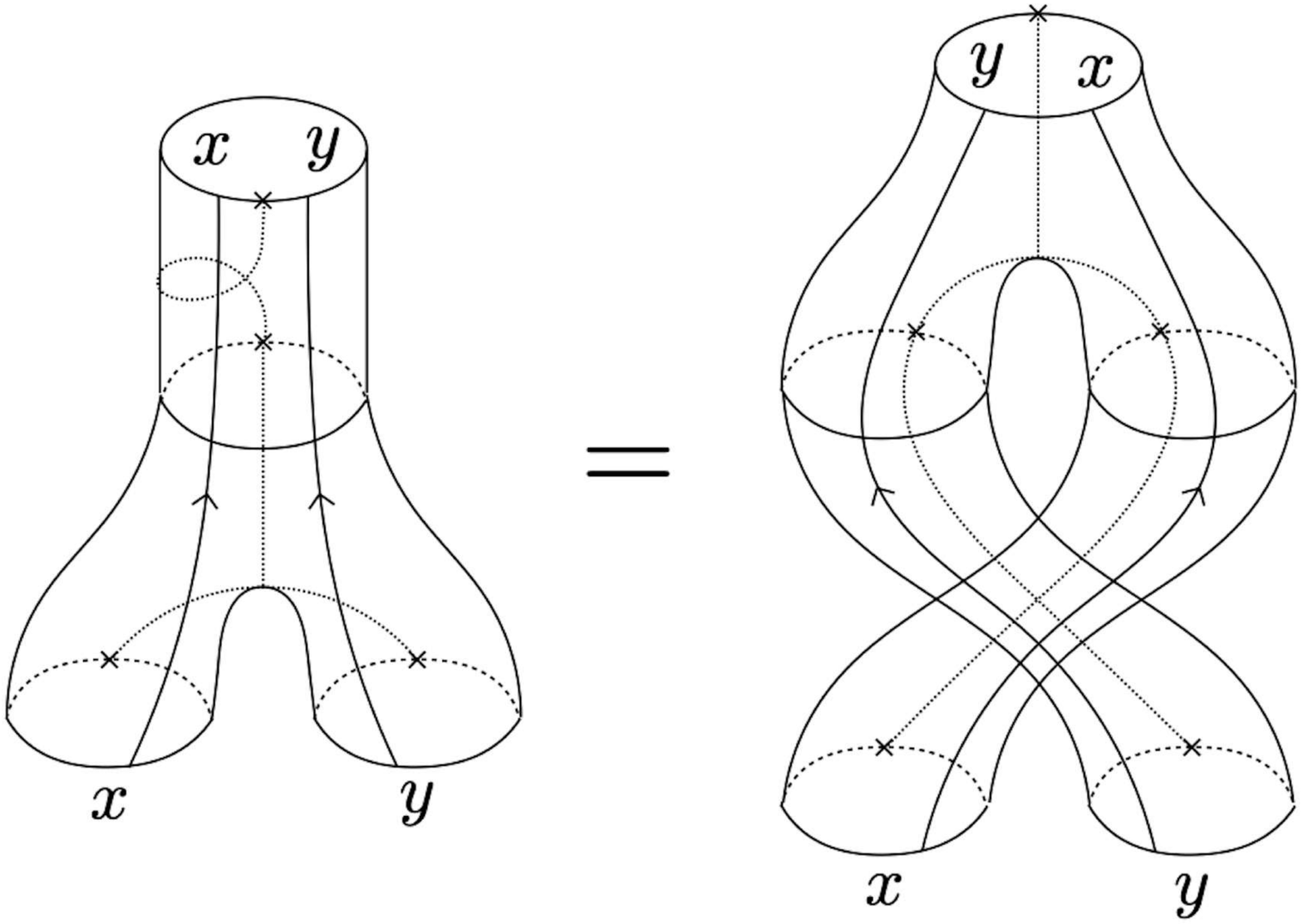}
\caption{The multiplication $M$ is commutative up to isomorphism $X_{xy}$.}
\label{fig: twisted commutativity}
\end{center}
\end{figure}
The multiplication $M$ is commutative up to the change of the base point:
\begin{equation}
M_{yx} \circ c_{V_x V_y} = X_{xy} \circ M_{xy},
\tag{O9}
\label{eq: twisted commutativity}
\end{equation}
where $c_{V_x V_y}: V_x \otimes V_y \rightarrow V_y \otimes V_x$ is the symmetric braiding of $\mathrm{Vec}$, which is defined by $c_{V_x V_y}(v_x \otimes v_y) := v_y \otimes v_x$ for any $v_x \in V_x$ and $v_y \in V_y$.
\item [Commutativity of topological point operators and the multiplication] (figure \ref{fig: commutativity of a morphism and the multiplication})\\
\begin{figure}
\begin{center}
\includegraphics[width = 6cm]{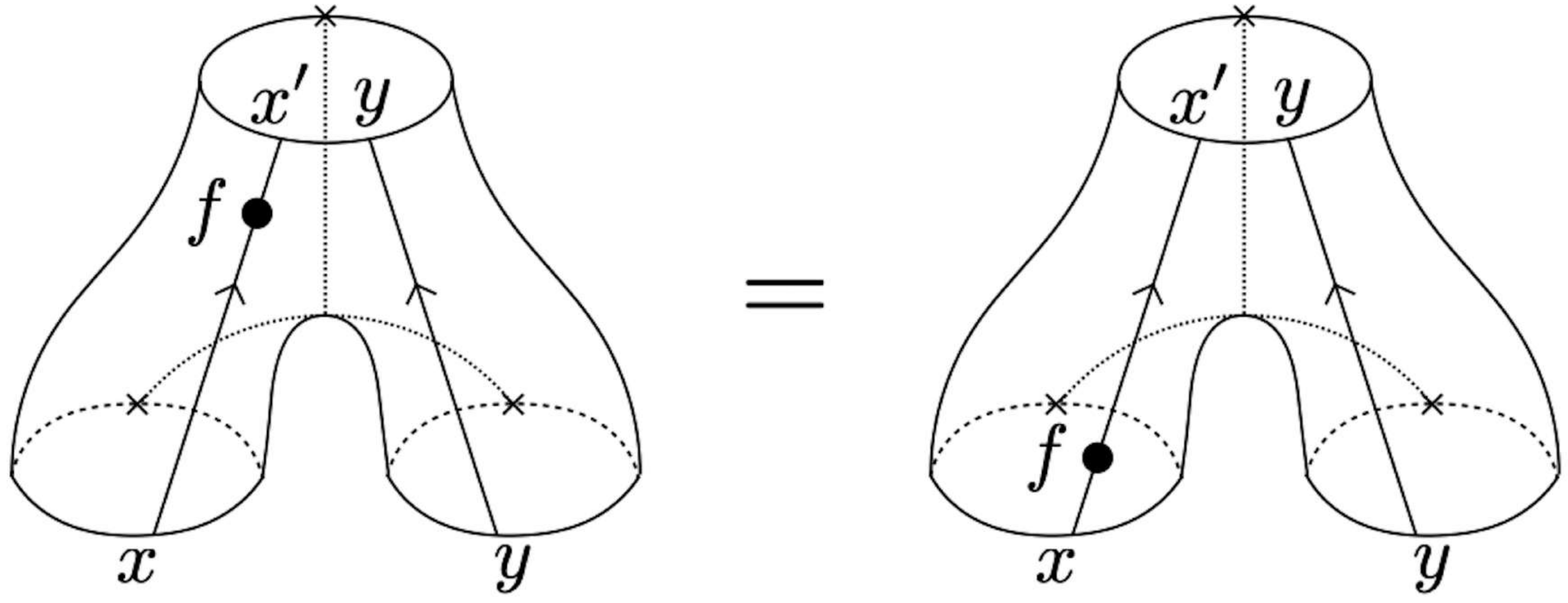}
\caption{The transition amplitude does not depend on whether a topological point operator $f \in \mathop{\mathrm{Hom}}(x, x^{\prime})$ is inserted before or after the multiplication $M_{xy}$. The same holds for $g \in \mathop{\mathrm{Hom}}(y, y^{\prime})$.}
\label{fig: commutativity of a morphism and the multiplication}
\end{center}
\end{figure}
Topological point operators commute with the multiplication $M$:
\begin{equation}
\begin{aligned}
M_{x^{\prime}y} \circ (Z(f) \otimes \mathrm{id}_{V_y}) & = Z(f \otimes \mathrm{id}_y) \circ M_{xy}, \quad \forall f \in \mathop{\mathrm{Hom}}(x, x^{\prime}),\\
M_{xy^{\prime}} \circ (\mathrm{id}_{V_x} \otimes Z(g)) & = Z(\mathrm{id}_x \otimes g) \circ M_{xy}, \quad \forall g \in \mathop{\mathrm{Hom}}(y, y^{\prime}).
\end{aligned}
\tag{O10}
\label{eq: commutativity of a morphism and the multiplication}
\end{equation}
\item [The uniqueness of the multiplication] (figure \ref{fig: uniqueness of the multiplication})\\
\begin{figure}[t]
\begin{center}
\includegraphics[width = 6cm]{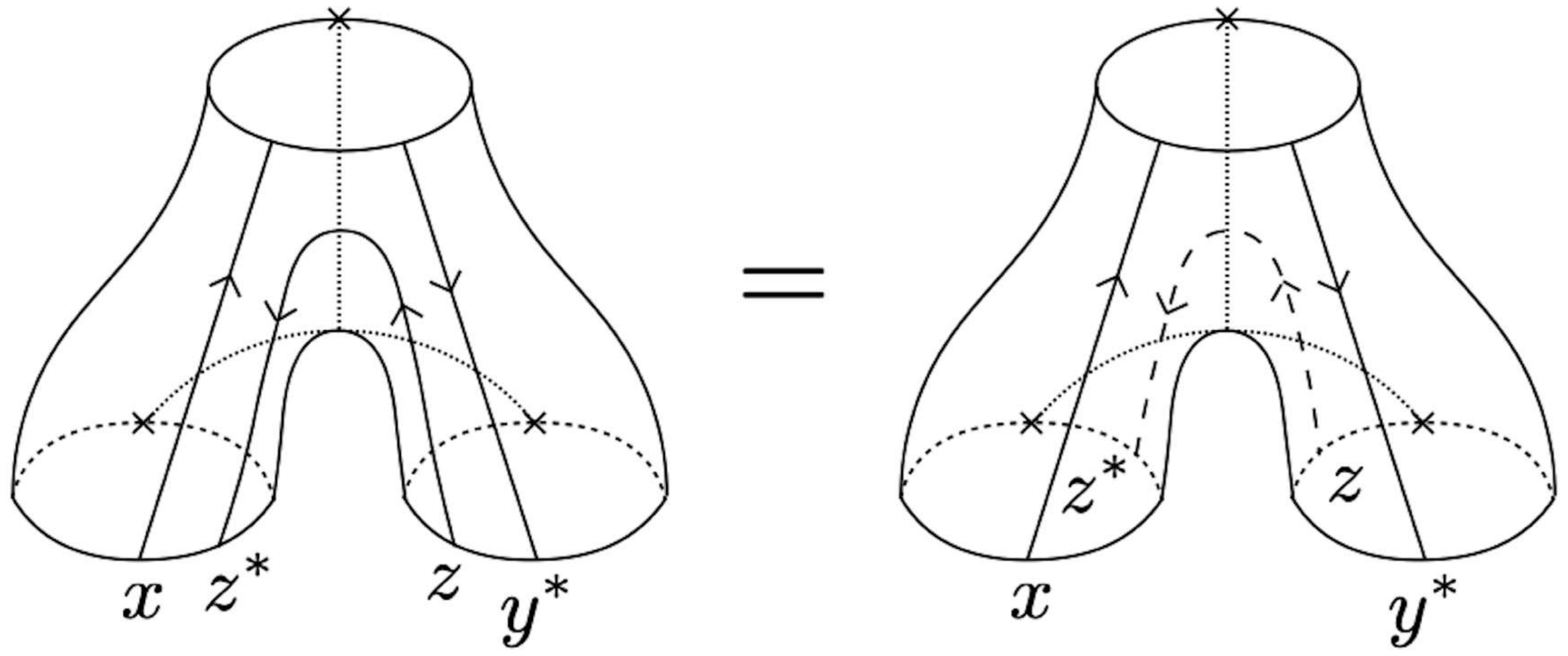}
\caption{
The auxiliary line does not intersect with topological defects on the left-hand side, while it intersects with topological defect $z$ three times on the right-hand side.
They give the same transition amplitude.}
\label{fig: uniqueness of the multiplication}
\end{center}
\end{figure}
The multiplication $M$ uniquely determines a map for topologically equivalent configurations of topological defects on the pants diagram:
\begin{equation}
\begin{aligned}
& Z((\mathrm{id}_x \otimes \mathrm{ev}_z^L) \otimes \mathrm{id}_{y^*}) \circ \mathcal{A}_{(x \otimes z^*) \otimes (z \otimes y^*) \rightarrow (x \otimes (z^* \otimes z)) \otimes y^*} \circ M_{x \otimes z^*, z \otimes y^*}\\
= ~ & Z(\mathrm{id}_{x \otimes y^*} \otimes \mathrm{ev}_z^R) \circ \mathcal{A}_{(z^* \otimes x) \otimes (y^* \otimes z) \rightarrow (x \otimes y^*) \otimes (z \otimes z^*)} \circ M_{z^* \otimes x, y^* \otimes z} \circ (X_{x, z^*} \otimes X_{z, y^*}).
\end{aligned}
\tag{O11}
\label{eq: uniqueness of the multiplication}
\end{equation}
\item [Consistency on the torus] (figure \ref{fig: consistency on the torus})\\
\begin{figure}
\begin{center}
\includegraphics[width = 7cm]{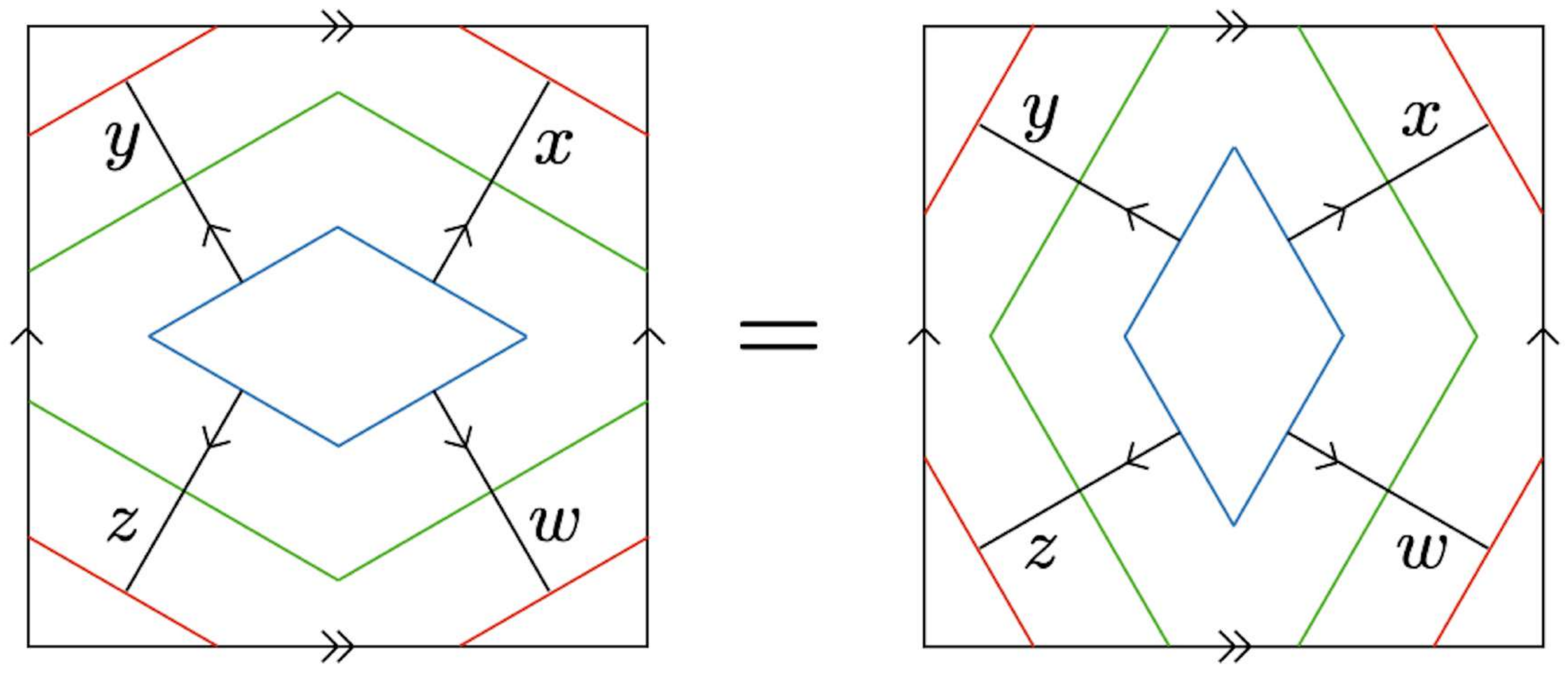}
\caption{The transition amplitude for a twice-punctured torus does not depend on a time function on it. The blue lines represent the initial time-slice and the red lines represent the final time-slice. A time-slice at an intermediate time, which consists of two disjoint circles, is represented by the green lines. Two different ways to split the initial circle into the intermediate circles give rise to the same transition amplitude. The auxiliary lines are omitted in the above figure.}
\label{fig: consistency on the torus}
\end{center}
\end{figure}
Different ways to decompose a twice-punctured torus give the same transition amplitude:
\begin{equation}
\begin{aligned}
& \mathcal{A}_{(y \otimes x) \otimes (w \otimes z) \rightarrow (x \otimes w) \otimes (z \otimes y)} \circ M_{y \otimes x, w \otimes z} \circ (X_{xy} \otimes X_{zw}) \circ \Delta_{x \otimes y, z \otimes w}\\
= ~ & M_{x \otimes w, z \otimes y} \circ (X_{wx} \otimes X_{yz}) \circ \Delta_{w \otimes x, y \otimes z} \circ \mathcal{A}_{(x \otimes y) \otimes (z \otimes w) \rightarrow (w \otimes x) \otimes (y \otimes z)}.
\end{aligned}
\tag{O12}
\label{eq: consistency on the torus}
\end{equation}
\end{description}
In summary, a two-dimensional bosonic oriented TQFT with fusion category symmetry is given by a quadruple $(Z, X, M, i)$ that satisfies the consistency conditions (\ref{eq: composition on a cylinder})--(\ref{eq: consistency on the torus}).\footnote{In \cite{BT2018}, there is an additional consistency condition called the cyclic symmetry of the multiplication. However, this condition turns out to be satisfied as a consequence of (\ref{eq: non-degenerate}) and (\ref{eq: associativity constraint}).}

\subsection{Bosonic fusion category SPT phases without time-reversal symmetry}
\label{sec: Bosonic fusion category SPT phases without time-reversal symmetry}
In this section, we classify bosonic SPT phases with fusion category symmetry $\mathcal{C}$ by solving the consistency conditions (\ref{eq: composition on a cylinder})--(\ref{eq: consistency on the torus}).
To focus on the solutions that describe SPT phases, we further impose the condition that the ground state is unique on a circle without topological defects, i.e. $\mathop{\mathrm{dim}} V_1 = 1$.
More generally, the dimension of the vector space $V_x$ for an SPT phase is equal to the quantum dimension of the object $x$
\begin{equation}
\mathop{\mathrm{dim}} V_x = \mathop{\mathrm{dim}} x,
\label{eq: dim = qdim}
\end{equation}
which can be seen from modular invariance of the partition function on a torus \cite{CLSWY2019}.

For SPT phases, the unit $i: \mathbb{C} \rightarrow V_1$ is an isomorphism because it is a non-zero linear map between one-dimensional vector spaces.
In particular, $i$ is unitary if the partition function on a sphere is unity:
\begin{equation}
i^{\dagger} \circ i = 1.
\label{eq: unitary unit}
\end{equation}
In the following, we assume that this condition is satisfied.
Furthermore, we assume that the multiplication $M_{xy}: V_{x} \otimes V_y \rightarrow V_{x \otimes y}$ is also unitary:
\begin{equation}
M_{xy}^{\dagger} = M_{xy}^{-1} \Leftrightarrow \adjincludegraphics[valign = c, width = 1.4cm]{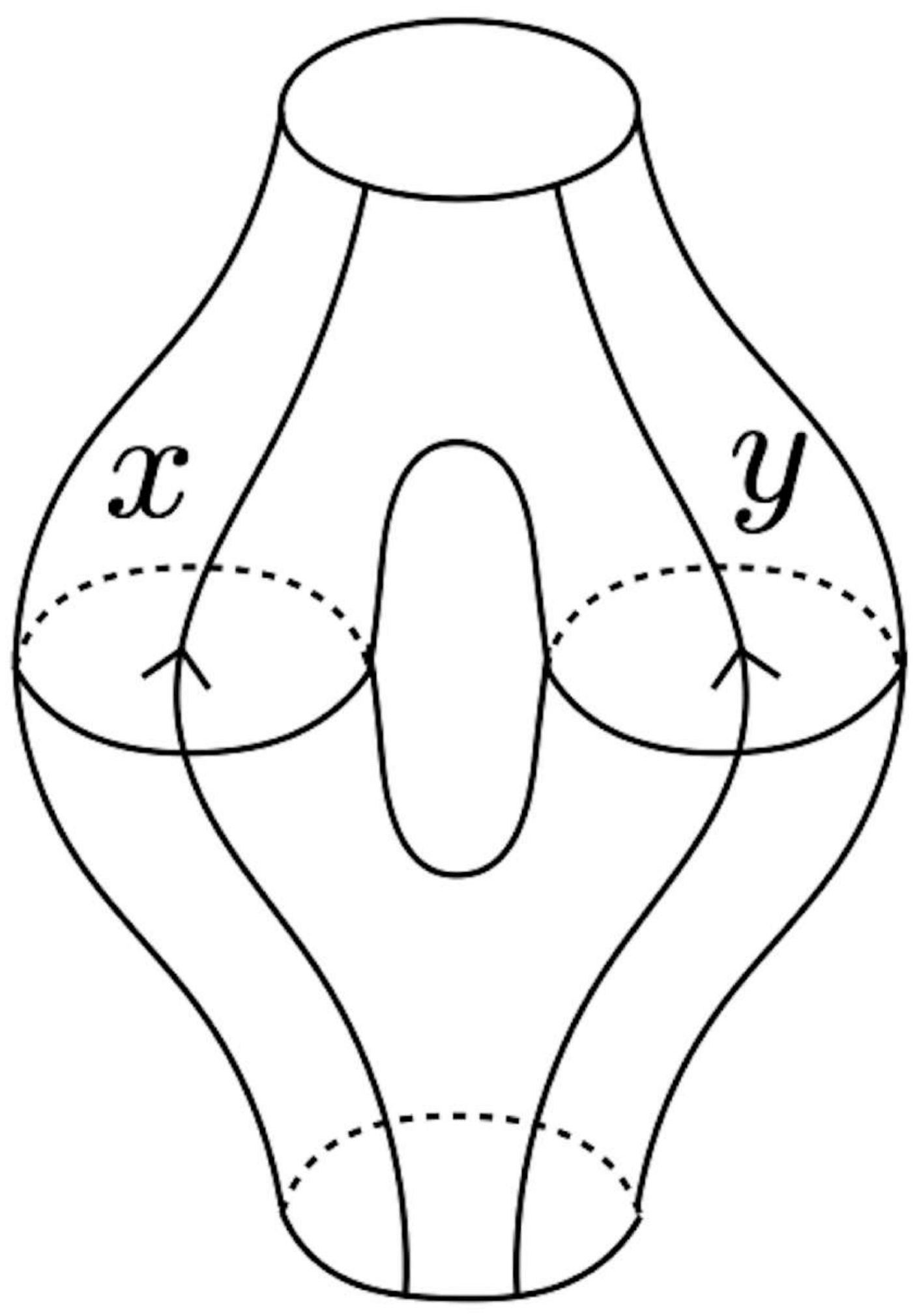} ~ = ~ \adjincludegraphics[valign = c, width = 1.4cm]{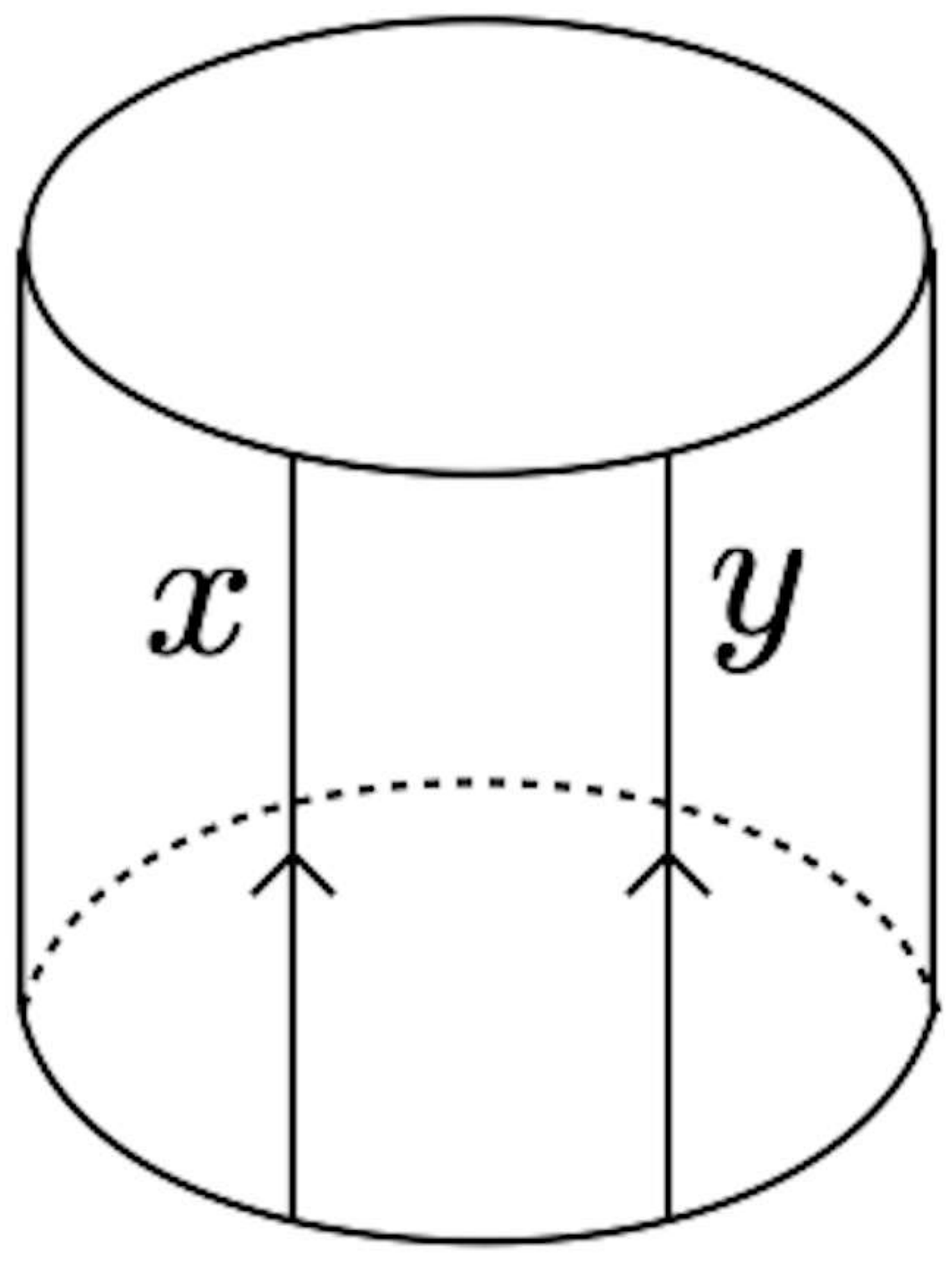}.
\label{eq: unitary M}
\end{equation}
This equation reduces to eq. (\ref{eq: dim = qdim}) if we form a loop of $x$ by choosing $y = x^{*}$ and close the punctures with a cap and its adjoint.

Now, we solve the consistency conditions (\ref{eq: composition on a cylinder})--(\ref{eq: consistency on the torus}) under the constraints (\ref{eq: unitary unit}) and (\ref{eq: unitary M}).
We first notice that $Z: \mathcal{C} \rightarrow \mathrm{Vec}$ is a functor by (\ref{eq: composition on a cylinder}), where an object $x$ is mapped to a vector space $V_x$ and a morphism $f \in \mathop{\mathrm{Hom}}(x, y)$ is mapped to a linear map $Z(f): V_x \rightarrow V_y$.
We recall that the functor $Z$ is additive in objects in the sense that the direct sum of objects $x$ and $y$ is mapped to the direct sum of vector spaces $V_x$ and $V_y$.
This functor is also $\mathbb{C}$-linear in morphisms due to (\ref{eq: C-linear}).
Furthermore, (\ref{eq: unit constraint}) and (\ref{eq: associativity constraint}) indicate that the triple $(Z, M, i)$ is a monoidal functor where the naturality of the isomorphism $M_{xy}$ follows from (\ref{eq: commutativity of a morphism and the multiplication}).
Thus, $(Z, M, i)$ is a tensor functor from $\mathcal{C}$ to $\mathrm{Vec}$, namely a fiber functor.
In this way, we can extract the data of a fiber functor from an SPT phase with fusion category symmetry.
Conversely, we can show that the data of a fiber functor $(Z, M, i)$ is sufficient to construct a solution of (\ref{eq: composition on a cylinder})--(\ref{eq: consistency on the torus}) as we will see below.
In other words, the other consistency conditions are automatically satisfied when $(Z, M, i)$ is a fiber functor.

Since $M_{xy}$ is unitary (\ref{eq: unitary M}), the isomorphism $X_{xy}$ is determined by (\ref{eq: twisted commutativity}) as
\begin{equation}
X_{xy} = M_{yx} \circ c_{V_x V_y} \circ M_{xy}^{-1}.
\label{eq: X = McM^{-1}}
\end{equation}
This satisfies (\ref{eq: C2}) because $c_{V_x V_y}$ is the symmetric braiding of $\mathrm{Vec}$.
Furthermore, the naturality of $M_{xy}$ and $c_{V_x V_y}$ implies that $X_{xy}$ is also natural (\ref{eq: C4}).
The non-degeneracy of the pairing (\ref{eq: non-degenerate}) is guaranteed by the equality
\begin{equation}
\adjincludegraphics[valign = c, width = 1.25cm]{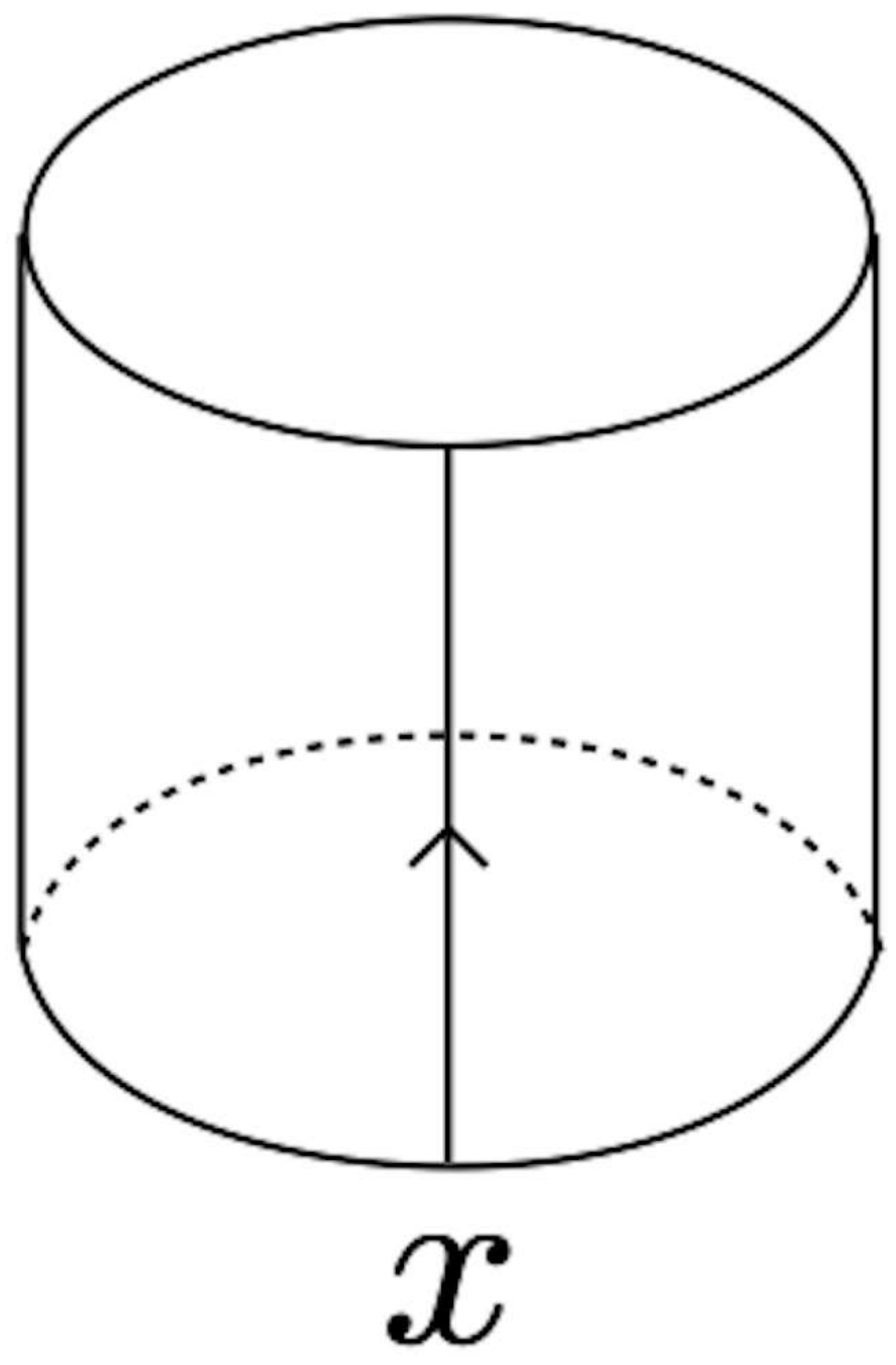} ~ = ~ 
\adjincludegraphics[valign = c, width = 2.1cm]{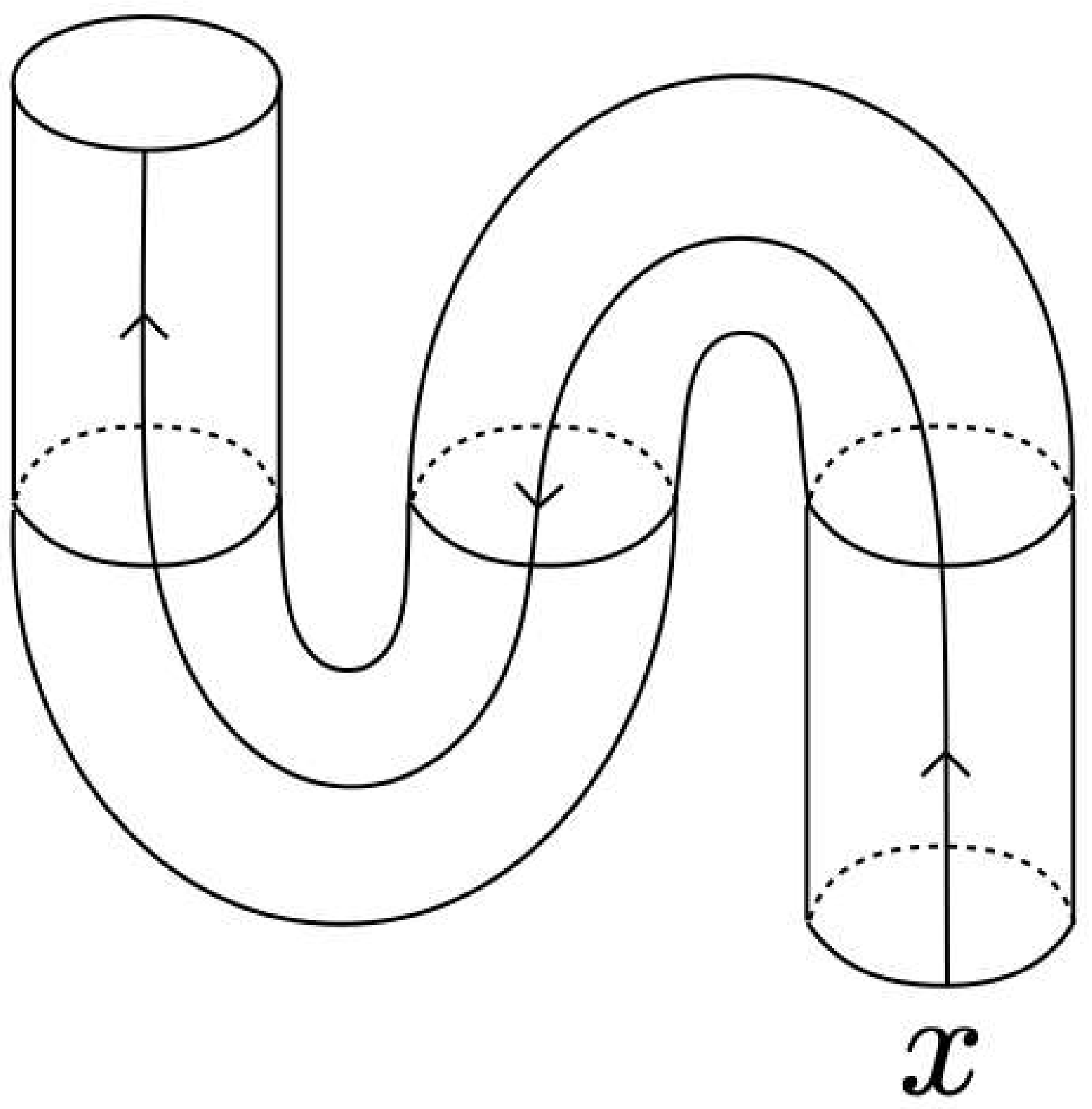},
\end{equation}
which follows from the Frobenius relation represented by the following commutative diagram:
\begin{equation}
\begin{tikzcd}[row sep = 12pt]
V_1 \otimes V_x \arrow[r, "M_{1x}"] \arrow[dd, "Z(\mathrm{coev}_x^L)"'] & V_{1 \otimes x} \arrow[r, "Z(l_x)"] \arrow[rd, "Z(\mathrm{coev}_x^L \otimes \mathrm{id}_x)"'] & V_x \arrow[r, "Z(r_x^{-1})"] & V_{x \otimes 1} \arrow[r, "M_{x1}^{-1}"] & V_x \otimes V_1\\
& & V_{x \otimes x^* \otimes x} \arrow[ru, "Z(\mathrm{id}_x \otimes \mathrm{ev}_x^L)"'] & & \\
V_{x \otimes x^*} \otimes V_x \arrow[rru, "M_{x \otimes x^*, x}"] \arrow[rr, "M_{x, x^*}^{-1}"'] & & V_x \otimes V_{x^*} \otimes V_x \arrow[rr, "M_{x^*, x}"'] & & V_x \otimes V_{x^* \otimes x} \arrow[llu, "M_{x, x^* \otimes x}"'] \arrow[uu, "Z(\mathrm{ev}_x^L)"']
\end{tikzcd}
\end{equation}

To proceed further, we notice that a linear map from $V_{x_1 \otimes y_1 \otimes \cdots} \otimes V_{x_2 \otimes y_2 \otimes \cdots} \otimes \cdots$ to $V_{x^{\prime}_1 \otimes y^{\prime}_1 \otimes \cdots} \otimes V_{x^{\prime}_2 \otimes y^{\prime}_2 \otimes \cdots} \otimes \cdots$ that consists only of $M$ and $c$ is uniquely determined by the permutation of the topological defect $(x_1, y_1, \cdots, x_2, y_2, \cdots) \rightarrow (x^{\prime}_1, y^{\prime}_1, \cdots, x^{\prime}_2, y^{\prime}_2, \cdots)$ and the change of the parentheses.
The same holds for a linear map that consists only of $M$, $c$, $Z(\alpha)$, and $X$ because $Z(\alpha)$ and $X$ can be expressed in terms of $M$ and $c$ due to (\ref{eq: associativity constraint}) and (\ref{eq: twisted commutativity}).
This shows (\ref{eq: coherence}) and (\ref{eq: consistency on the torus}).
This also reduces the consistency condition (\ref{eq: uniqueness of the multiplication}) to the following commutative diagram:
\begin{equation}
\begin{tikzcd}
& V_1 &\\
V_{z^* \otimes z} \arrow[ru, "Z(\mathrm{ev}_z^L)"] & \mathbb{C} \arrow[u, "i"] & V_{z \otimes z^*} \arrow[lu, "Z(\mathrm{ev}_z^R)"'] \\
V_{z^*} \otimes V_z \arrow[u, "M_{z^*, z}"] \arrow[ru, "\mathrm{ev}_{V_z}^L"] \arrow[rr, "c_{V_{z^*} V_z}"'] & & V_z \otimes V_{z^*} \arrow[lu, "\mathrm{ev}_{V_z}^R"'] \arrow[u, "M_{z, z^*}"']
\end{tikzcd}
\label{eq: C10 commutative diagram}
\end{equation}
The commutativity of the left square in the above diagram defines the evaluation morphism $\mathrm{ev}_{V_z}^L: V_{z^{*}} \otimes V_z \rightarrow \mathbb{C}$, which is the non-degenerate pairing (\ref{eq: non-degenerate}).
Accordingly, we can consider $V_{z^*}$ as the dual vector space of $V_z$.
If we denote a basis of $V_z$ as $\{ e_i \mid i = 1, 2, \cdots, \mathop{\mathrm{dim}} V_z \}$, the dual basis $\{ e^i \mid i = 1, 2, \cdots, \mathop{\mathrm{dim}} V_{z^*} = \mathop{\mathrm{dim}} V_z \}$ of $V_{z^*}$ is defined by
\begin{equation}
\mathrm{ev}_{V_z}^L(e^i \otimes e_j) = \delta^i_j,
\label{eq: ev dual basis}
\end{equation}
where $\delta^i_j$ is the Kronecker delta.
Equation (\ref{eq: zigzag axiom}) implies that the coevaluation morphism $\mathrm{coev}_{V_z}^L: \mathbb{C} \rightarrow V_z \otimes V_{z^*}$ is given by the embedding
\begin{equation}
\mathrm{coev}_{V_z}^L (1) = \sum_{i = 1}^{\mathop{\mathrm{dim}} V_z} e_i \otimes e^i,
\label{eq: coev dual basis}
\end{equation}
whose adjoint agrees with the right evaluation morphism $\mathrm{ev}_{V_z}^R$ defined by the commutativity of the right square in the diagram (\ref{eq: C10 commutative diagram}): 
\begin{equation}
\mathrm{ev}_{V_z}^R (e_i \otimes e^j) = \delta_i^j.
\label{eq: ev R dual basis}
\end{equation}
From eqs. (\ref{eq: ev dual basis}) and (\ref{eq: ev R dual basis}), we find that the bottom triangle in (\ref{eq: C10 commutative diagram}) is commutative.

In summary, the solutions of the consistency conditions (\ref{eq: composition on a cylinder})--(\ref{eq: consistency on the torus}) for fusion category SPT phases are in one-to-one correspondence with fiber functors of the fusion category $\mathcal{C}$.
To classify fusion category SPT phases, we regard naturally isomorphic fiber functors as the same SPT phase because a natural isomorphism between fiber functors corresponds to a change of the bases of the vector spaces.
Therefore, bosonic SPT phases with fusion category symmetry $\mathcal{C}$ are classified by isomorphism classes of fiber functors of $\mathcal{C}$.

\section{Unoriented TQFTs with fusion category symmetry}
\label{sec: Unoriented TQFTs with fusion category symmetry}

\subsection{Consistency conditions of unoriented TQFTs}
\label{sec: Consistency conditions of unoriented TQFTs}
A two-dimensional unoriented TQFT with finite group symmetry is formulated in \cite{KT2017, Swe2013} as a set of consistency conditions on algebraic data.\footnote{A two-dimensional unoriented TQFT without internal symmetry is formulated in \cite{TT2006}. Quantum field theories on unoriented manifolds are of interest in recent studies on topological phases of matter \cite{HSCRL2014, CHMR2015, HCR2016, CTR2016, BBCJW2019, Wit2016c, TY2017a, TY2017b, Bha2017, SR2017, SSR2017, SSR2017b, SSGR2018, Kob2019, Tur2020}.}
In this section, we reformulate 2d unoriented TQFTs to incorporate fusion category symmetries.
We assume that the total symmetry splits into time-reversal symmetry and finite internal symmetry.
In this case, we can treat orientation-reversing defects and symmetry defects separately.

\begin{figure}
\begin{center}
\includegraphics[width = 5cm]{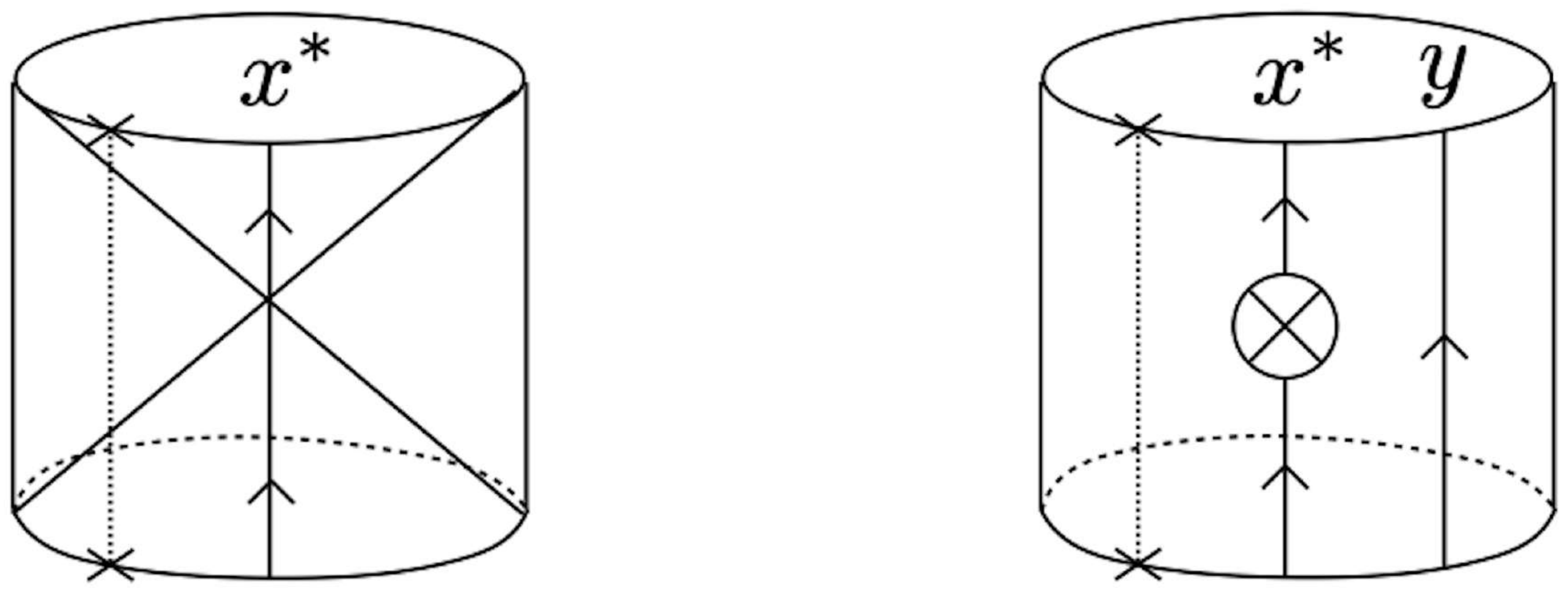}
\caption{Every unoriented surface can be decomposed into oriented surfaces and the above two unoriented surfaces. The left diagram represents the orientation-reversing diffeomorphism of a circle, which induces an isomorphism between the vector spaces $V_x$ and $V_{x^{*}}$. The right diagram represents a cylinder with a cross-cap, which induces a linear map from $V_{x \otimes y}$ to $V_{x^{*} \otimes y}$.}
\label{fig: unoriented amplitudes}
\end{center}
\end{figure}
\begin{figure}
\begin{center}
\includegraphics[width = 1.8cm]{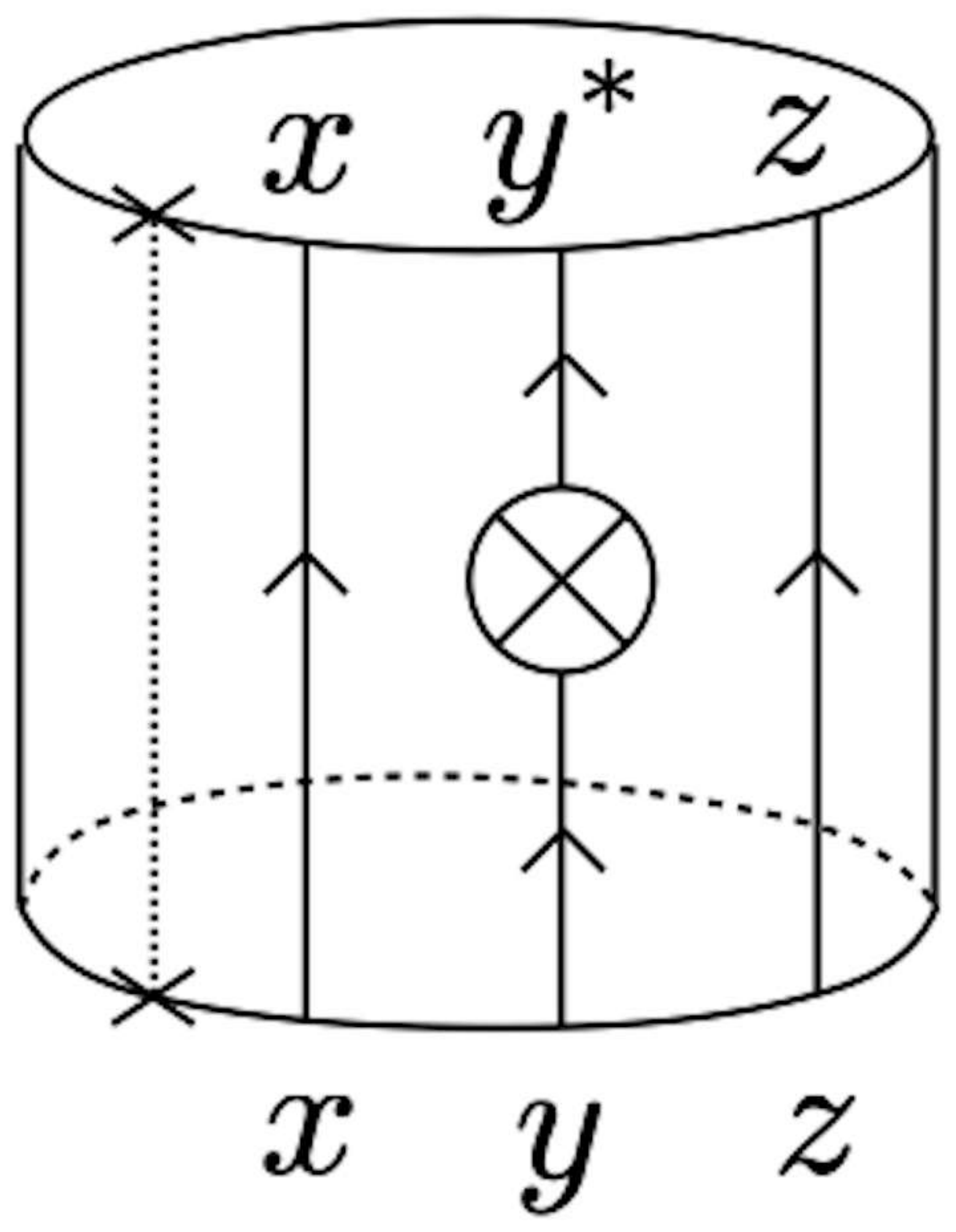}
\caption{The transition amplitude for the above cylinder with a cross-cap is a linear map from $V_{(x \otimes y) \otimes z}$ to $V_{(x \otimes y^*) \otimes z}$, which is given by $\mathcal{A}_{y^* \otimes (z \otimes x) \rightarrow (x \otimes y^*) \otimes z} \circ \theta_{y, z \otimes x} \circ \mathcal{A}_{(x \otimes y) \otimes z \rightarrow y \otimes (z \otimes x)}$.}
\label{fig: variant of theta}
\end{center}
\end{figure}

The algebraic data of an unoriented TQFT consist of the following linear maps in addition to the algebraic data of an oriented TQFT: the orientation-reversing isomorphism $\phi_x: V_x \rightarrow V_{x^{*}}$ and the cross-cap amplitude $\theta_{x, y}: V_{x \otimes y} \rightarrow V_{x^{*} \otimes y}$, see also figure \ref{fig: unoriented amplitudes}.
In particular, $\phi$ represents the linear action of $CP$ symmetry, which we assume to be unitary, rather than the anti-linear action of time-reversal symmetry.
We can avoid anti-linear maps by using $CPT$ symmetry \cite{KT2017}.
A variant of the cross-cap amplitude as shown in figure \ref{fig: variant of theta} can be computed by composing the cross-cap amplitude $\theta$ and the generalized associators $\mathcal{A}$.

A general unoriented surface can be decomposed into the following building blocks and their adjoints: a cylinder, a cap, a pair of pants, and a cylinder with a cross-cap.
Correspondingly, the transition amplitude for a general unoriented surface is given by the composition of the transition amplitudes for these elements that appear in the decomposition.
However, the decomposition is not unique in general.
Therefore, we need to impose consistency conditions so that the transition amplitude does not depend on a decomposition.
In the following, we list the consistency conditions that involve $\phi$ and $\theta$.
\begin{description}
\item[Invariance of the unit]~\\
The unit $i: \mathbb{C} \rightarrow V_1$ is invariant under the action of the orientation-reversing isomorphism
\begin{equation}
\adjincludegraphics[valign = c, width = 1.5cm]{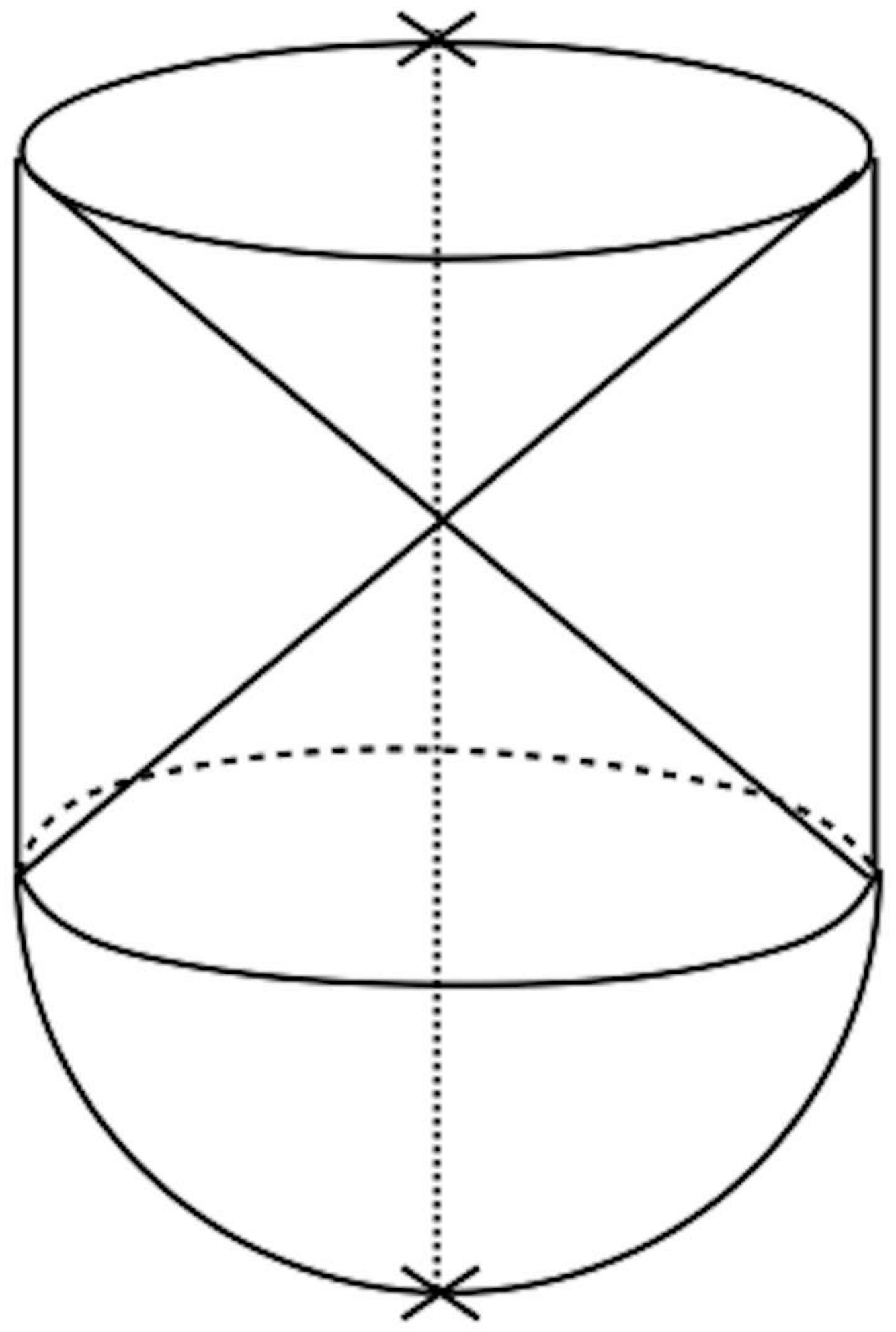} ~ = ~ 
\adjincludegraphics[valign = c, width = 1.5cm]{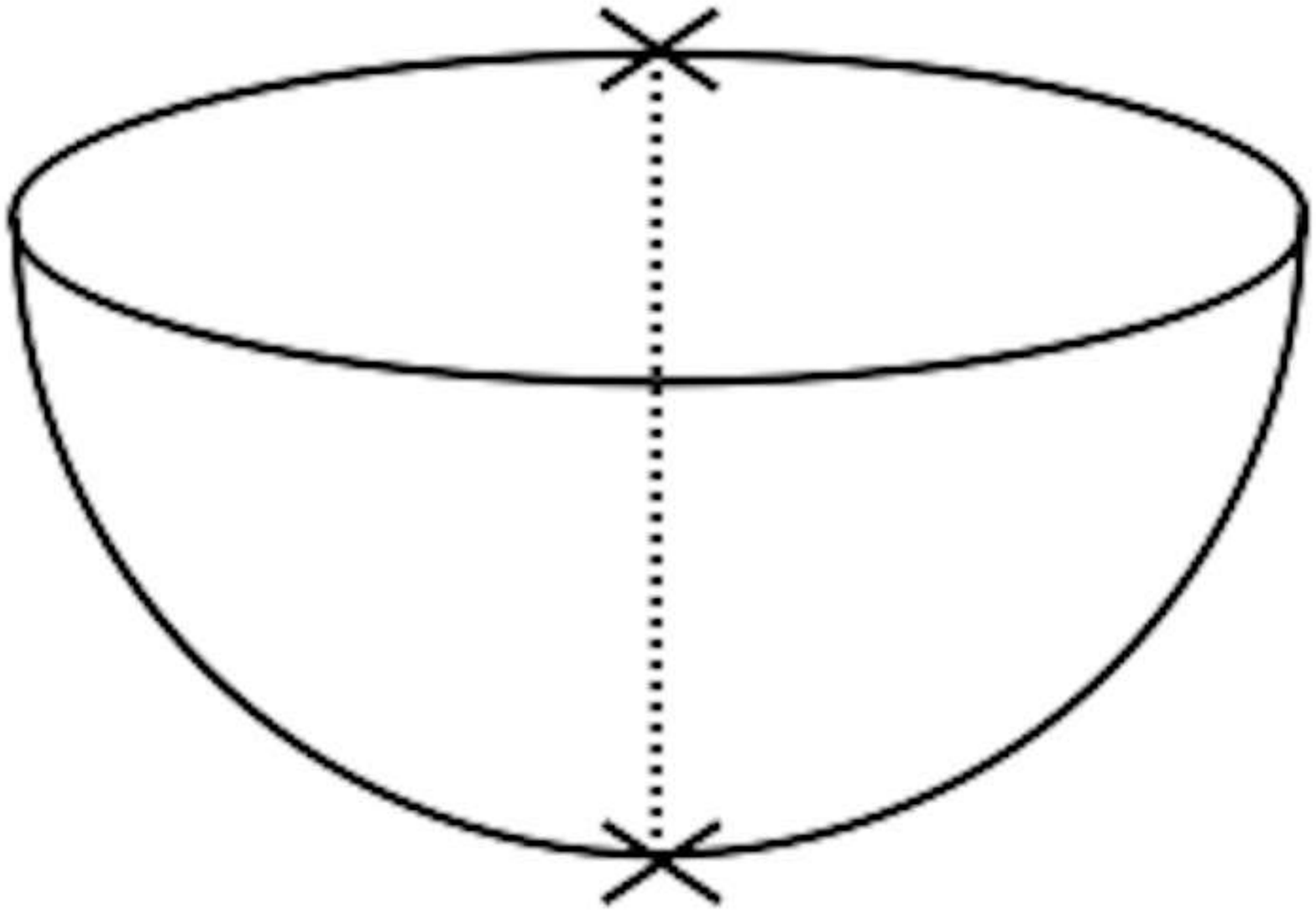}.
\tag{U1}
\label{eq: invariant unit}
\end{equation}

\item[Involution]~\\
The orientation-reversing isomorphism is involutive up to pivotal structure $a_x: x \rightarrow x^{**}$
\begin{equation}
\adjincludegraphics[valign = c, width = 1.5cm]{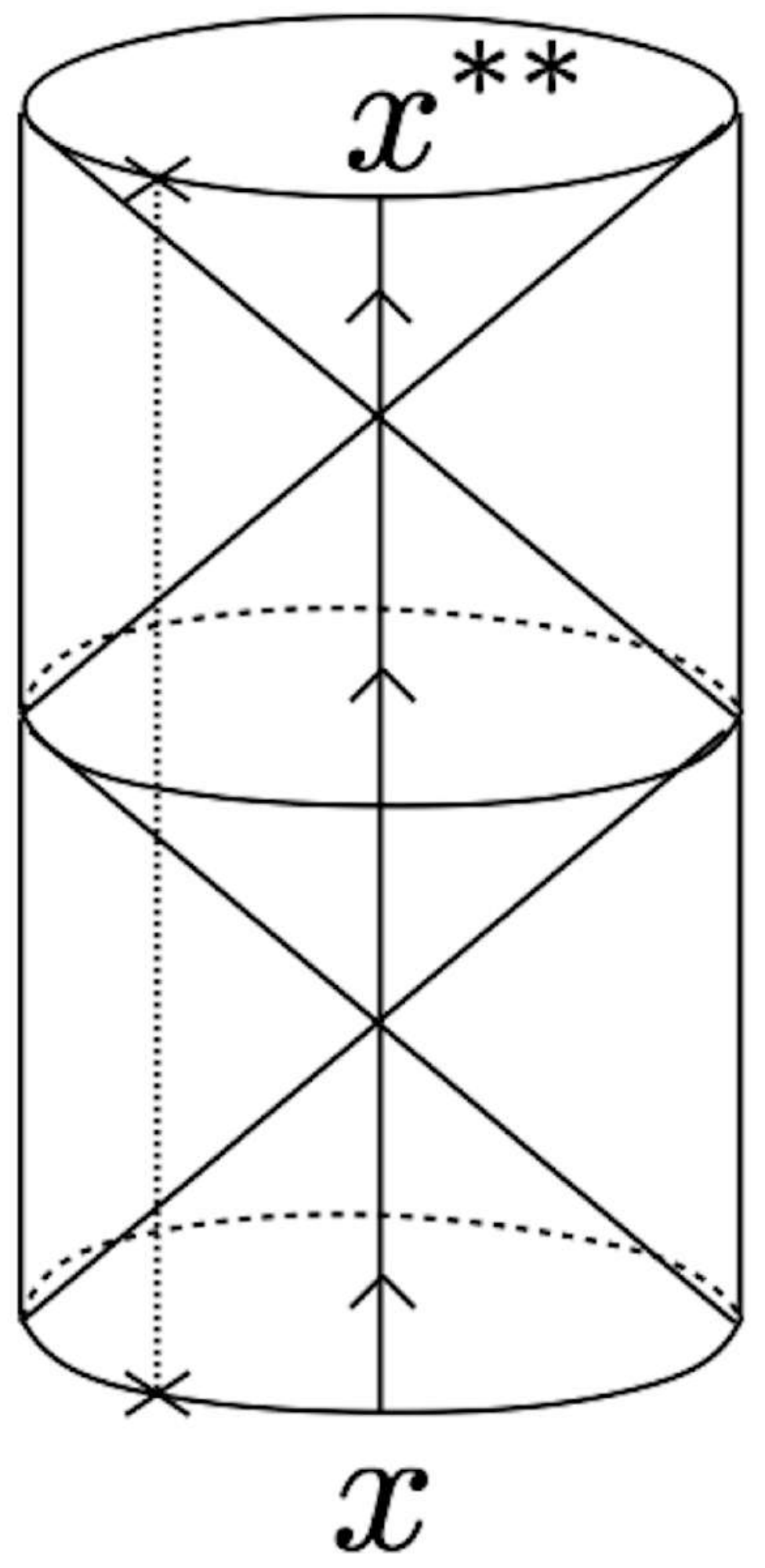} ~ = ~ \adjincludegraphics[valign = c, width = 1.5cm]{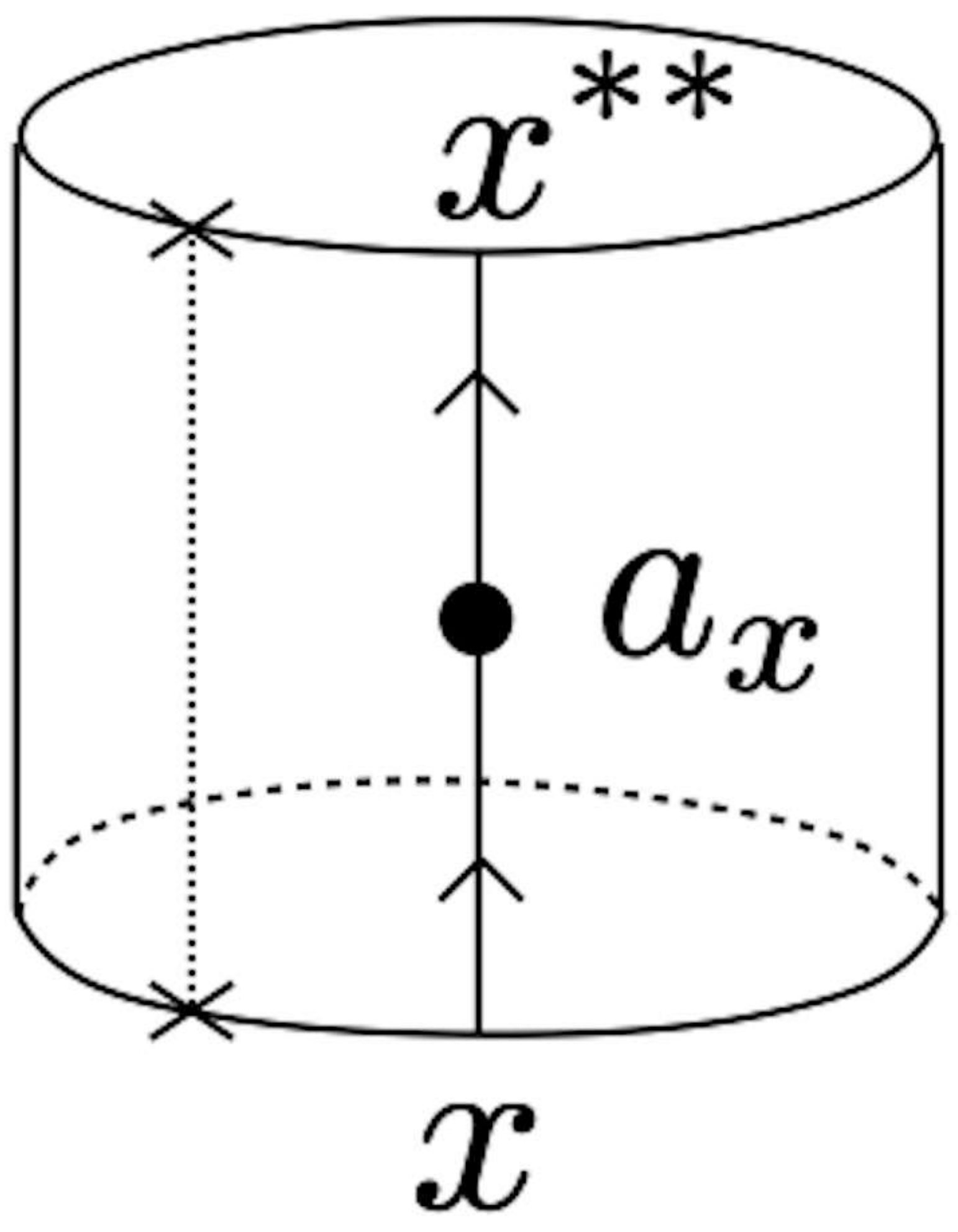}.
\tag{U2}
\label{eq: involution}
\end{equation}

\item[Commutativity of the orientation reversal and the change of the base point] ~\\
The orientation-reversing isomorphism commutes with the change of the base point:
\begin{equation}
\adjincludegraphics[valign = c, width = 1.5cm]{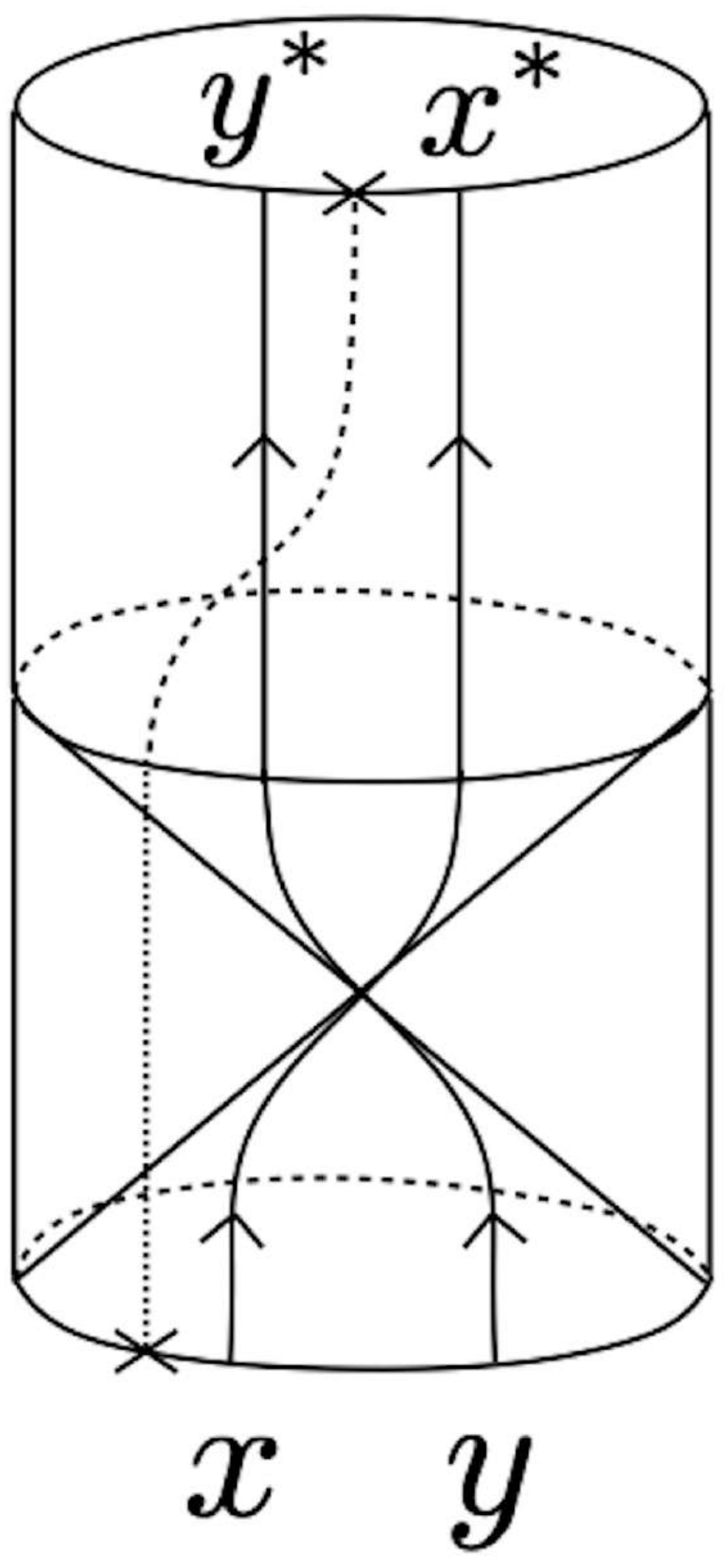} ~ = ~ \adjincludegraphics[valign = c, width = 1.5cm]{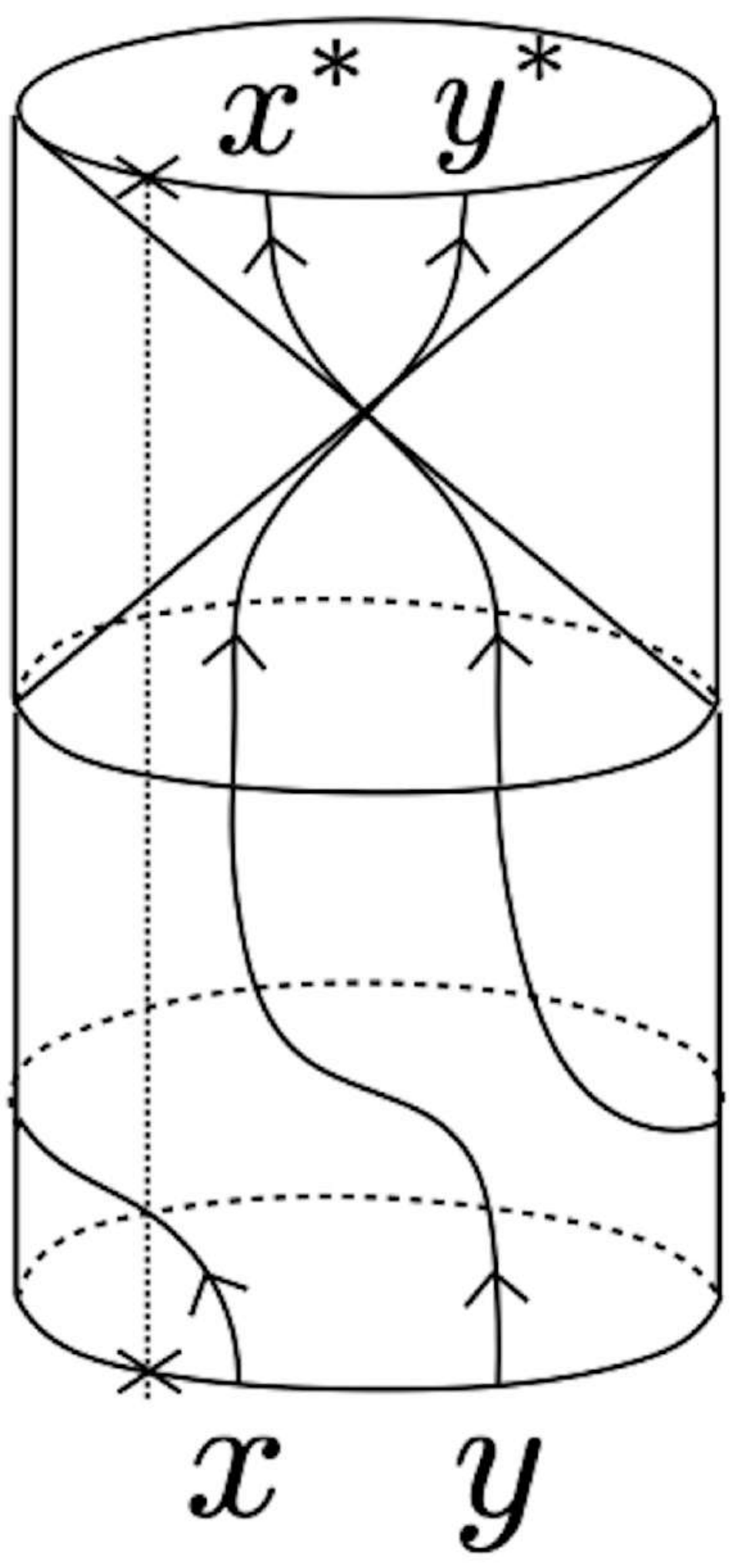}. 
\tag{U3}
\label{eq: commutativity of X and phi}
\end{equation}

\item[Orientation reversal of a topological point operator] ~\\
The orientation-reversing symmetry acts on a topological point operator $f \in \mathop{\mathrm{Hom}}(x, y)$ and turn it into another topological point operator $\overline{f} \in \mathop{\mathrm{Hom}}(x^*, y^*)$ such that
\begin{equation}
\adjincludegraphics[valign = c, width = 1.35cm]{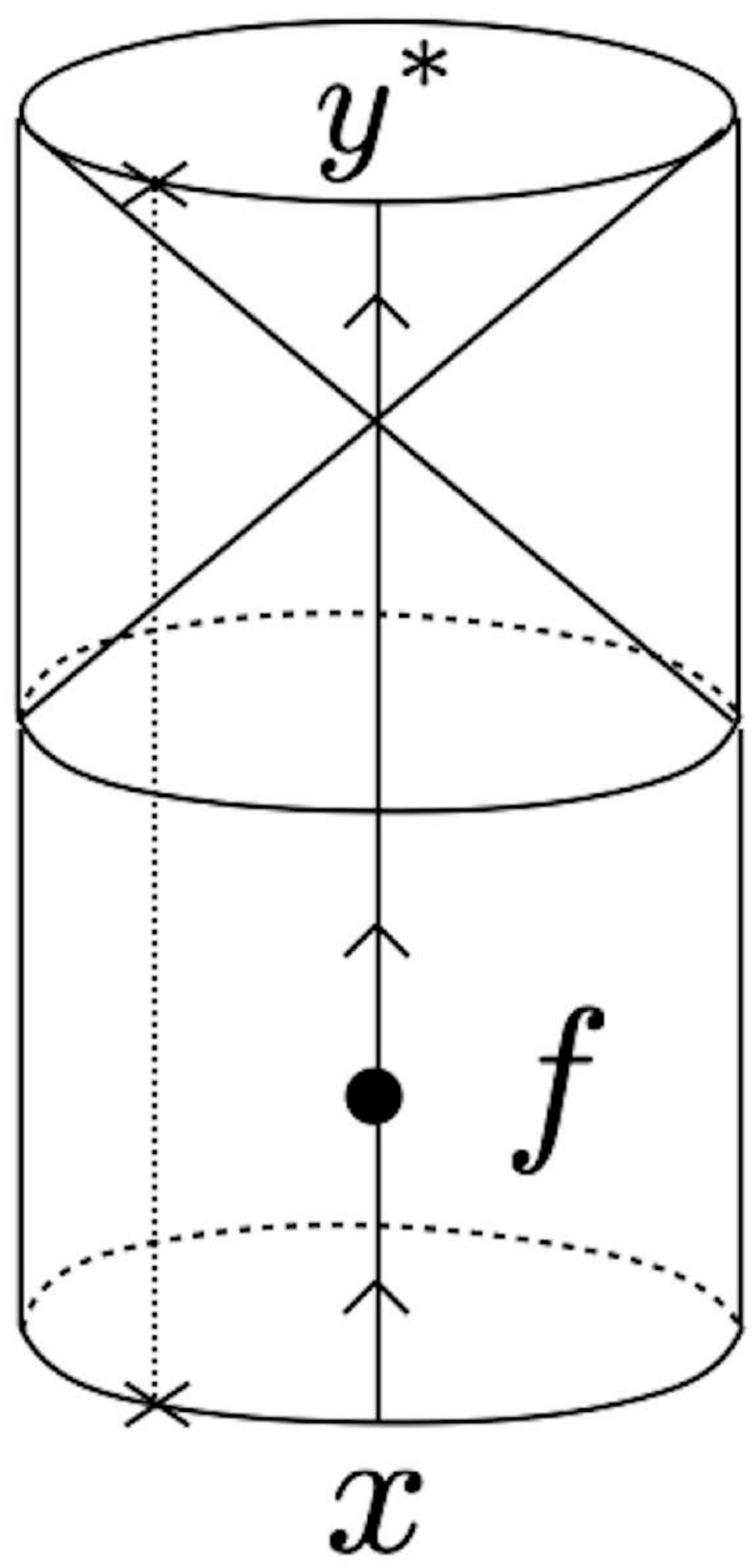} ~ = ~ \adjincludegraphics[valign = c, width = 1.35cm]{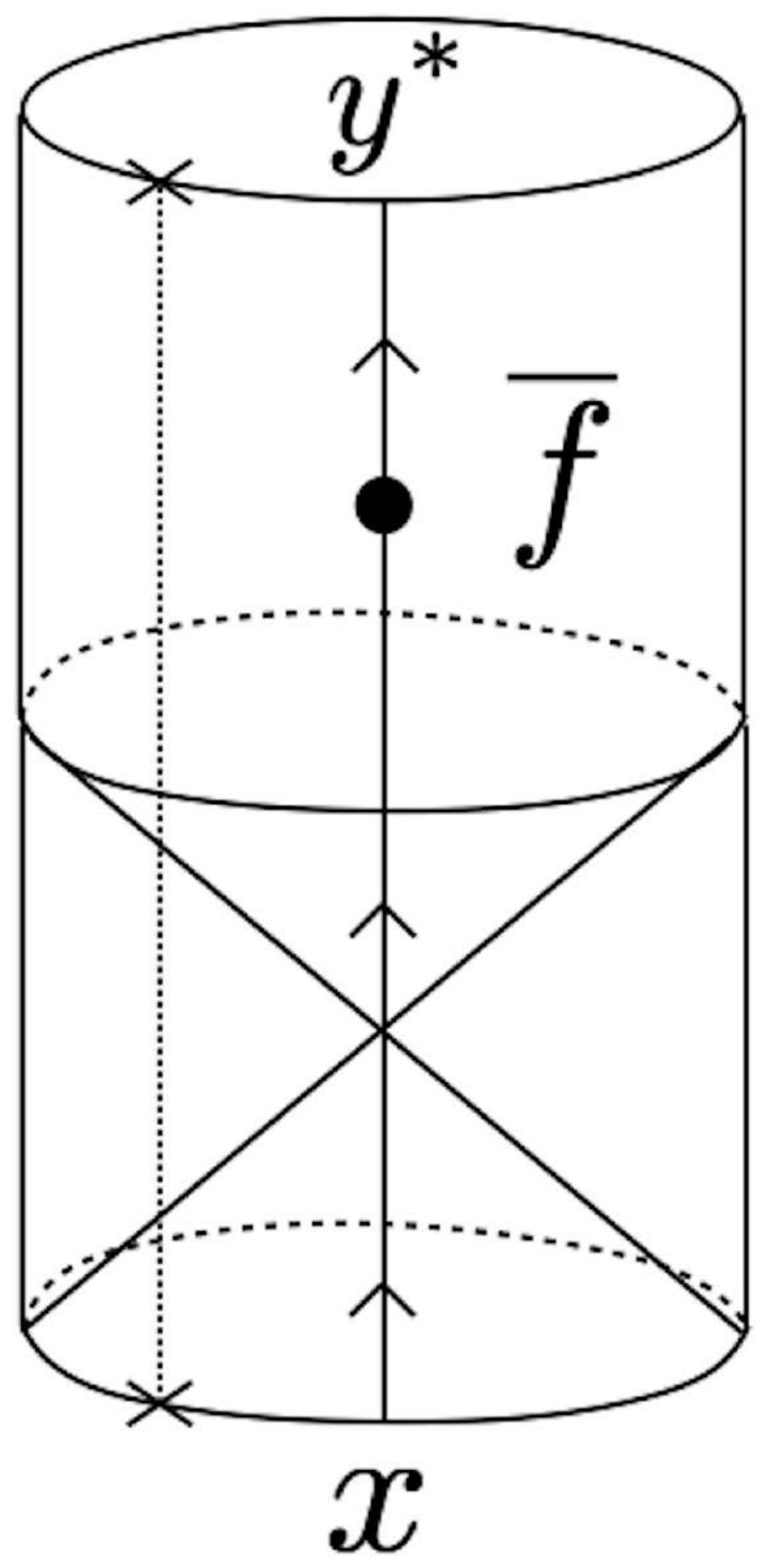}, \quad
\adjincludegraphics[valign = c, width = 1.5cm]{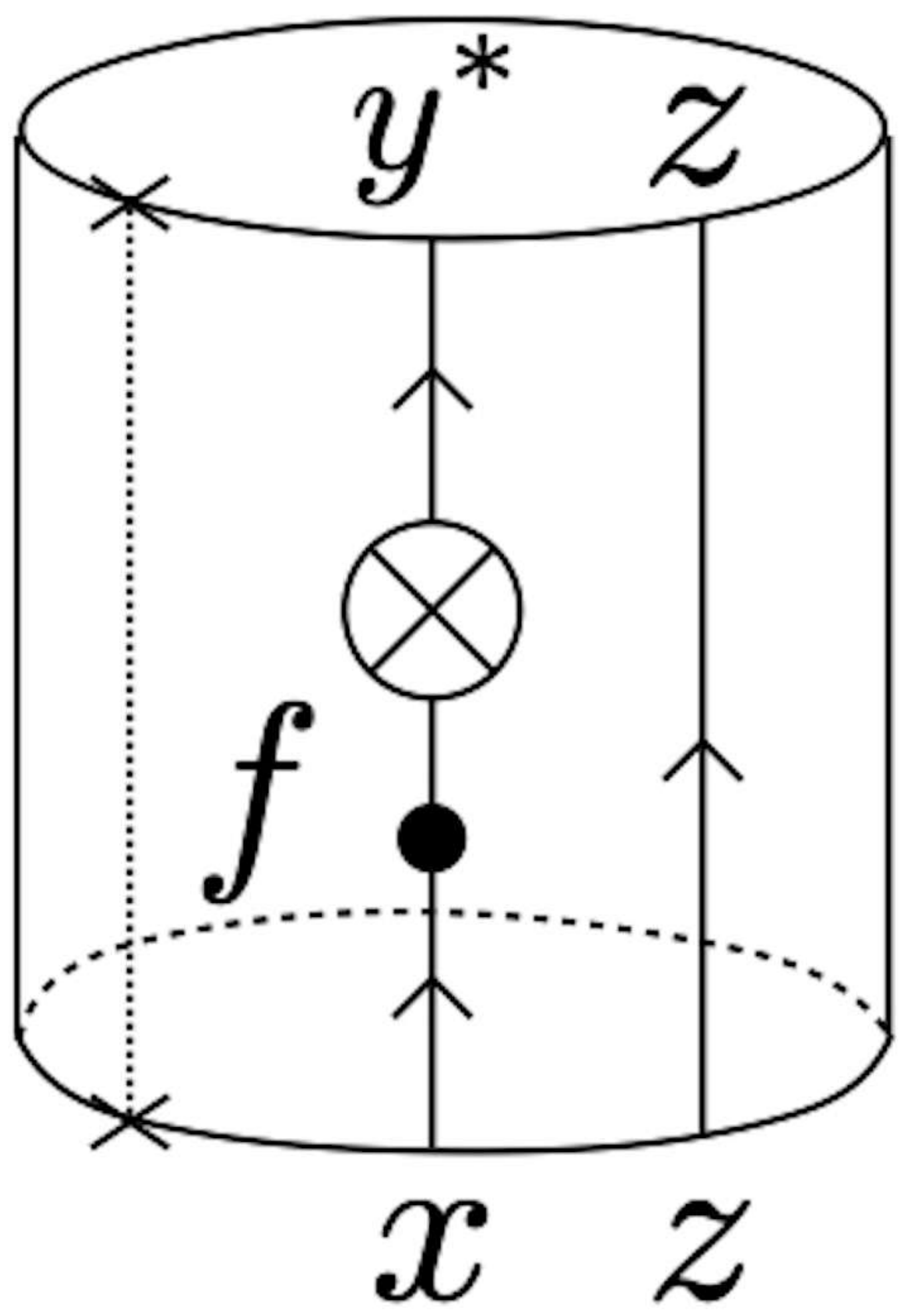} ~ = ~ \adjincludegraphics[valign = c, width = 1.5cm]{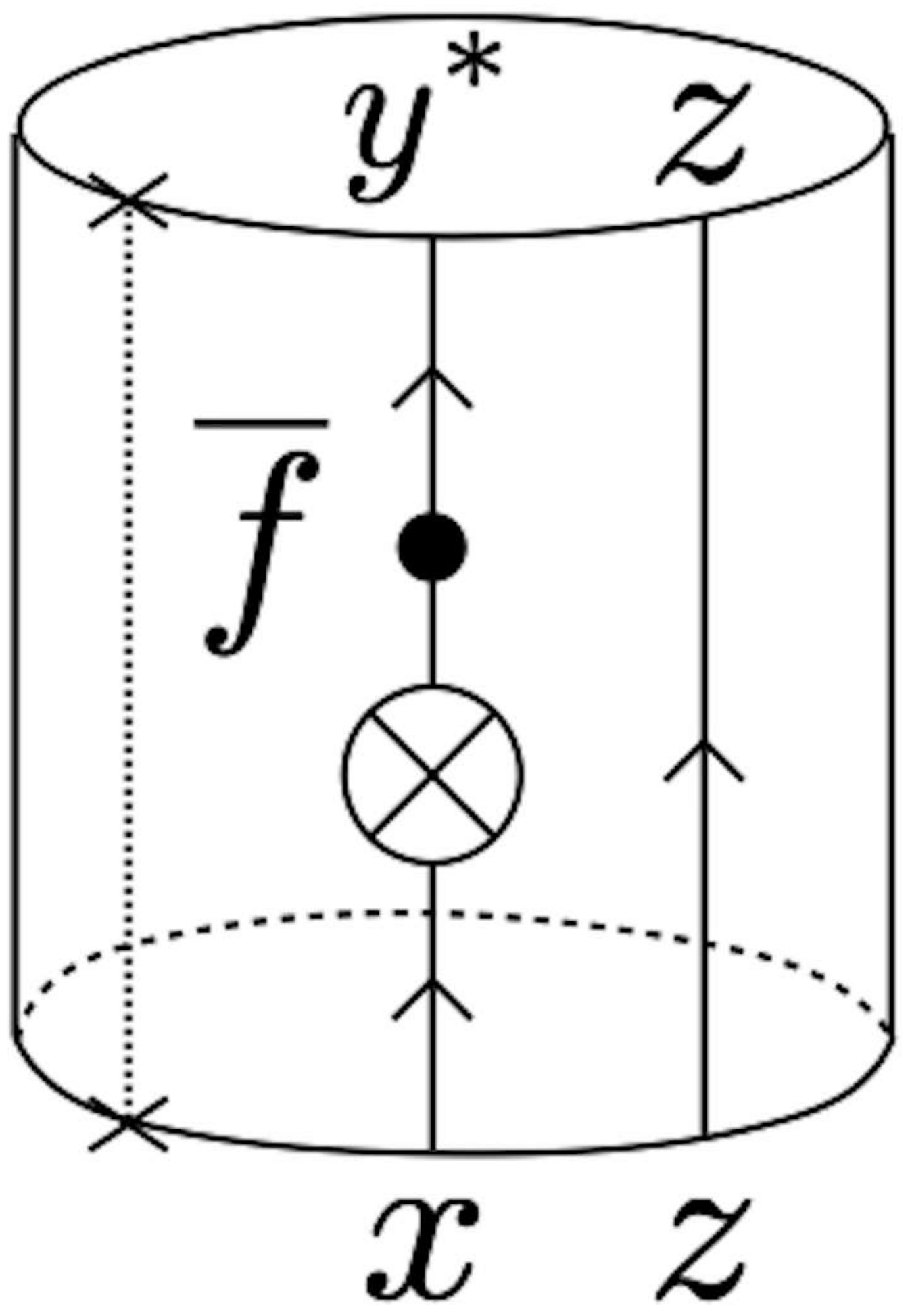}. 
\tag{U4}
\label{eq: orientation reversal of f}
\end{equation}

\item[Cross-cap on a folded topological defect] ~\\
We can move the position of a cross-cap along a folded topological defect line:
\begin{equation}
\adjincludegraphics[valign = c, width = 1.5cm]{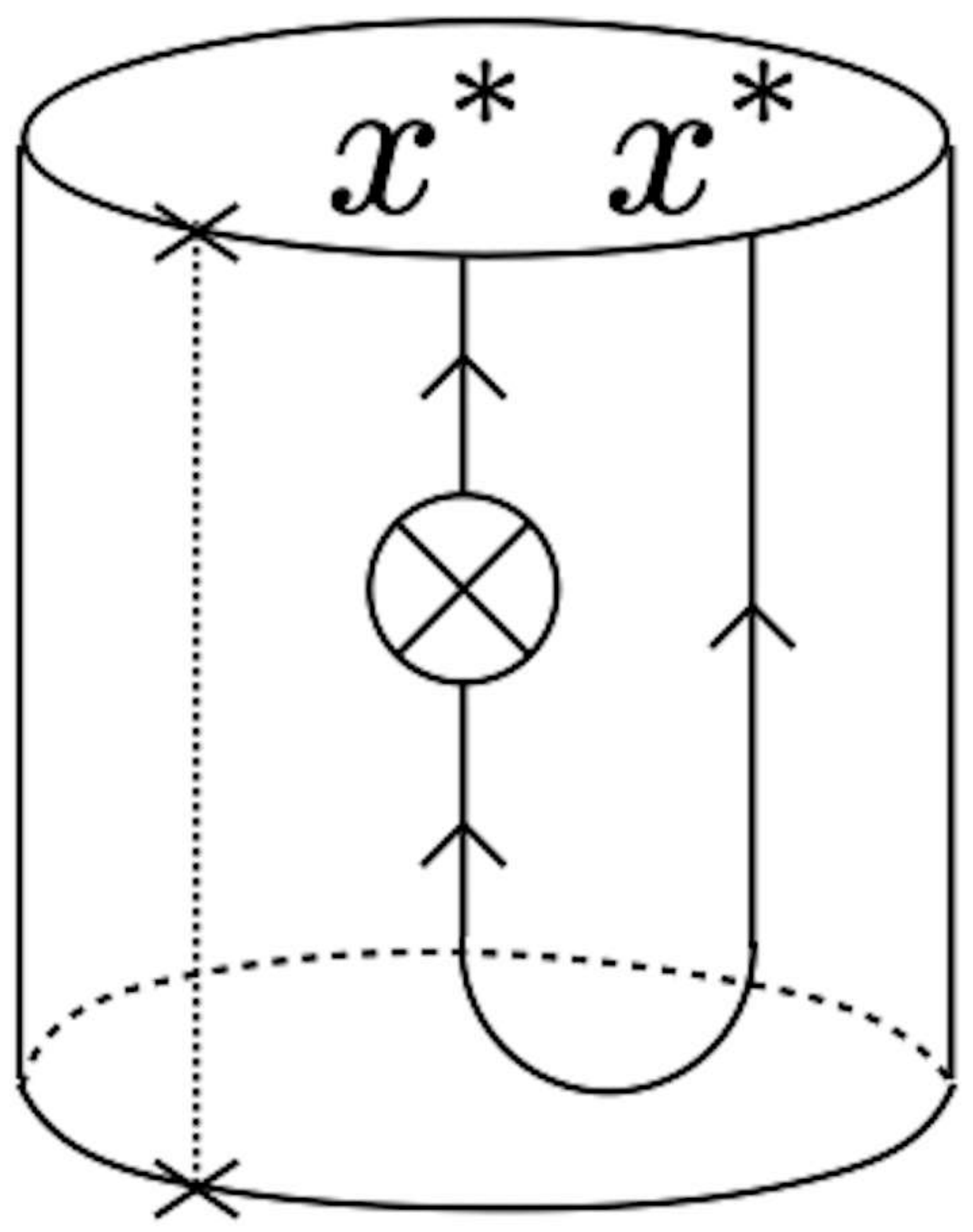} ~ = ~ \adjincludegraphics[valign = c, width = 1.5cm]{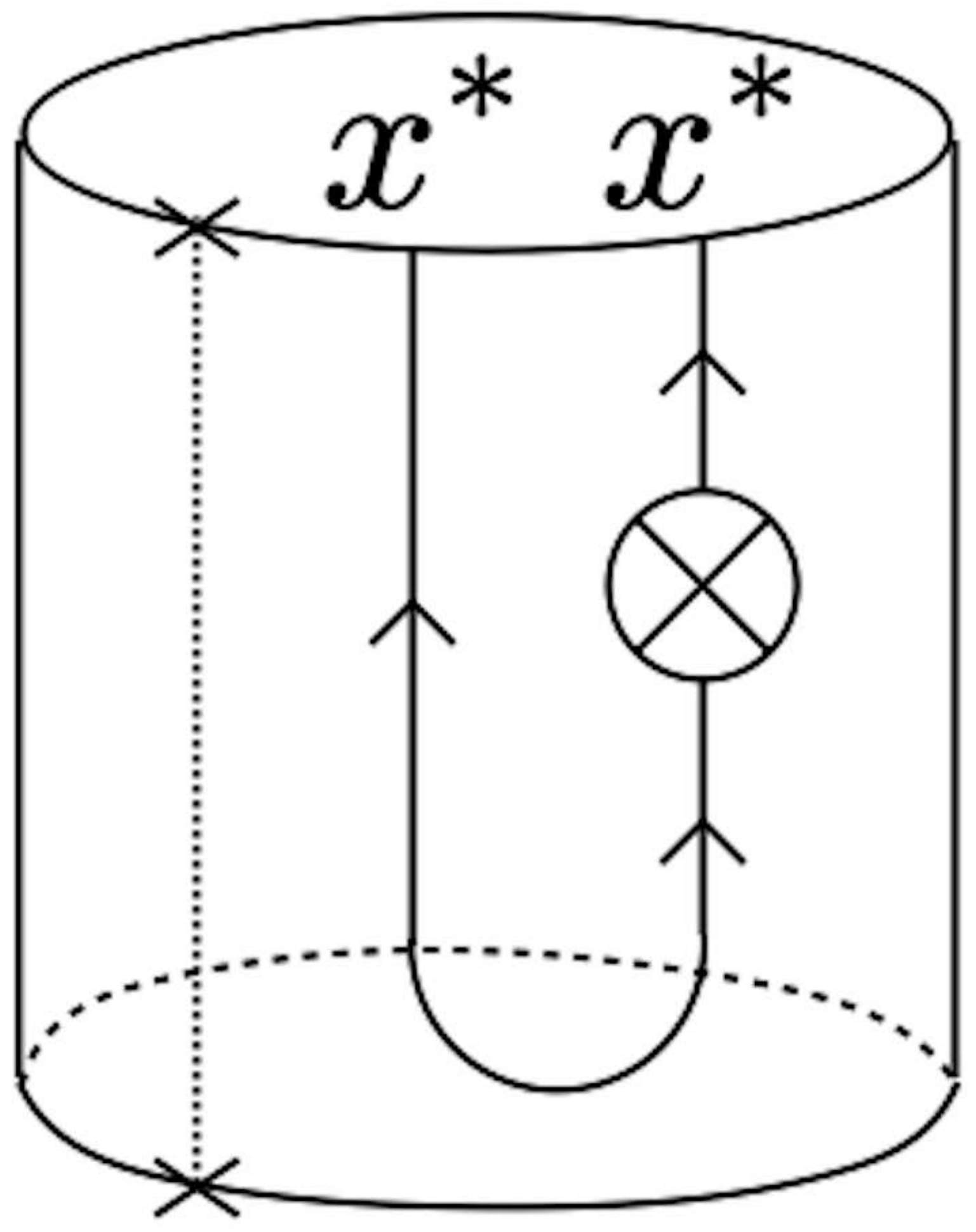}. 
\tag{U5}
\label{eq: compatibility of left-theta and right-theta}
\end{equation}

\item[Deformation of a topological defect through a cross-cap] ~\\
A topological defect that passes a cross-cap twice can avoid the cross-cap by a deformation:
\begin{equation}
\adjincludegraphics[valign = c, width = 1.5cm]{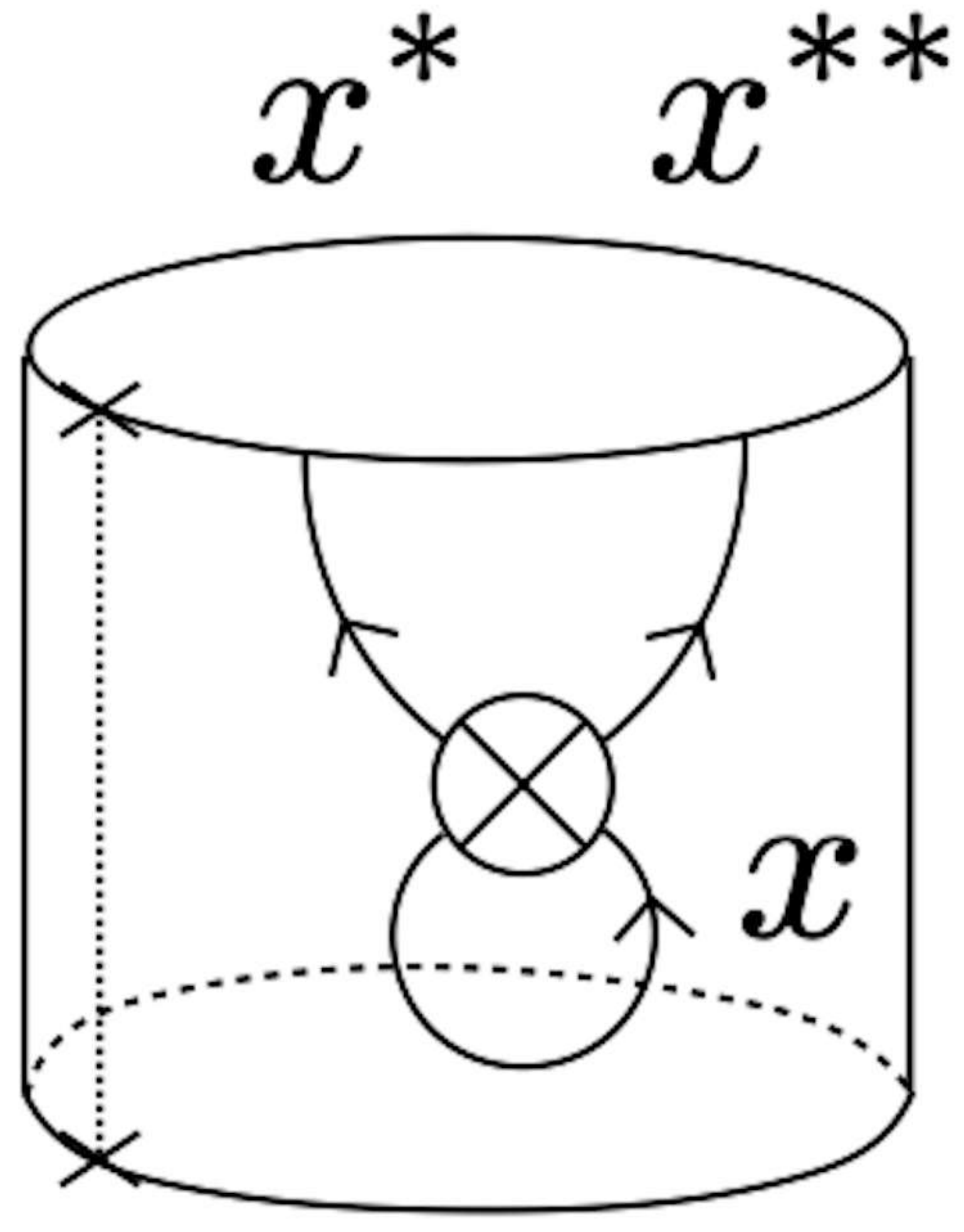} ~ = ~ \adjincludegraphics[valign = c, width = 1.5cm]{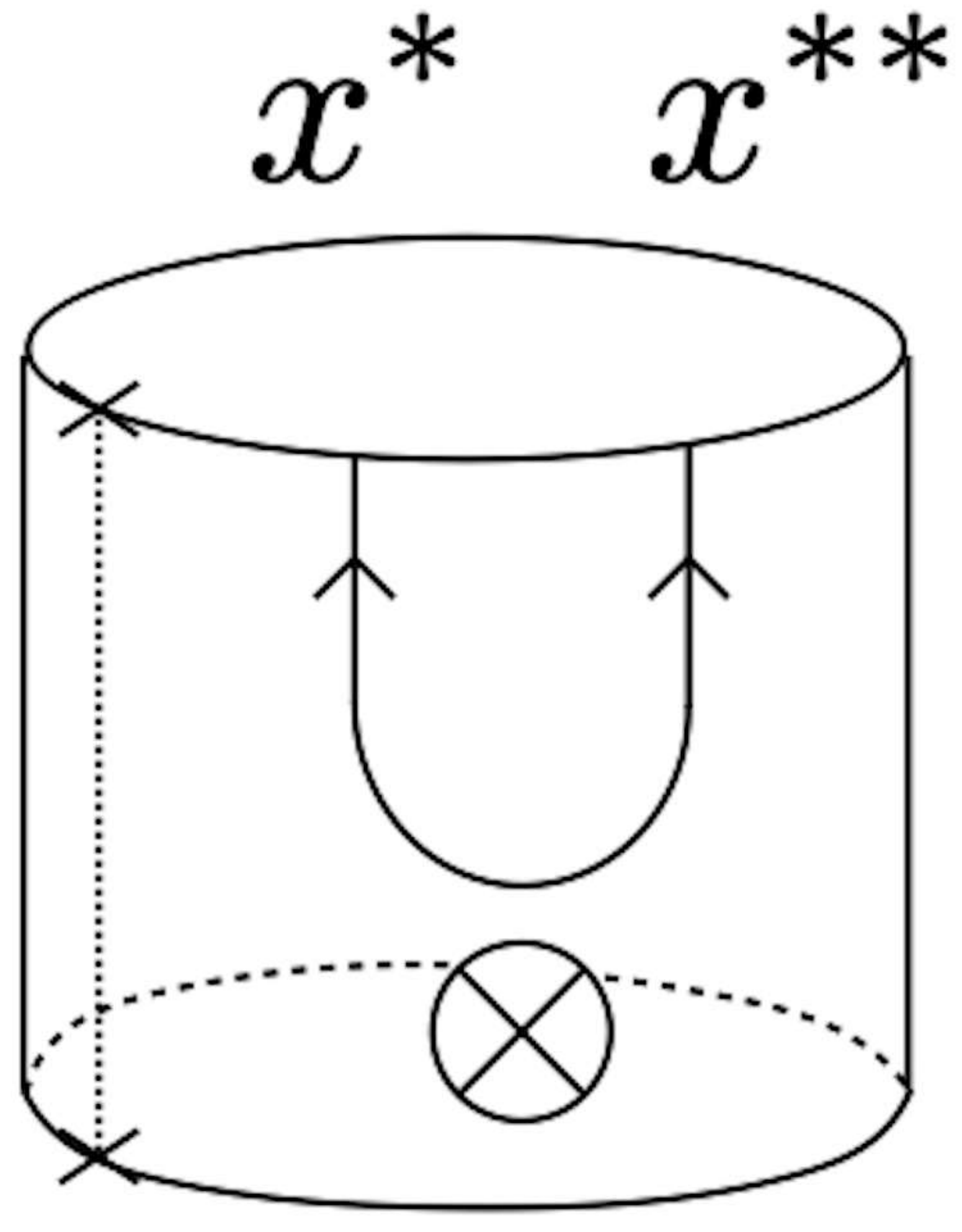}, \quad \quad
\adjincludegraphics[valign = c, width = 1.7cm]{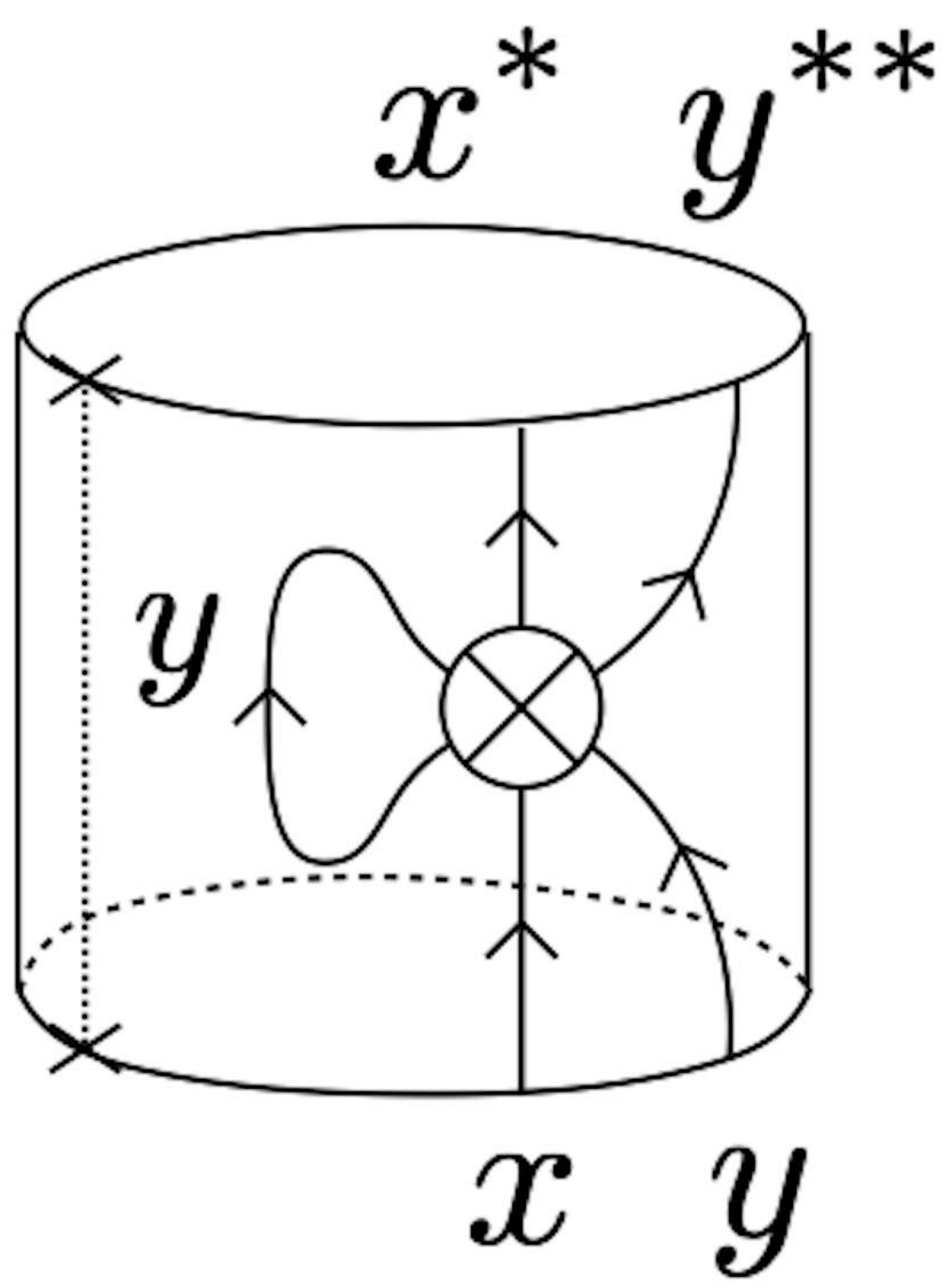} ~ = ~ \adjincludegraphics[valign = c, width = 1.5cm]{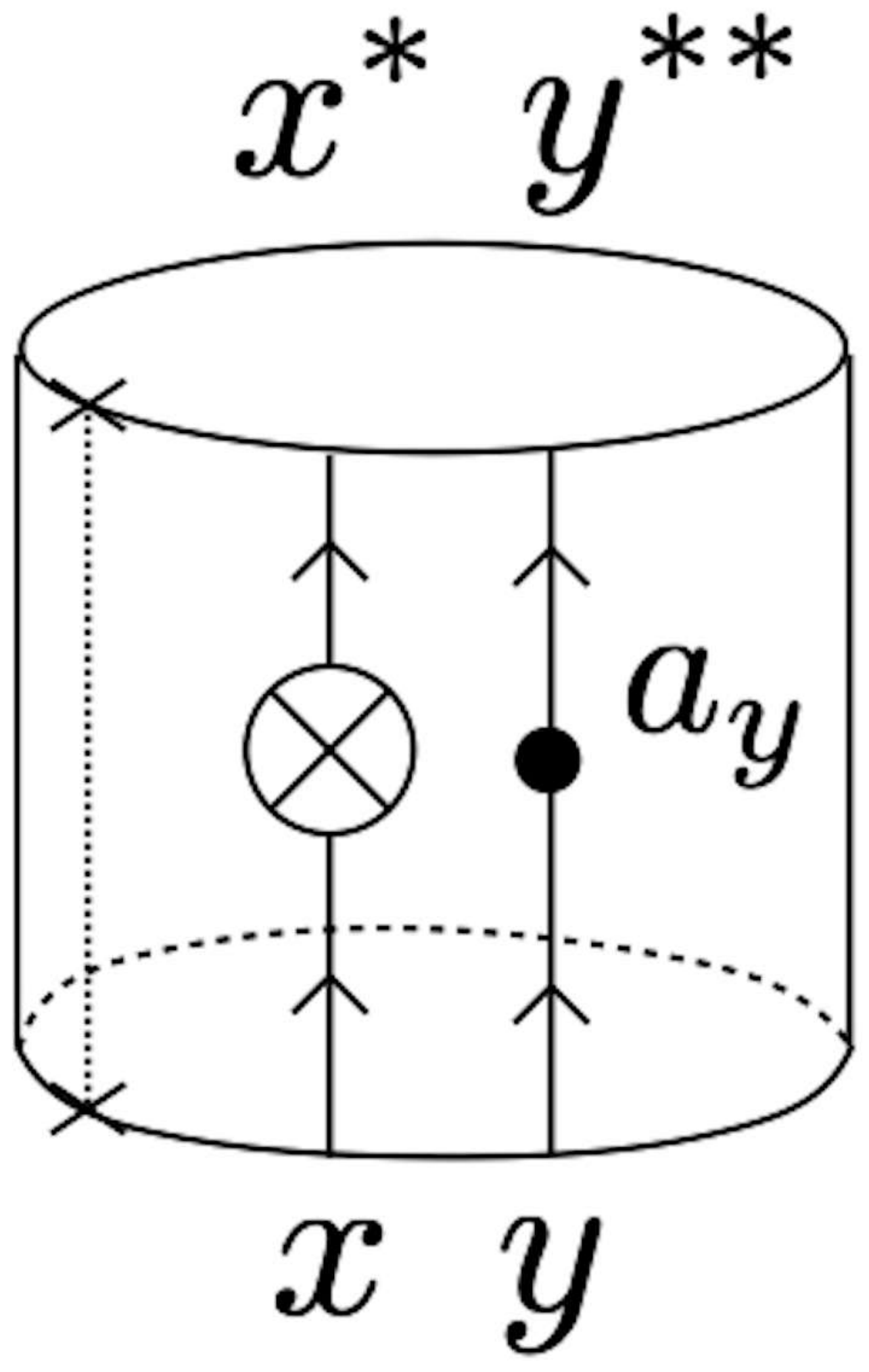}.
\tag{U6}
\label{eq: reconnection}
\end{equation}
We also have similar equations for the diagrams flipped horizontally.

\item[Commutativity of the cross-cap and the orientation reversal] ~\\
The orientation-reversing isomorphism commutes with the cross-cap amplitude:
\begin{equation}
\adjincludegraphics[valign = c, width = 1.35cm]{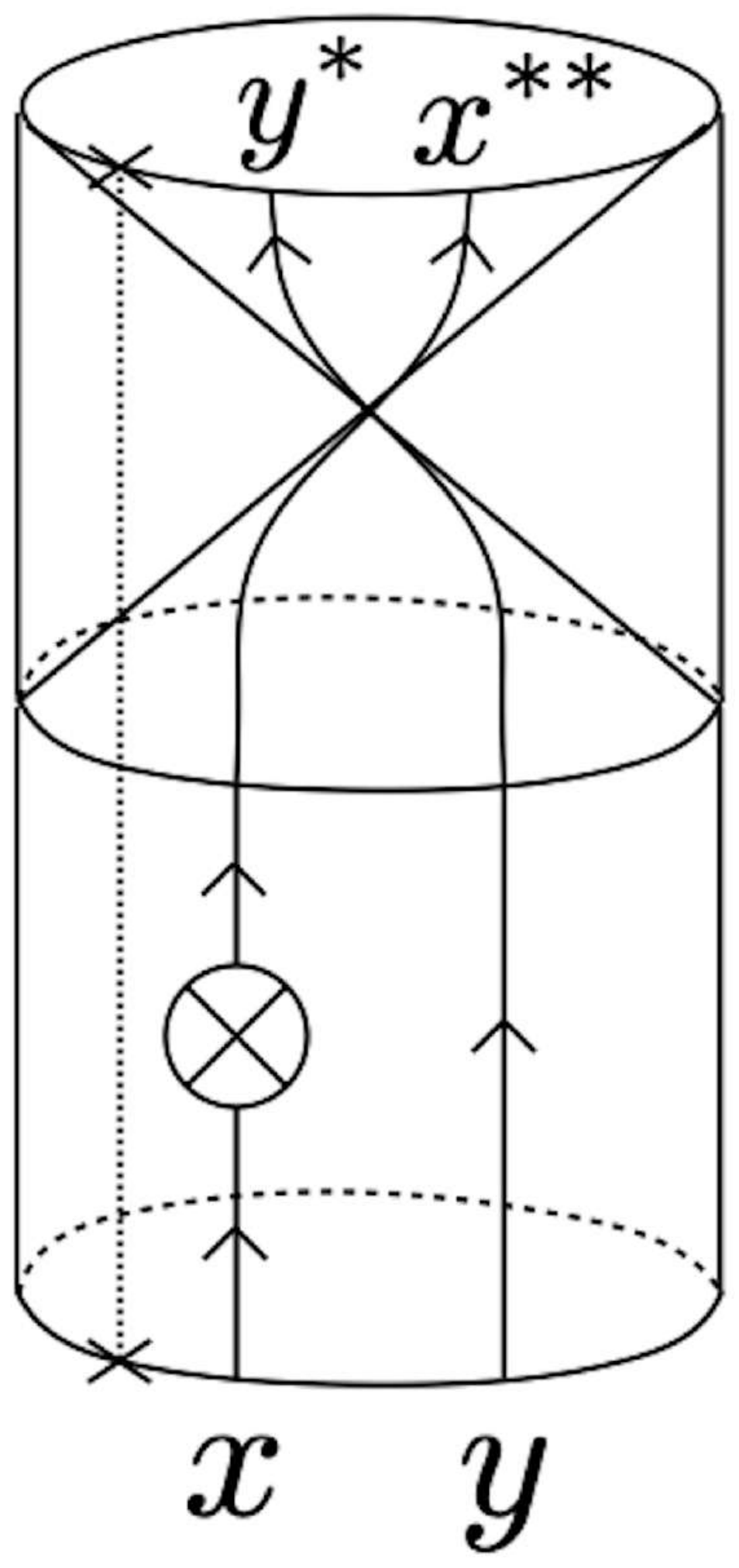} ~ = ~ \adjincludegraphics[valign = c, width = 1.35cm]{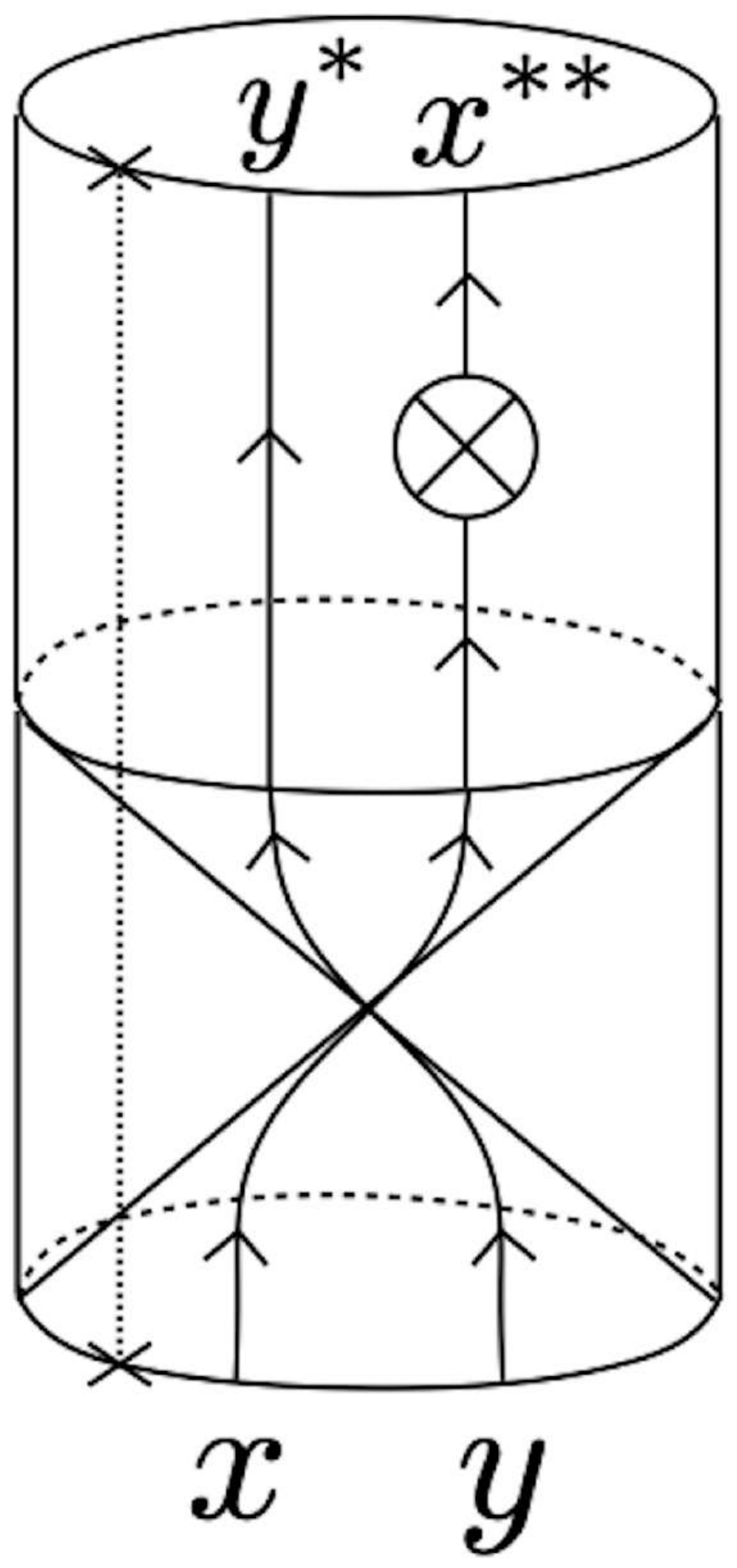}. 
\tag{U7}
\label{eq: compatibility of theta and phi}
\end{equation}

\item[M\"{o}bius identity] ~\\
Two different ways to compute the transition amplitude for the punctured M\"{o}bius strip lead to the same result:
\begin{equation}
\adjincludegraphics[valign = c, width = 1.5cm]{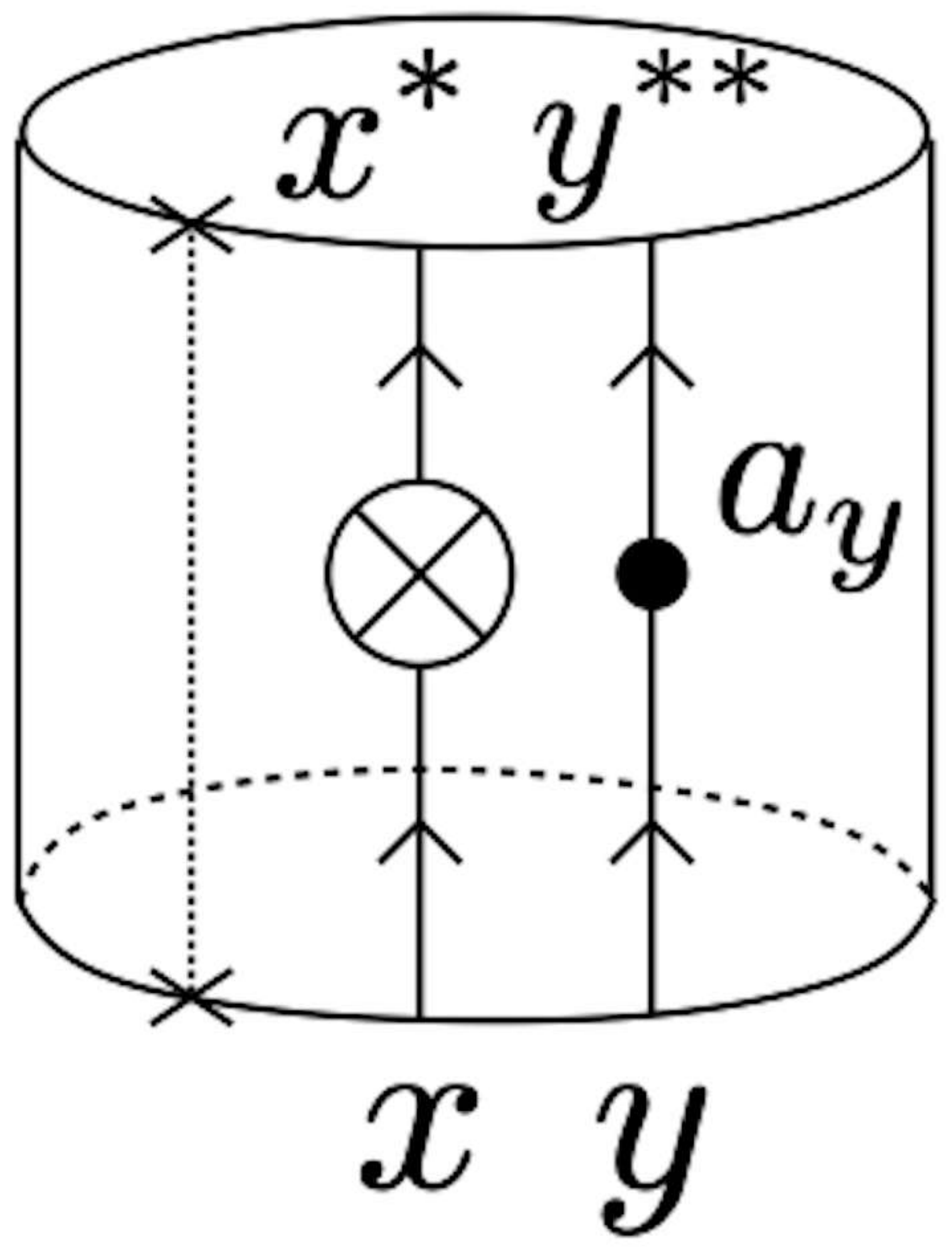} ~ = ~ \adjincludegraphics[valign = c, width = 1.35cm]{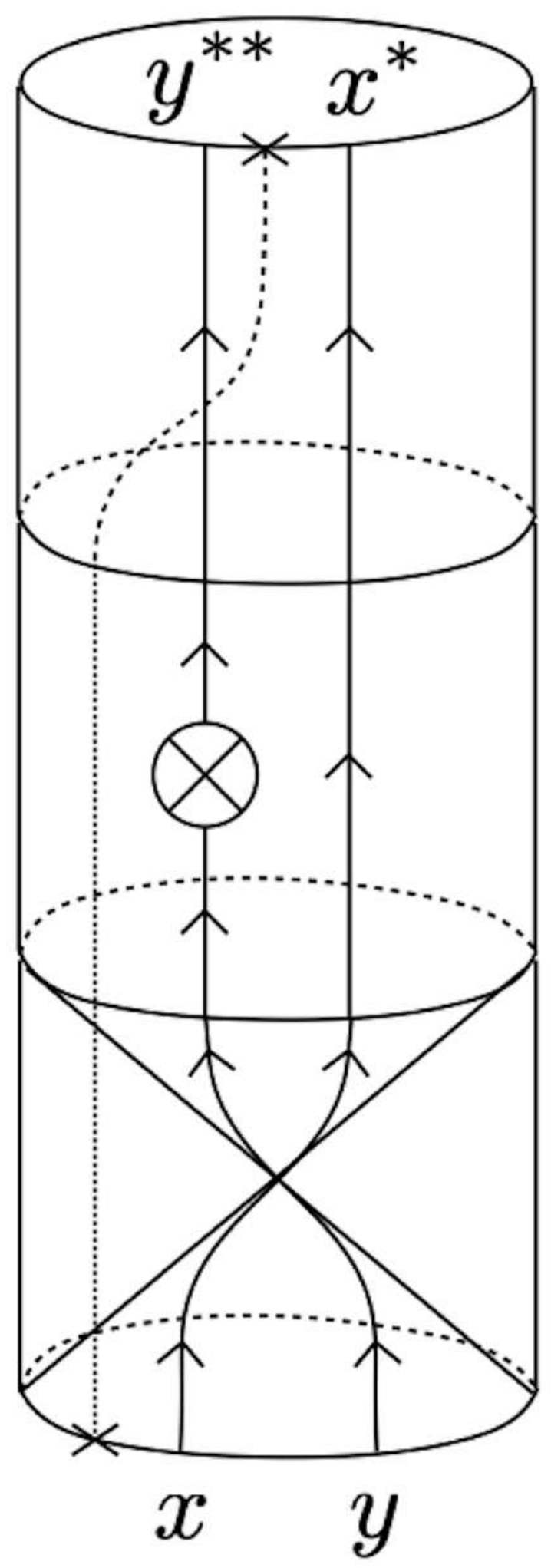}.
\tag{U8}
\label{eq: Mobius}
\end{equation}

\item[Equivariance of the multiplication] ~\\
The multiplication is equivariant with respect to the orientation-reversing isomorphism:
\begin{equation}
\adjincludegraphics[valign = c, width = 2.25cm]{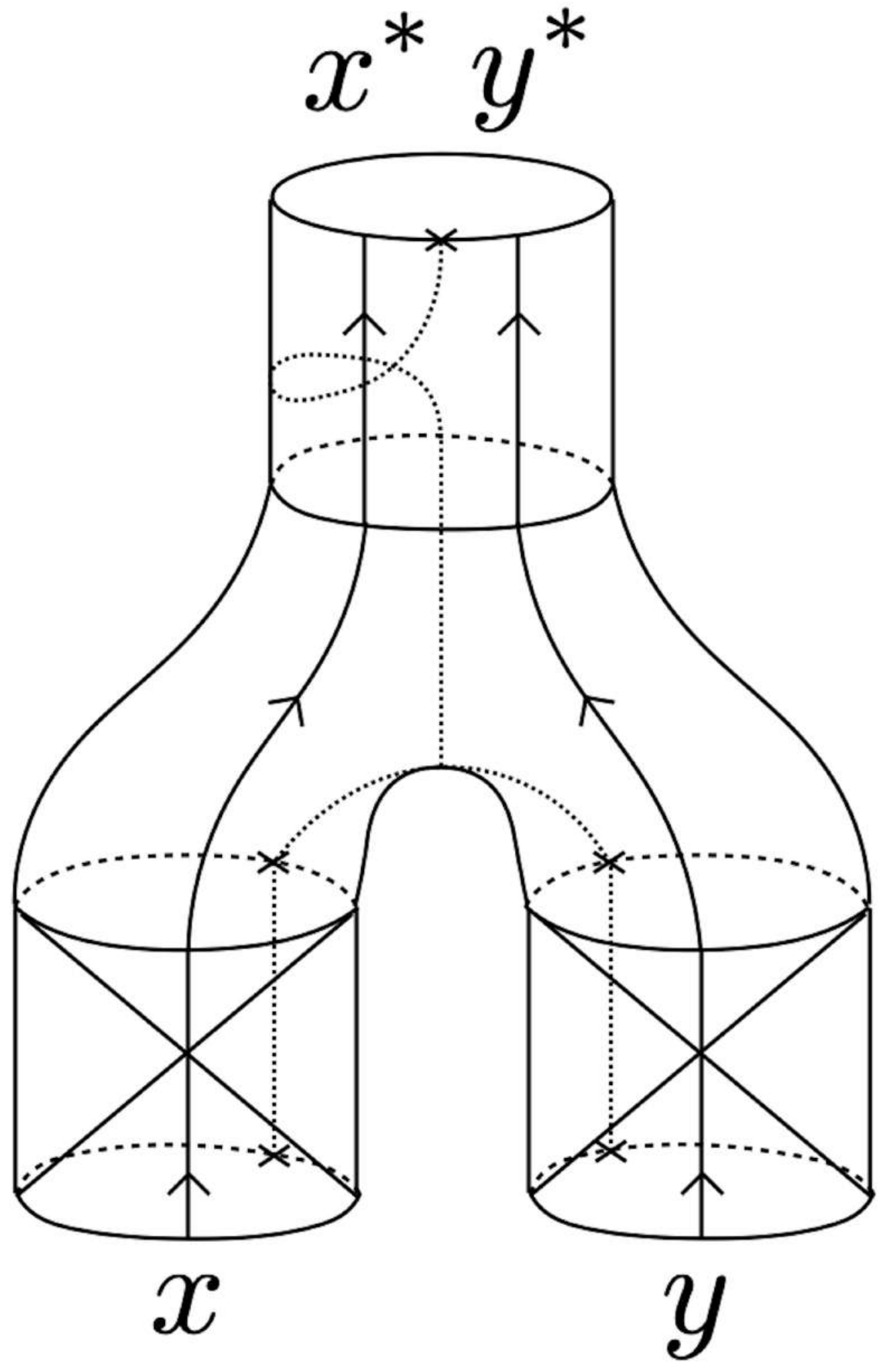} ~ = ~ \adjincludegraphics[valign = c, width = 2.25cm]{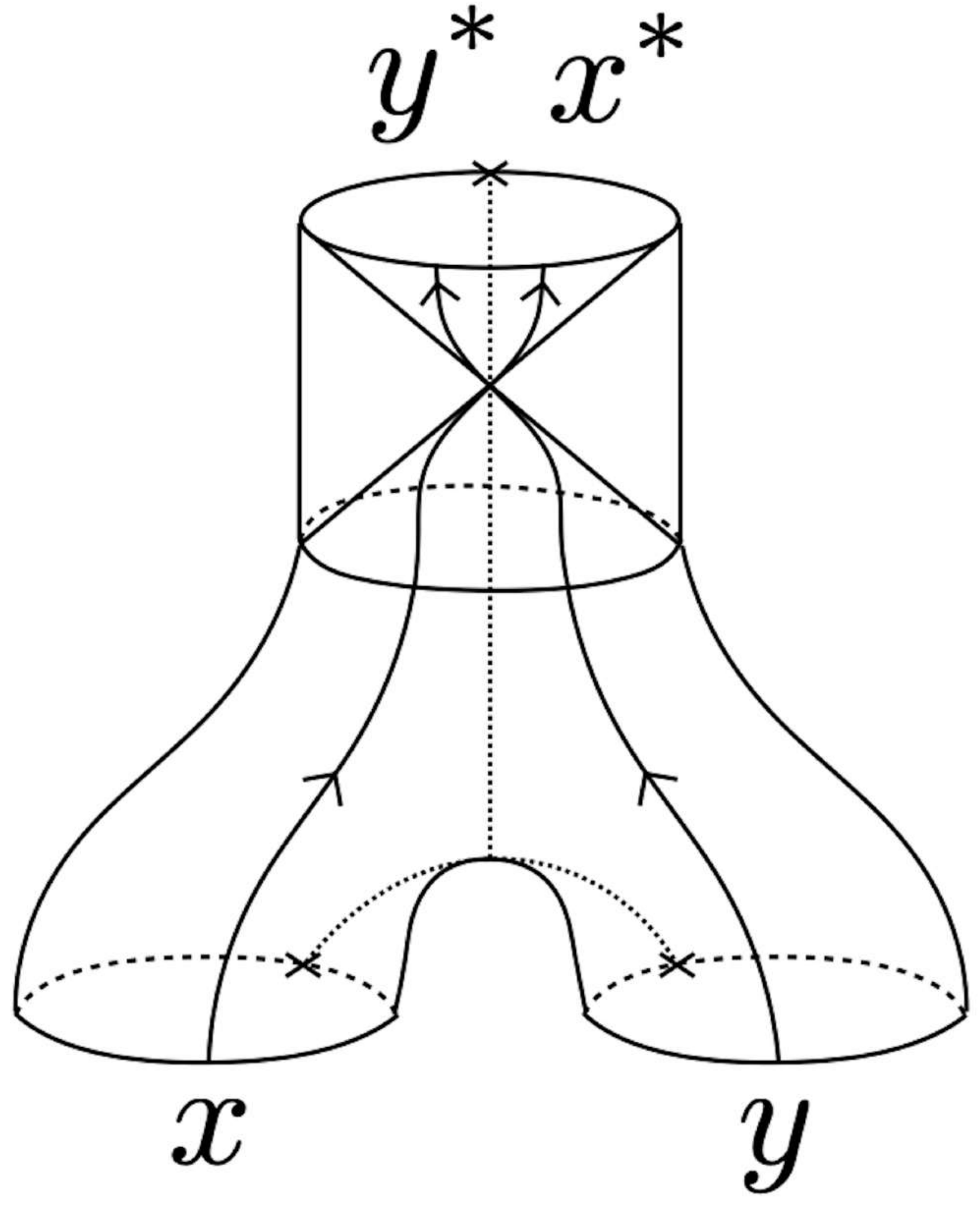}. 
\tag{U9}
\label{eq: equivariance}
\end{equation}

\item[Compatibility of the cross-cap amplitude and the multiplication] ~\\
We can move the position of a cross-cap through the pants diagram:
\begin{equation}
\adjincludegraphics[valign = c, width = 2.25cm]{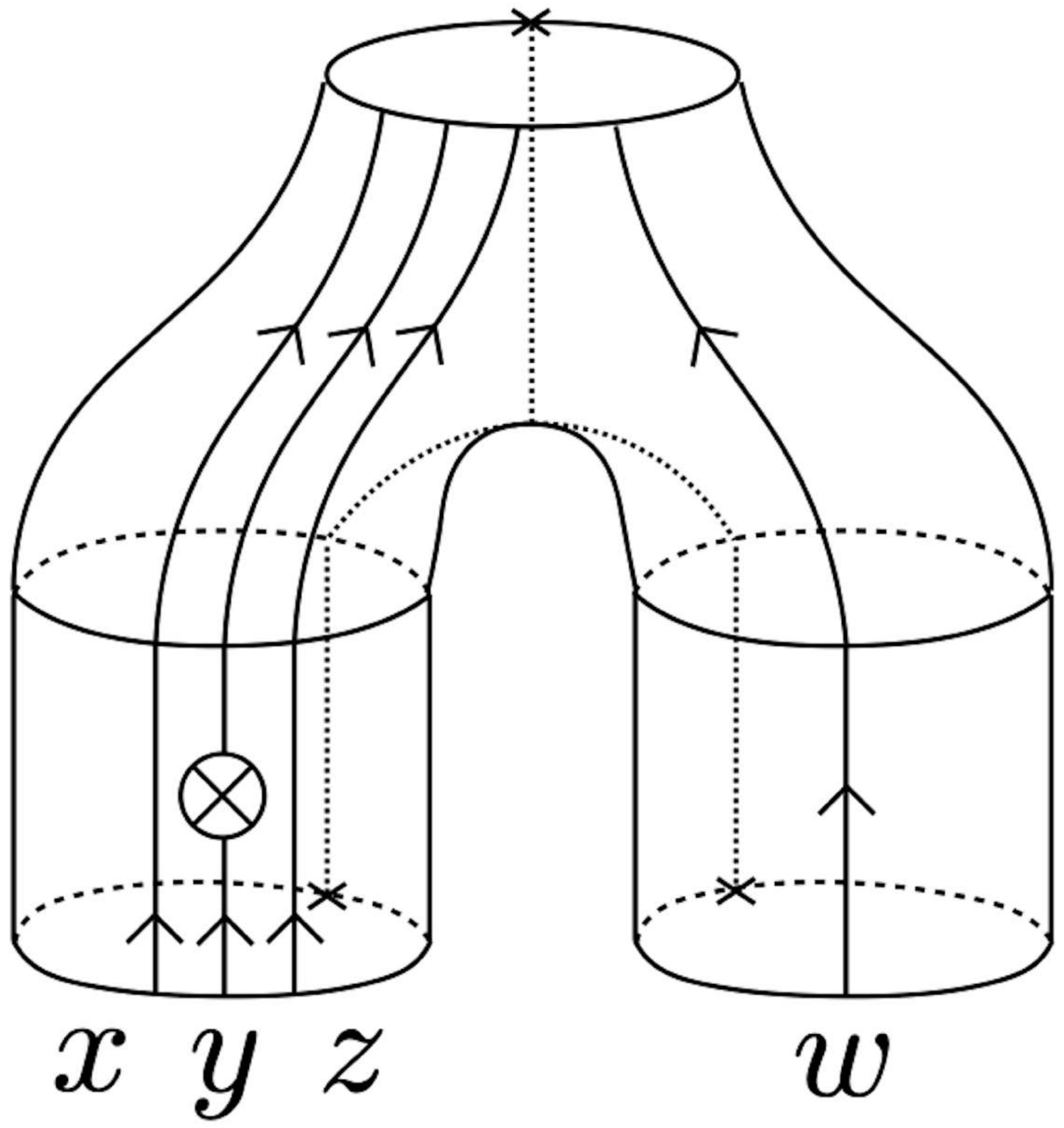} ~ = ~ \adjincludegraphics[valign = c, width = 2.25cm]{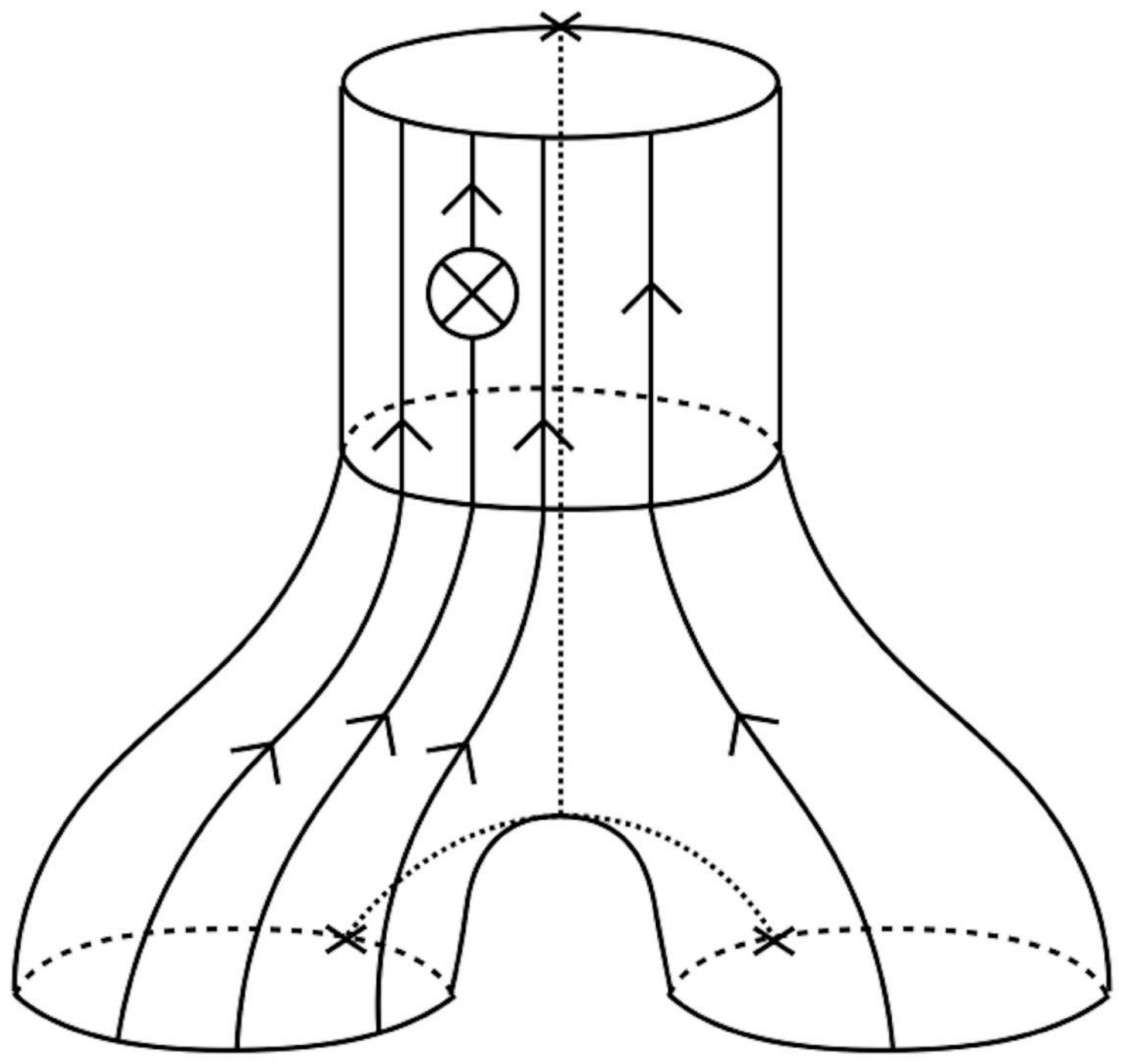}, \quad
\adjincludegraphics[valign = c, width = 2.25cm]{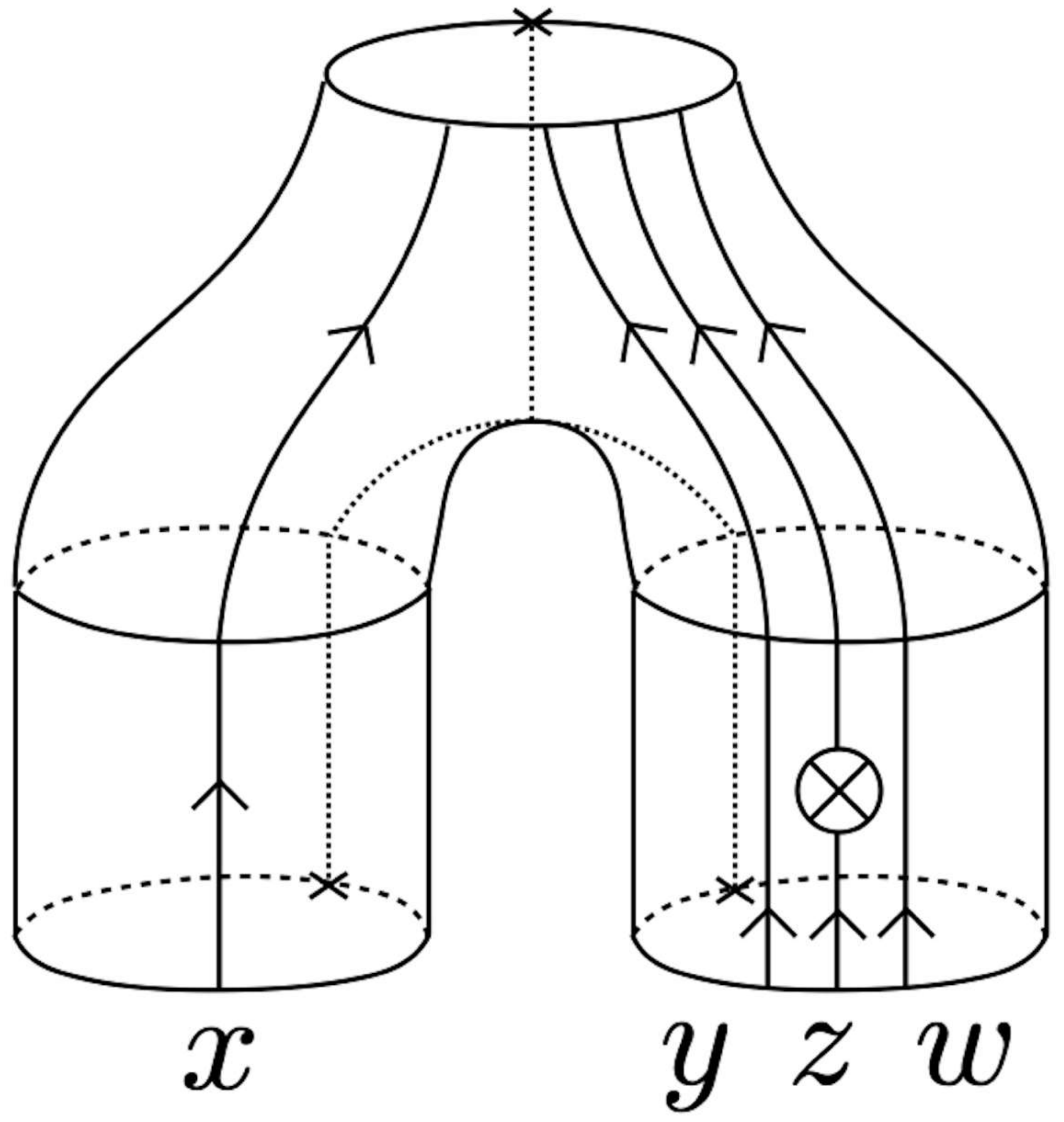} ~ = ~ \adjincludegraphics[valign = c, width = 2.25cm]{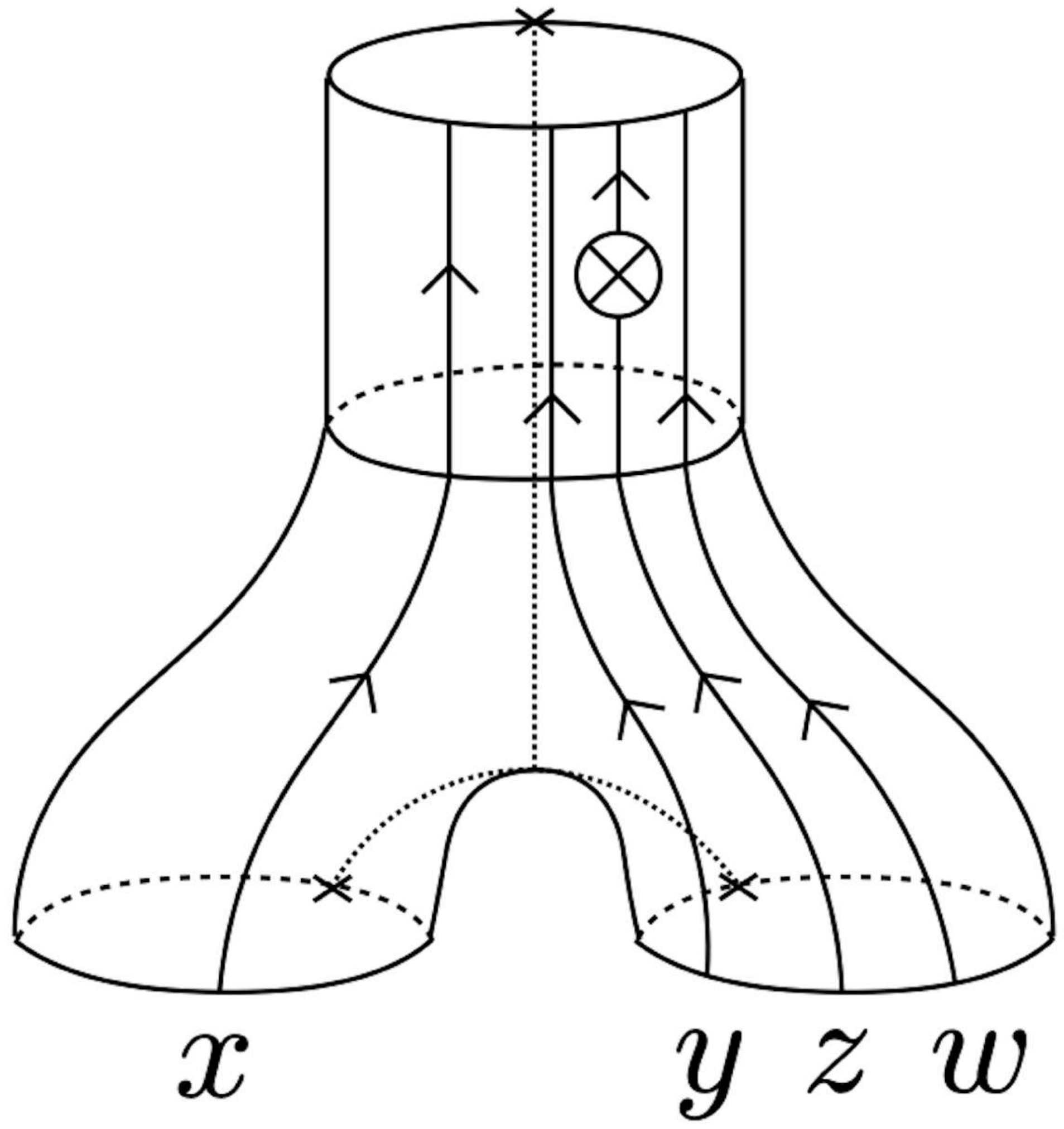}. 
\tag{U10}
\label{eq: compatibility of theta and M}
\end{equation}

\item[Klein identity] ~\\
Two different decompositions of the twice-punctured Klein bottle give rise to the same transition amplitude:
\begin{equation}
\adjincludegraphics[valign = c, width = 1.5cm]{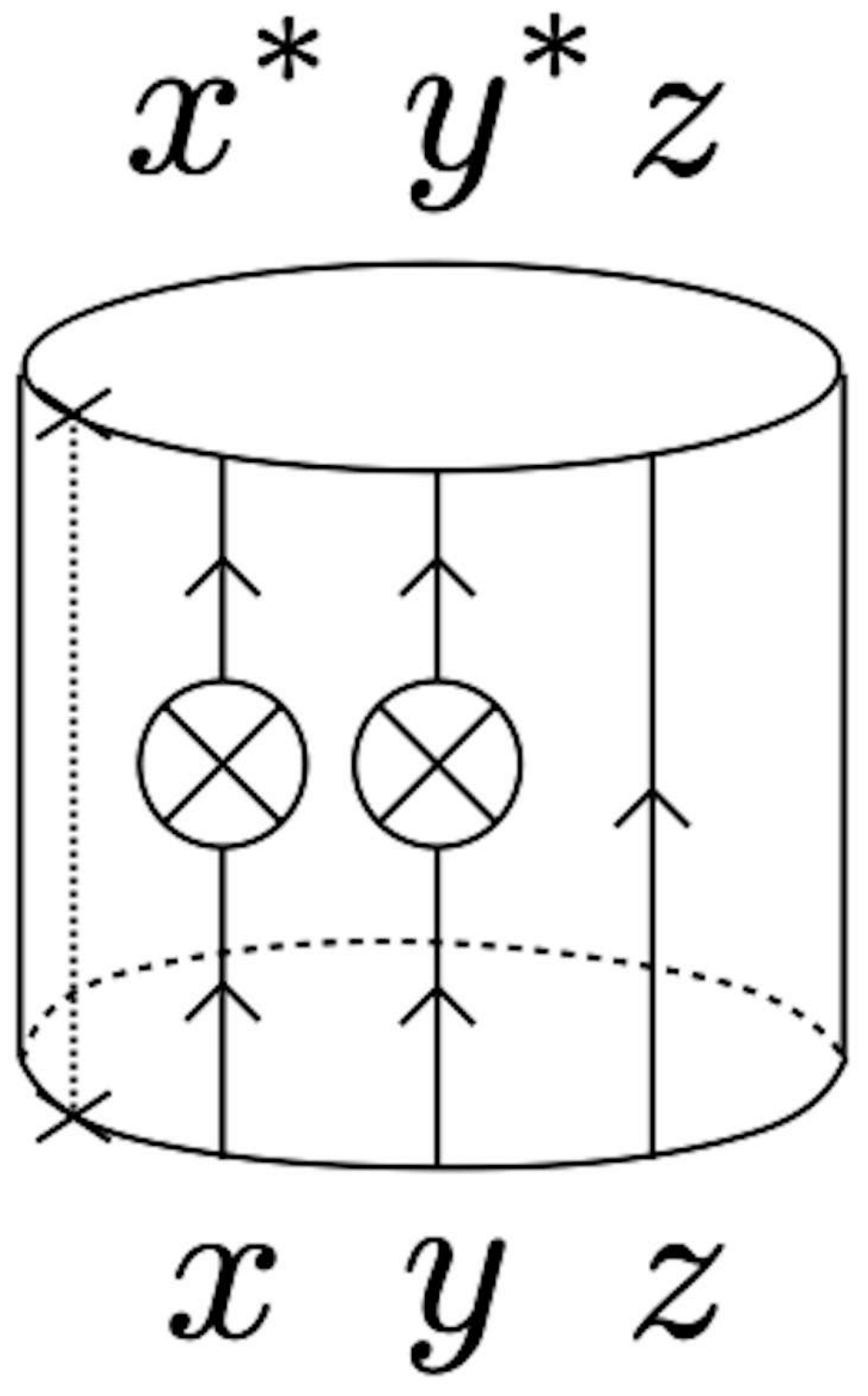} ~ = ~ \adjincludegraphics[valign = c, width = 2cm]{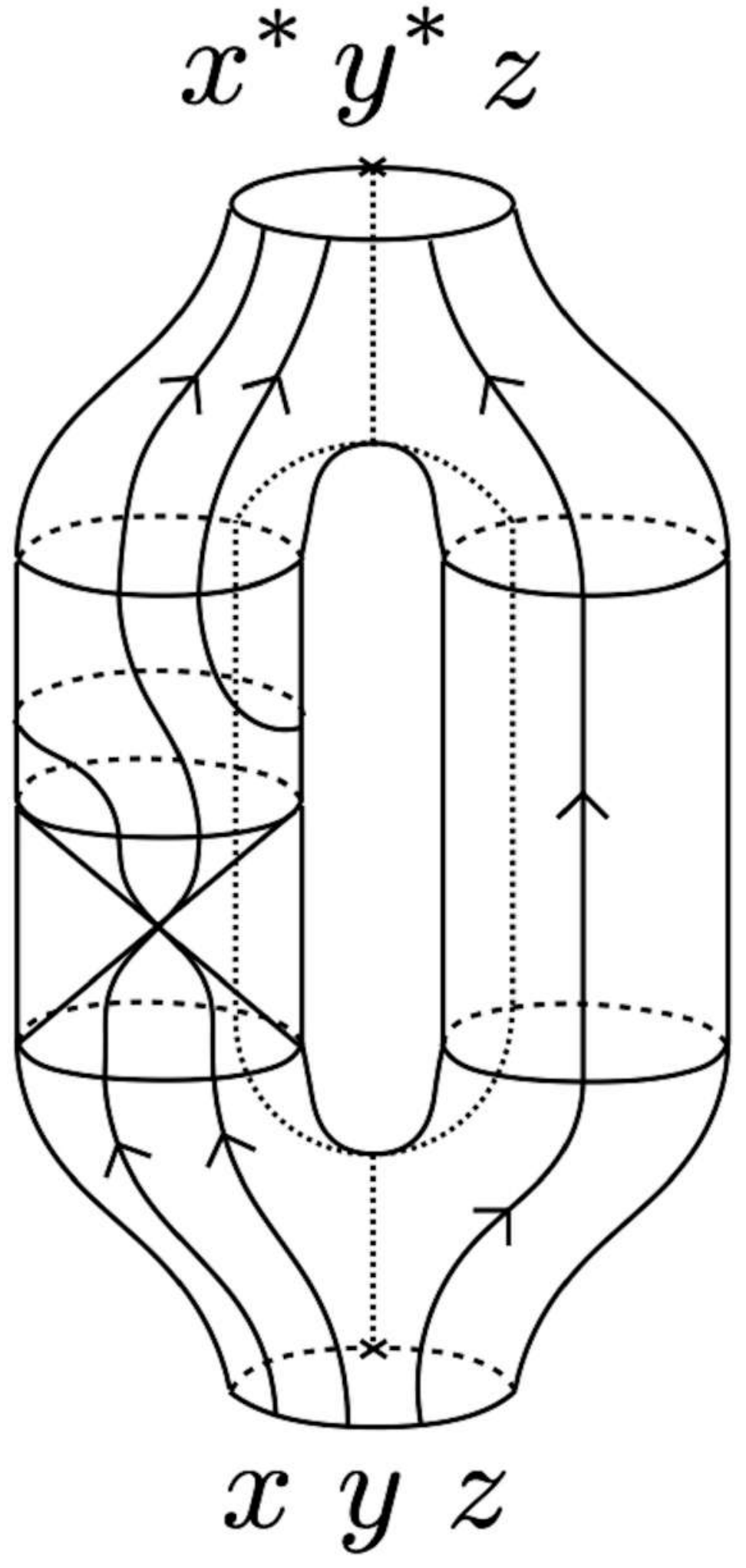}.
\tag{U11}
\label{eq: Klein}
\end{equation}
\end{description}

We need to check the consistency of the assignment of the transition amplitudes.
There are three types of consistency as follows:
\begin{enumerate}
\item consistency under changes of the time function,
\item consistency under deformations of topological defects on each basic element, and 
\item consistency under the composition of the orientation-reversing isomorphism. 
\end{enumerate}
Let us begin with the consistency of type 1.
The transition amplitudes must be invariant under changes of the time function.
As discussed in the appendix of \cite{KT2017}, we have two consistency conditions of this type for unoriented surfaces: one is the consistency on a pair of pants with a cross-cap and the other is the consistency on a twice-punctured Klein bottle.
These consistency conditions are given by (\ref{eq: compatibility of theta and M}) and (\ref{eq: Klein}) respectively.
Next, we consider the consistency of type 2.
Topologically equivalent configurations of topological defects on a basic element give rise to the same transition amplitude.
For this type of consistency conditions, it suffices to consider the consistency on a cylinder with a cross-cap because the consistency on the other basic elements is already considered in oriented TQFTs.
The consistency on a cylinder with a cross-cap is given by (\ref{eq: compatibility of left-theta and right-theta}) and (\ref{eq: reconnection}).
Finally, we have the consistency conditions of type 3 for each basic element as follows: (\ref{eq: involution}) and (\ref{eq: commutativity of X and phi}) for a cylinder, (\ref{eq: invariant unit}) for a cap, (\ref{eq: equivariance}) for a pair of pants, and (\ref{eq: compatibility of theta and phi}) and (\ref{eq: Mobius}) for a cylinder with a cross-cap.
The remaining equation (\ref{eq: orientation reversal of f}) is the consistency of topological point operators.
This completes the consistency conditions on transition amplitudes for unoriented surfaces.
Therefore, a two-dimensional bosonic unoriented TQFT with fusion category symmetry $\mathcal{C}$ is described by a sextuple $(Z, X, M, i, \phi, \theta)$ that satisfies (\ref{eq: composition on a cylinder})--(\ref{eq: consistency on the torus}) and (\ref{eq: invariant unit})--(\ref{eq: Klein}).
As we will see in appendix \ref{sec: Unoriented equivariant TQFTs}, this reduces to an unoriented equivariant TQFT when the symmetry is given by a finite group $\mathbb{Z}_2^T \times G$.

\subsection{Bosonic fusion category SPT phases with time-reversal symmetry}
\label{sec: Bosonic fusion category SPT phases with time-reversal symmetry}
In this section, we classify fusion category SPT phases with time-reversal symmetry by solving the consistency conditions (\ref{eq: invariant unit})--(\ref{eq: Klein}) under the constraints (\ref{eq: unitary unit}) and (\ref{eq: unitary M}).
We first recall that the triple $(Z, M, i): \mathcal{C} \rightarrow \mathrm{Vec}$ is a fiber functor and $X$ is given by eq. (\ref{eq: X = McM^{-1}}).
In particular, the multiplication $M_{x, y}$ is unitary.
Therefore, the cross-cap amplitude $\theta_{x, y}$ is determined by $\theta_{x, 1}$ via eq. (\ref{eq: compatibility of theta and M}) as follows:
\begin{equation}
\adjincludegraphics[valign = c, width = 1.5cm]{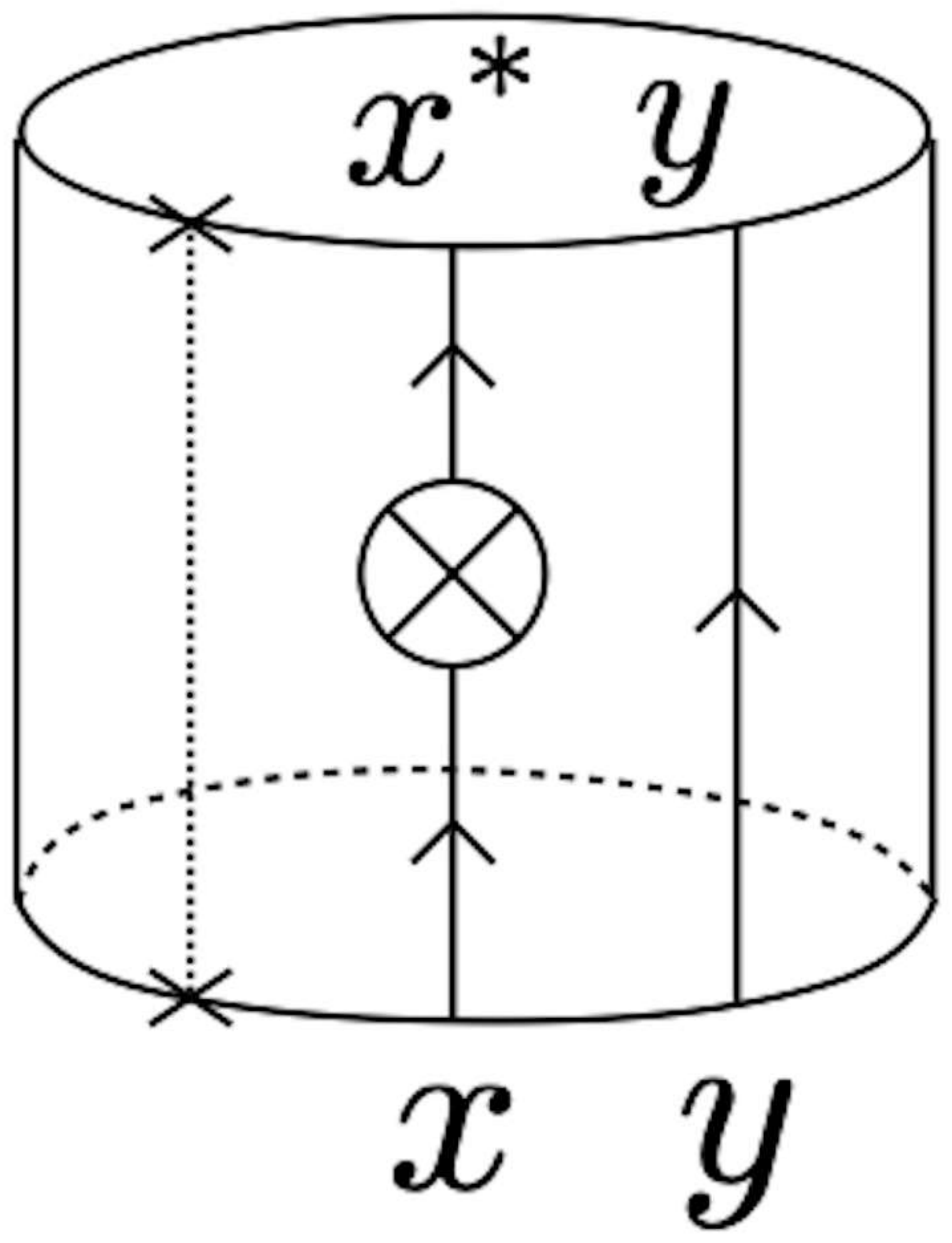} ~ = ~ \adjincludegraphics[valign = c, width =2cm]{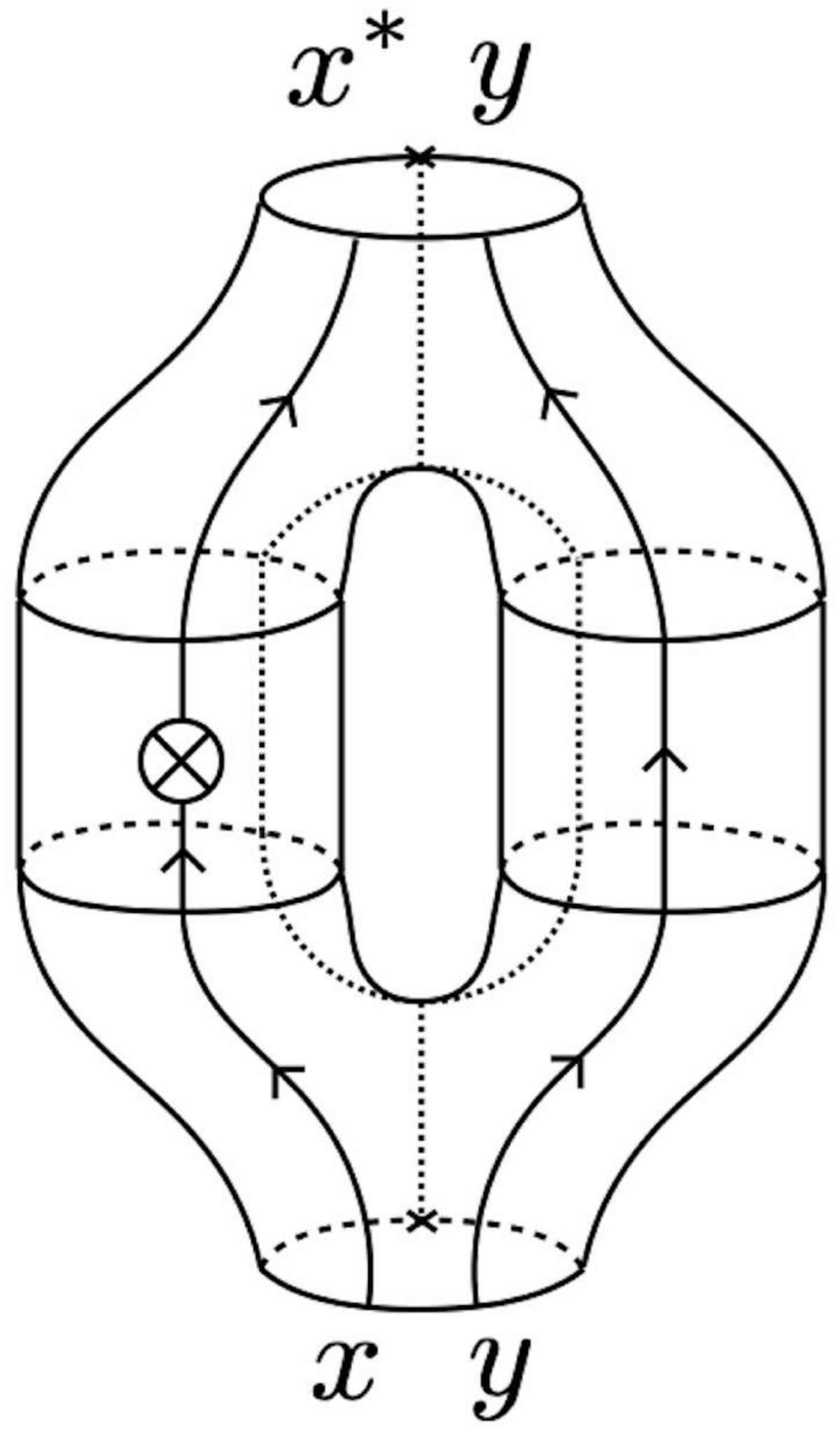}.
\label{eq: M-theta-M}
\end{equation}
Conversely, the consistency condition (\ref{eq: compatibility of theta and M}) follows from the above equation.
Furthermore, the Klein identity (\ref{eq: Klein}) reduces to
\begin{equation}
\adjincludegraphics[valign = c, width = 1.5cm]{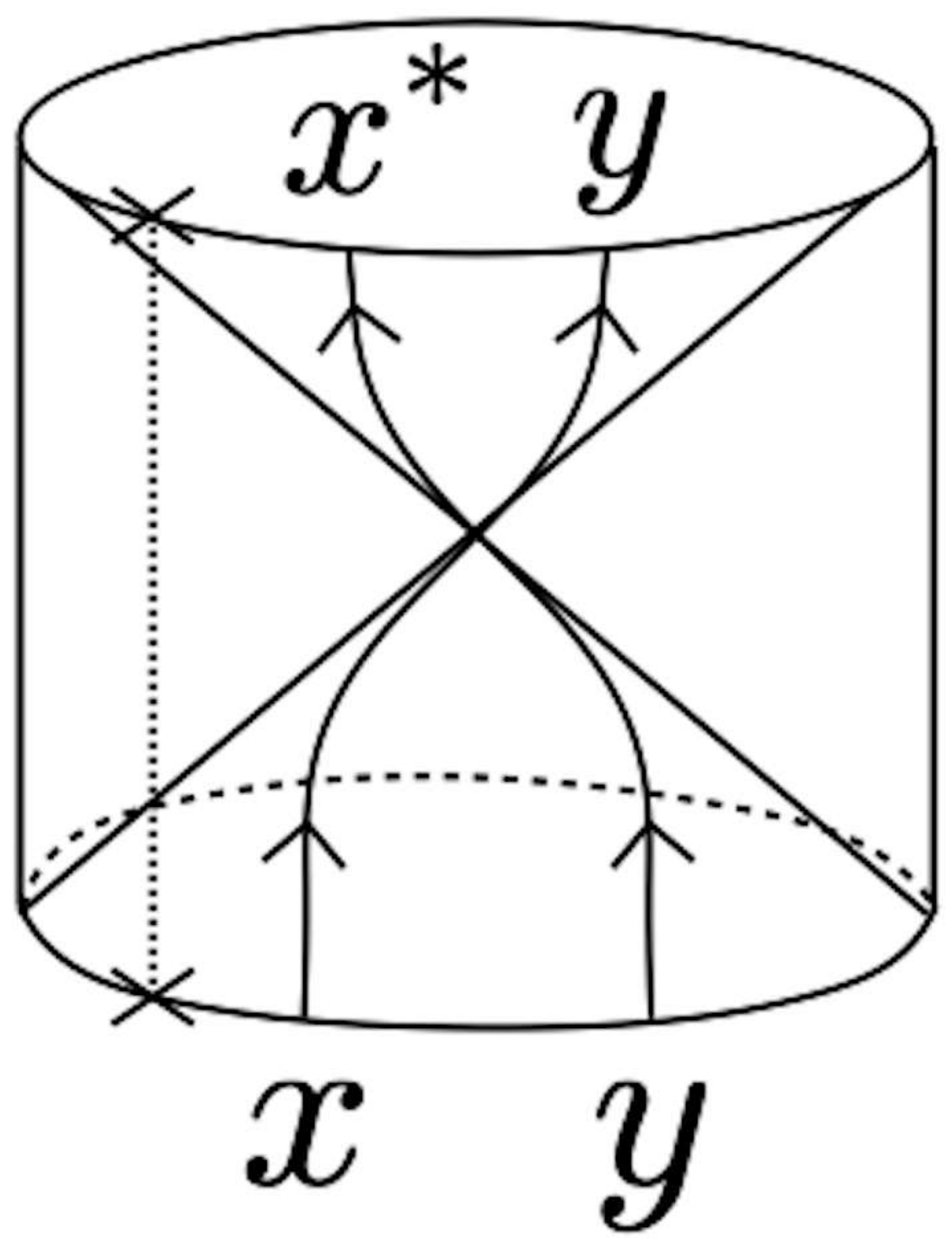} ~ = ~ \adjincludegraphics[valign = c, width = 1.5cm]{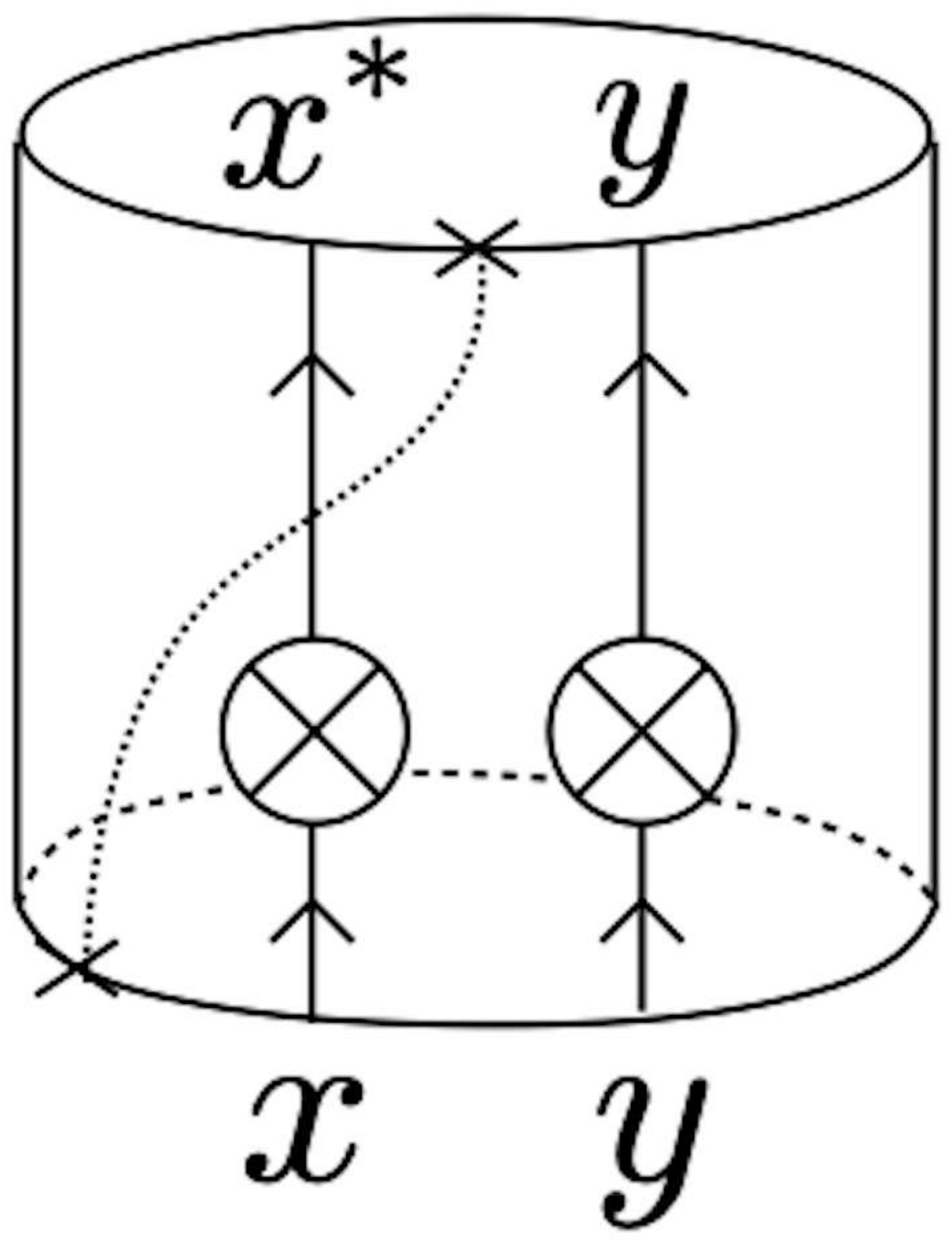}.
\label{eq: reduced Klein}
\end{equation}
This implies that the cross-cap amplitude $\theta_{x, y}$ is an isomorphism.
By using eqs. (\ref{eq: M-theta-M}) and (\ref{eq: reduced Klein}), the equivariance (\ref{eq: equivariance}) can be simplified as follows:
\begin{equation}
\adjincludegraphics[valign = c, width = 2.2cm]{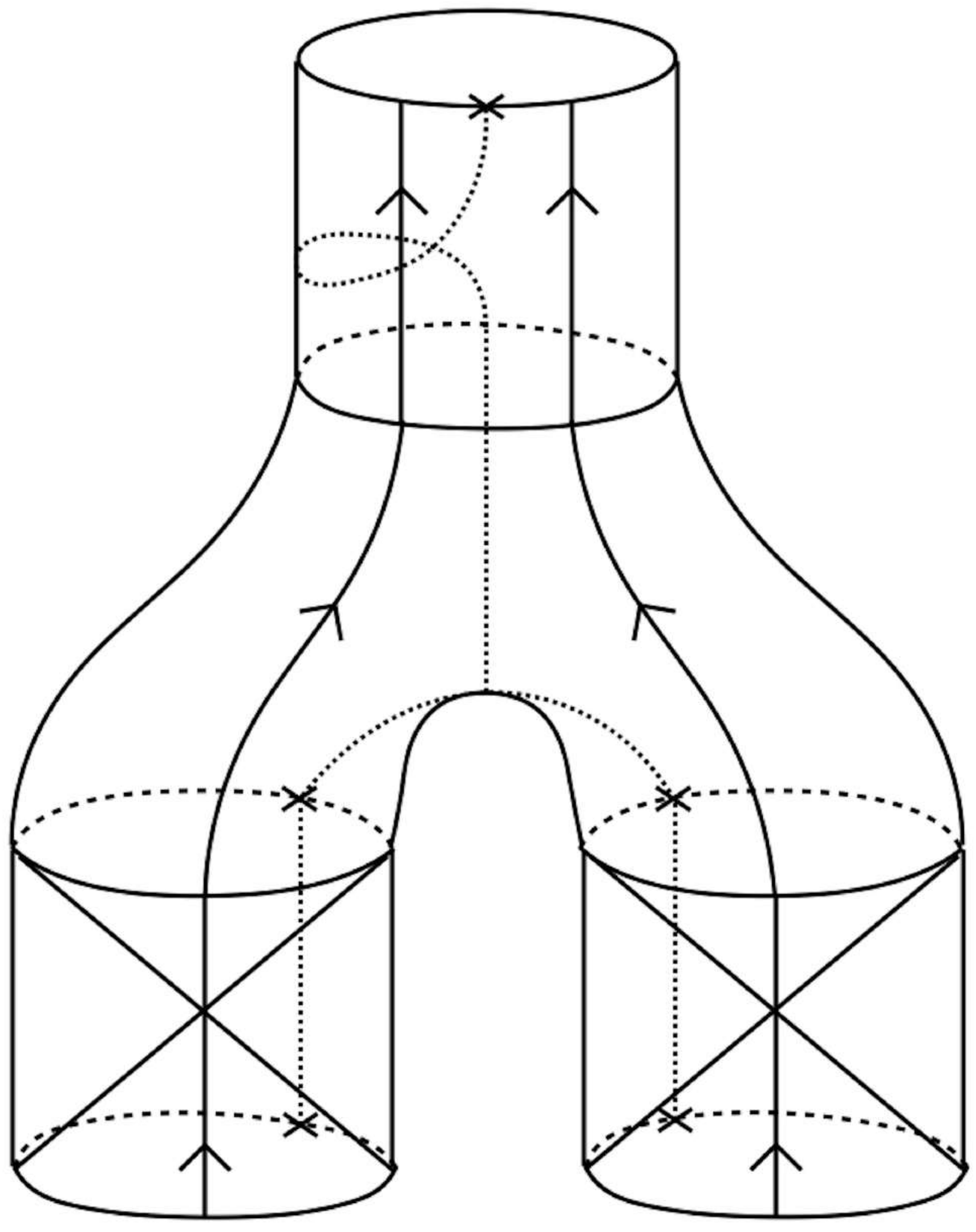} ~ = ~ \adjincludegraphics[valign = c, width = 2.2cm]{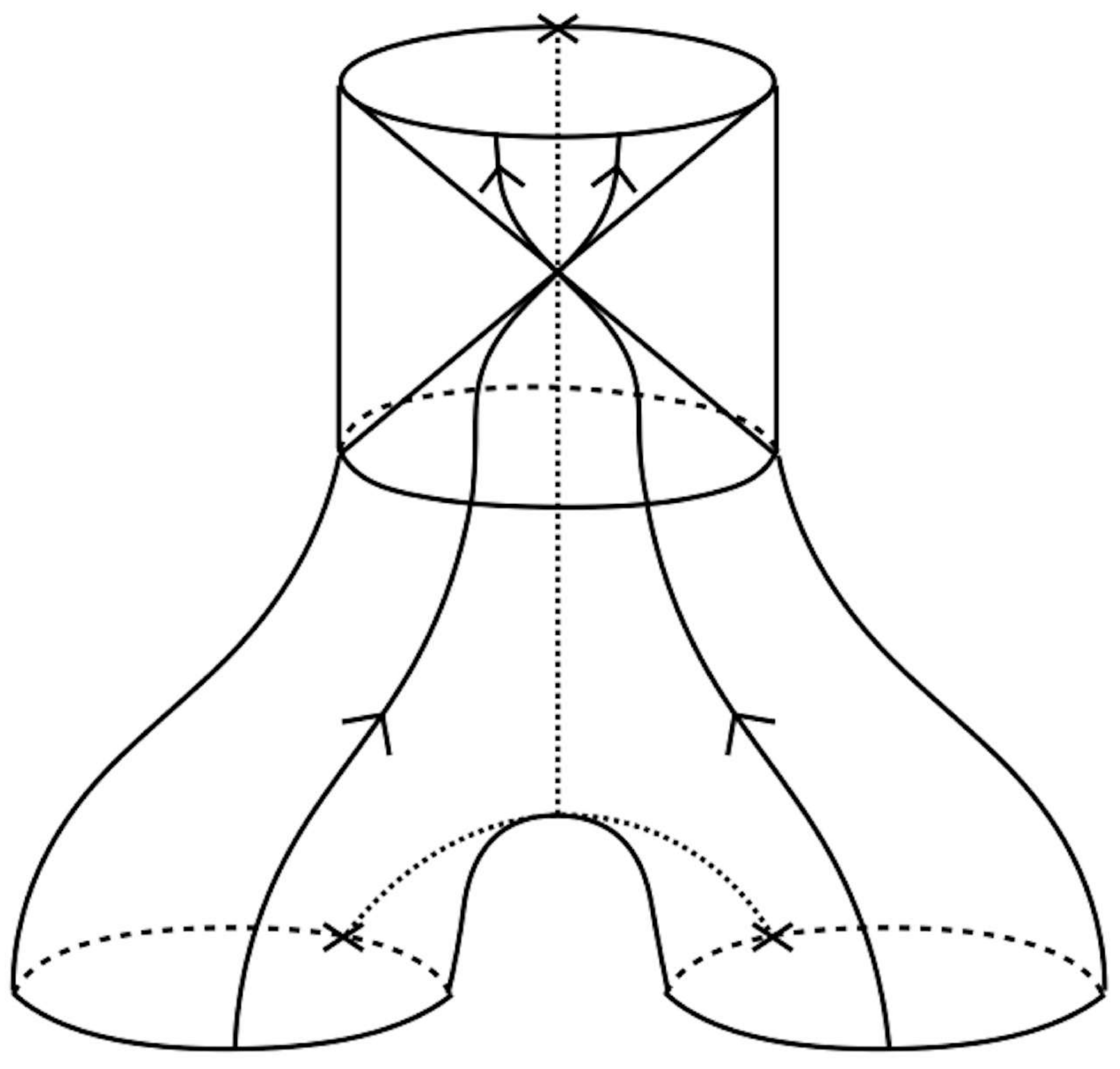}
\Leftrightarrow
\adjincludegraphics[valign = c, width = 1.5cm]{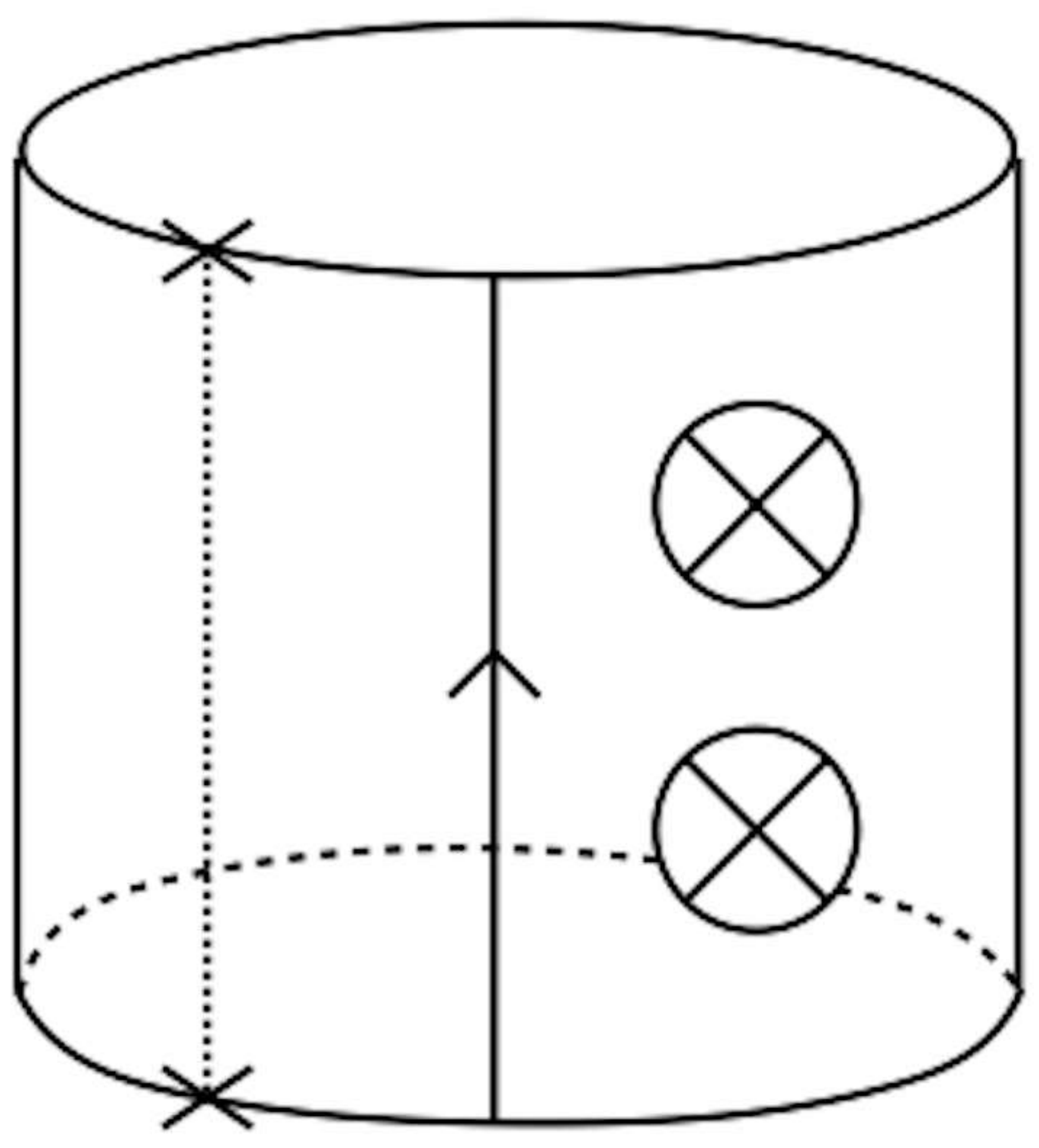} ~ = ~ \adjincludegraphics[valign = c, width = 1.5cm]{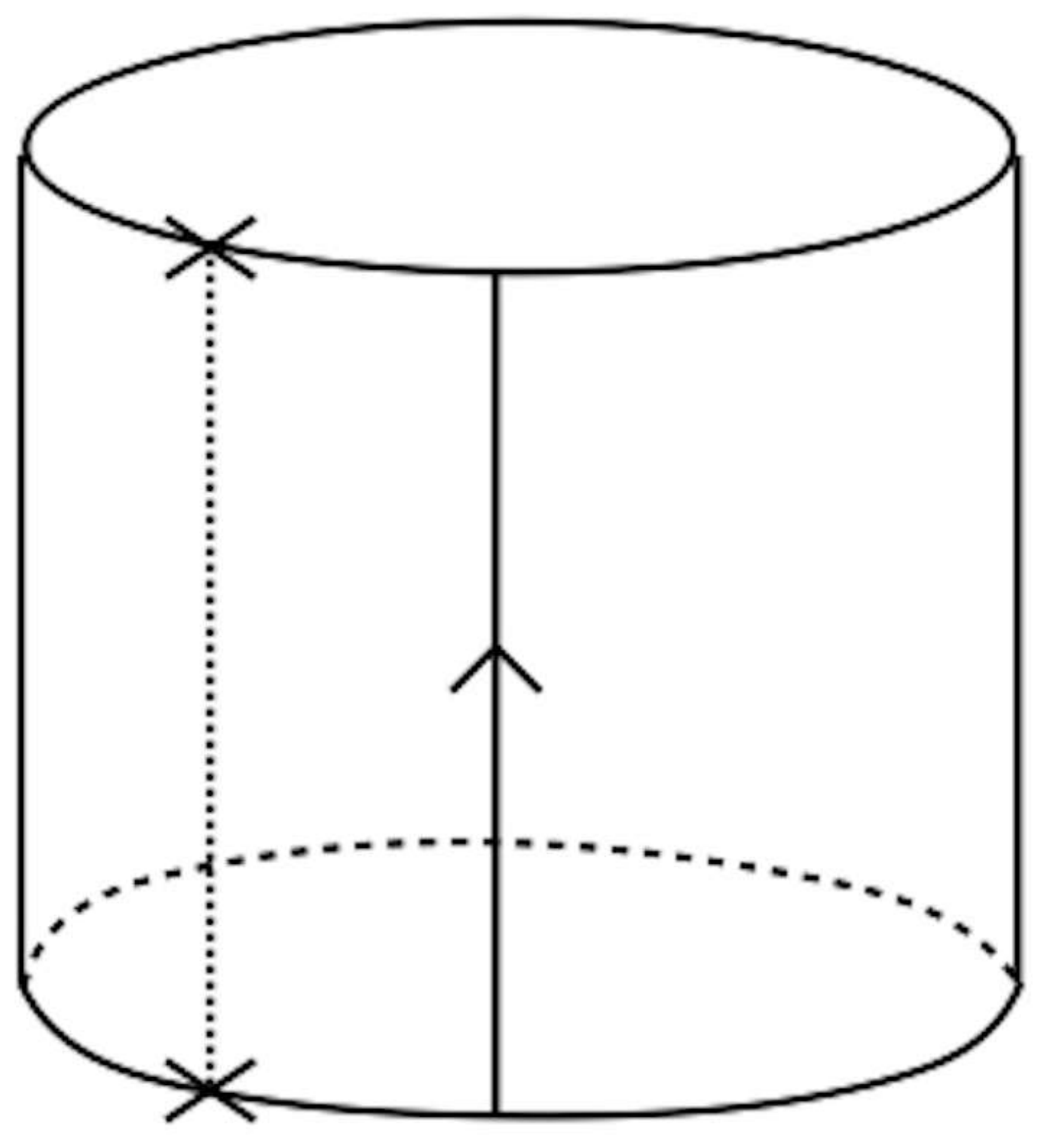}
\Leftrightarrow
\adjincludegraphics[valign = c, width = 1.5cm]{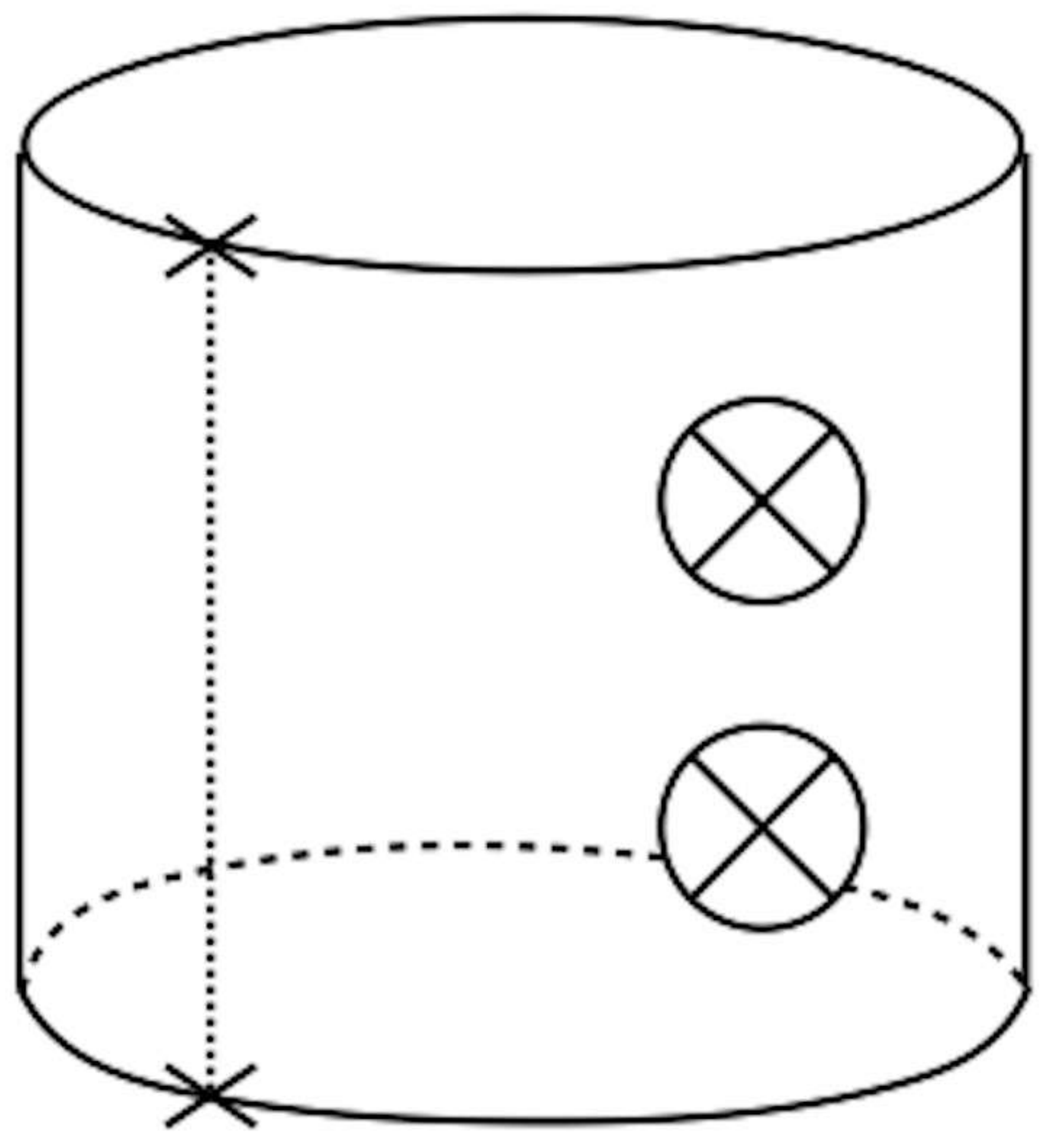} ~ = ~ \adjincludegraphics[valign = c, width = 1.5cm]{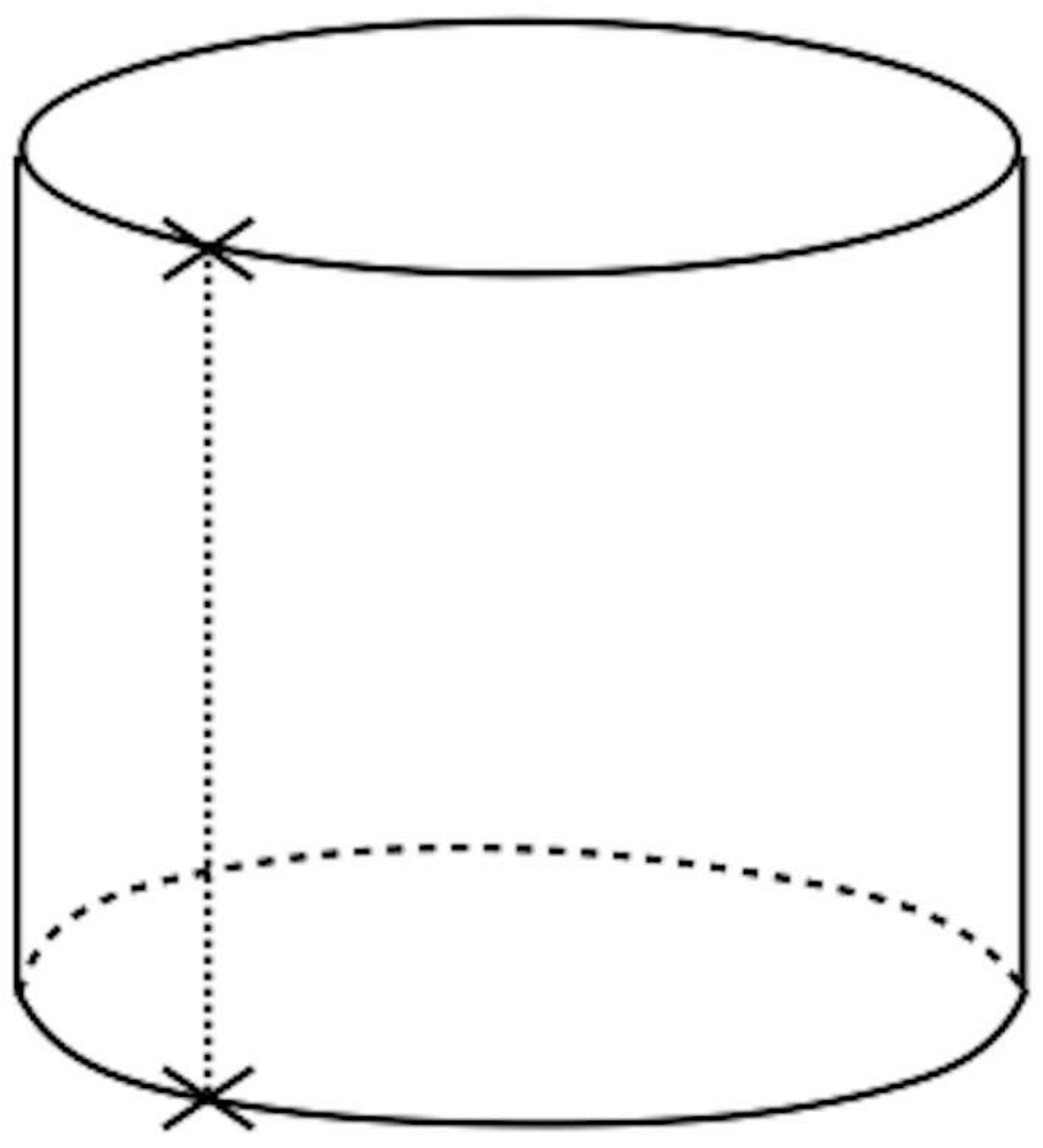}.
\label{eq: involutive cross-cap}
\end{equation}
More generally, we obtain
\begin{equation}
\adjincludegraphics[valign = c, width = 1.5cm]{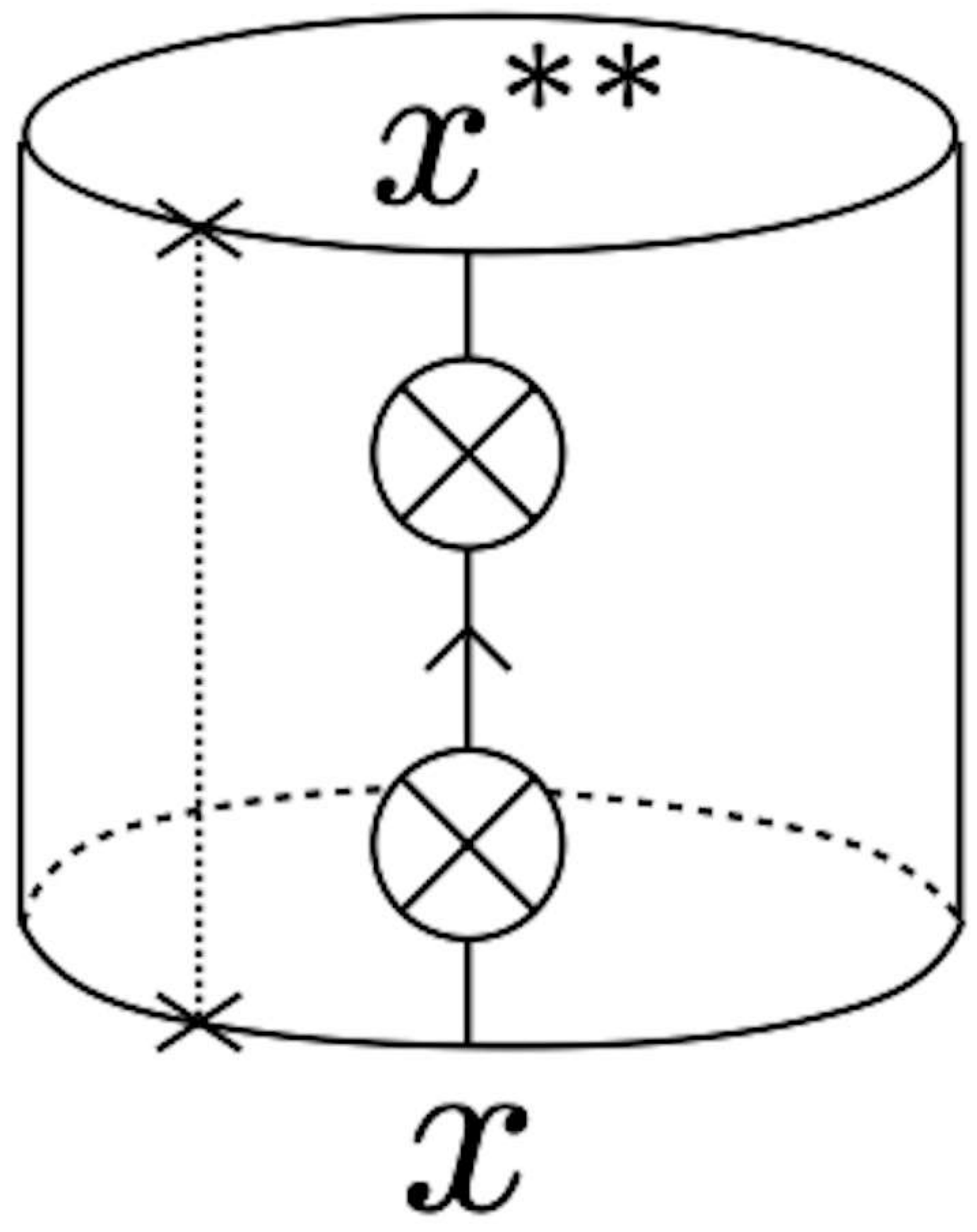} ~ = ~ \adjincludegraphics[valign = c, width = 1.5cm]{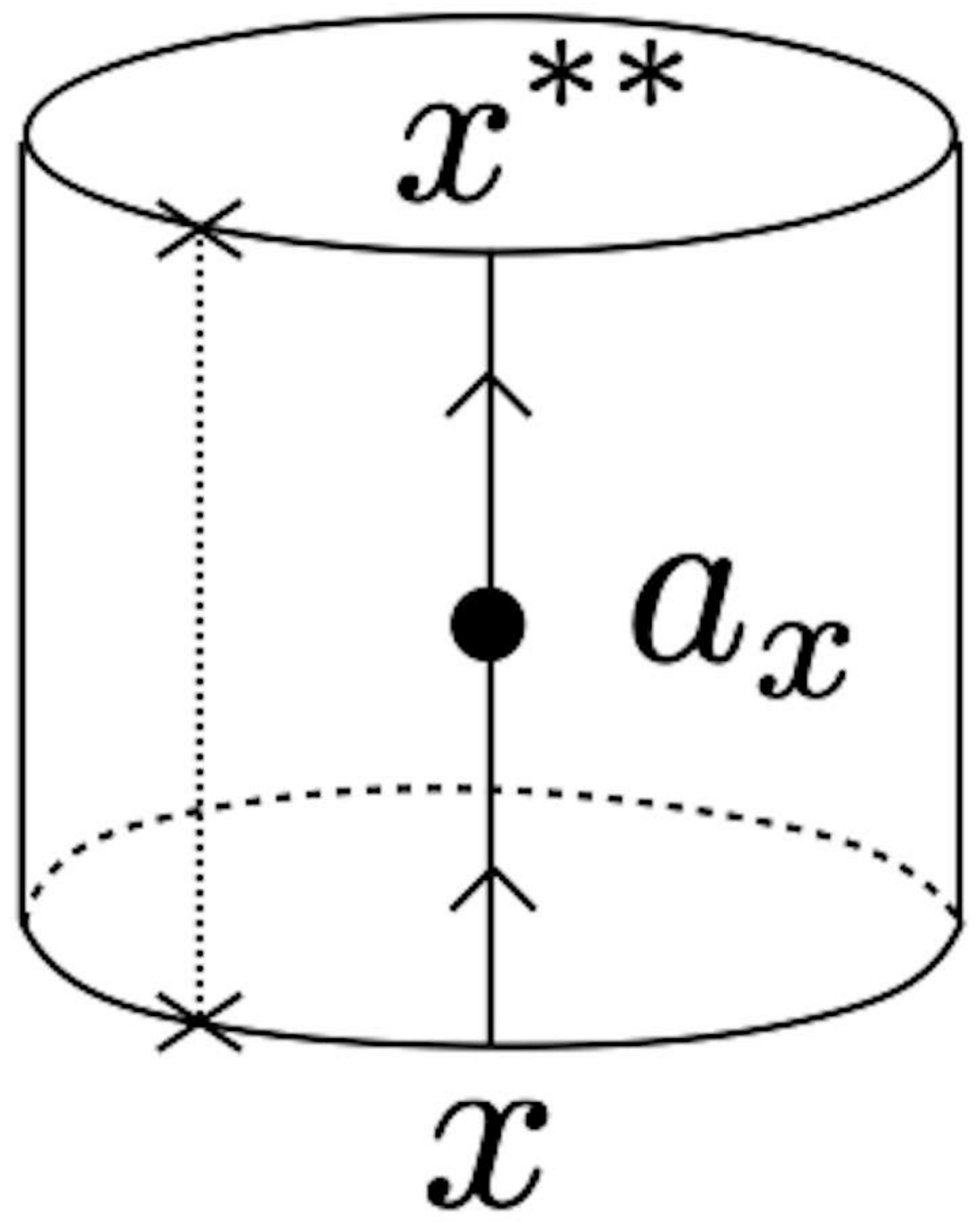}, 
\label{eq: involutive cross-cap2}
\end{equation}
from (\ref{eq: involution}).
The above equations indicate that the orientation-reversing isomorphism acts trivially on the vector space $V_1$, which shows (\ref{eq: invariant unit}).

As we will see below, it turns out that the remaining consistency conditions (\ref{eq: commutativity of X and phi})--(\ref{eq: Mobius}) follow from eqs. (\ref{eq: M-theta-M}), (\ref{eq: reduced Klein}), (\ref{eq: involutive cross-cap2}) and the consistency conditions of oriented TQFTs.
We first notice that (\ref{eq: commutativity of X and phi}), (\ref{eq: orientation reversal of f}), (\ref{eq: compatibility of theta and phi}), and (\ref{eq: Mobius}) immediately follow from eqs. (\ref{eq: M-theta-M}), (\ref{eq: reduced Klein}), and (\ref{eq: involutive cross-cap2}).
To show the consistency condition (\ref{eq: compatibility of left-theta and right-theta}), it is convenient to compose the cross-cap amplitude to both sides:
\begin{equation}
\adjincludegraphics[valign = c, width = 1.5cm]{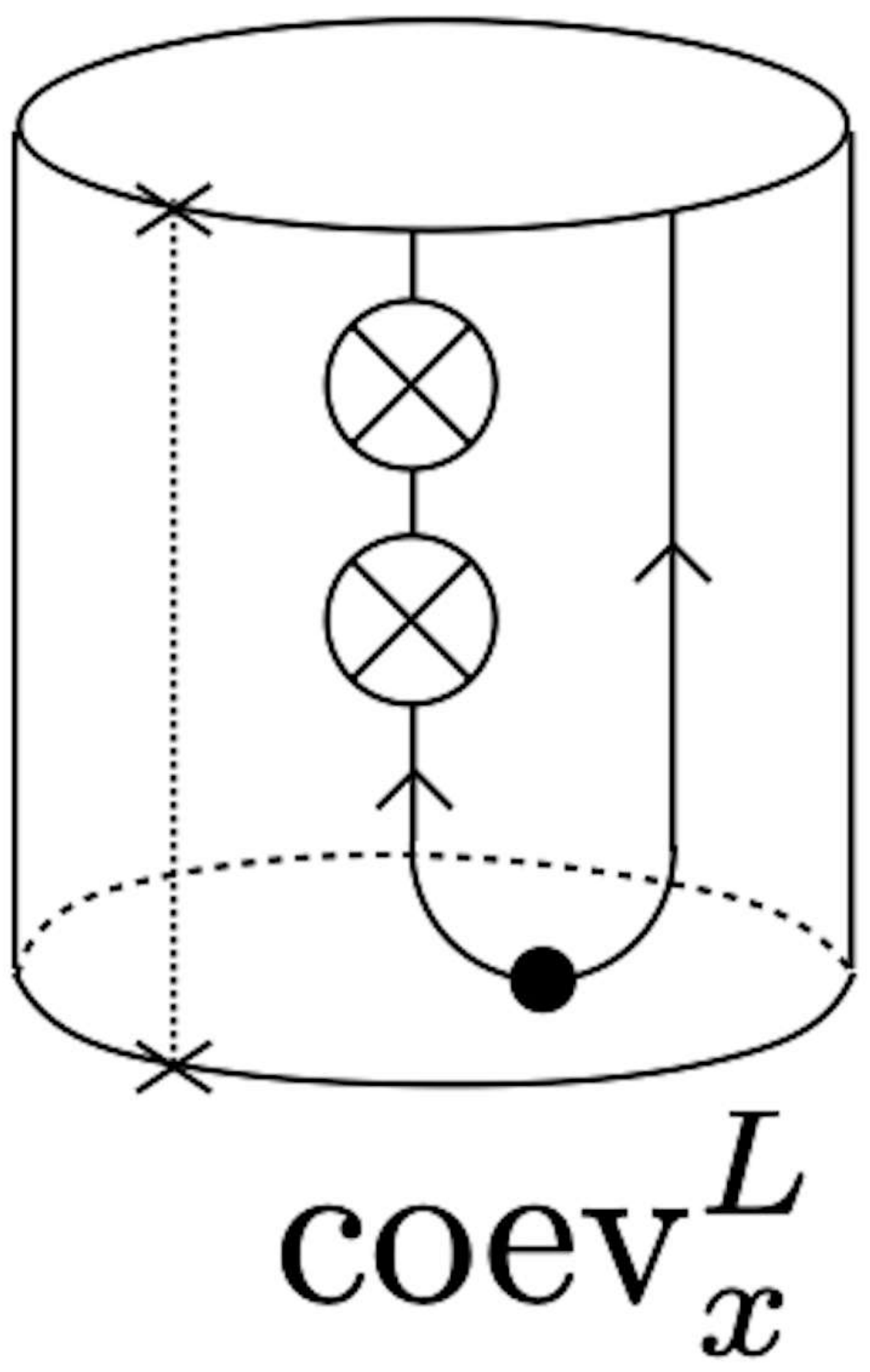} ~ = ~ 
\adjincludegraphics[valign = c, width = 1.5cm]{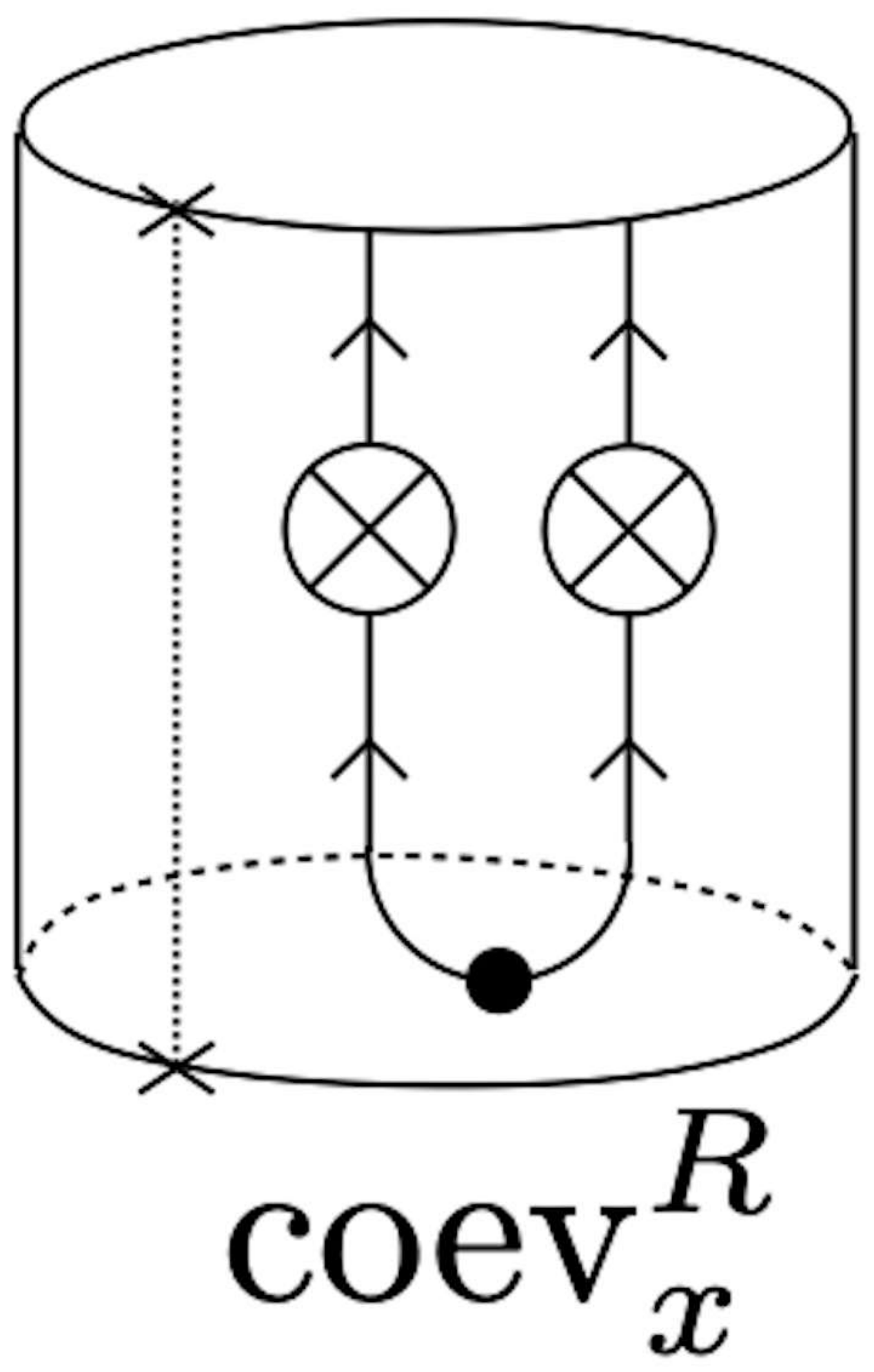}
\label{eq: U5 prime}
\end{equation}
This equation is equivalent to (\ref{eq: compatibility of left-theta and right-theta}) because the cross-cap amplitude is an isomorphism.
The left-hand side can be written as
\begin{equation}
\mathrm{(LHS)} = \adjincludegraphics[valign = c, width = 1.5cm]{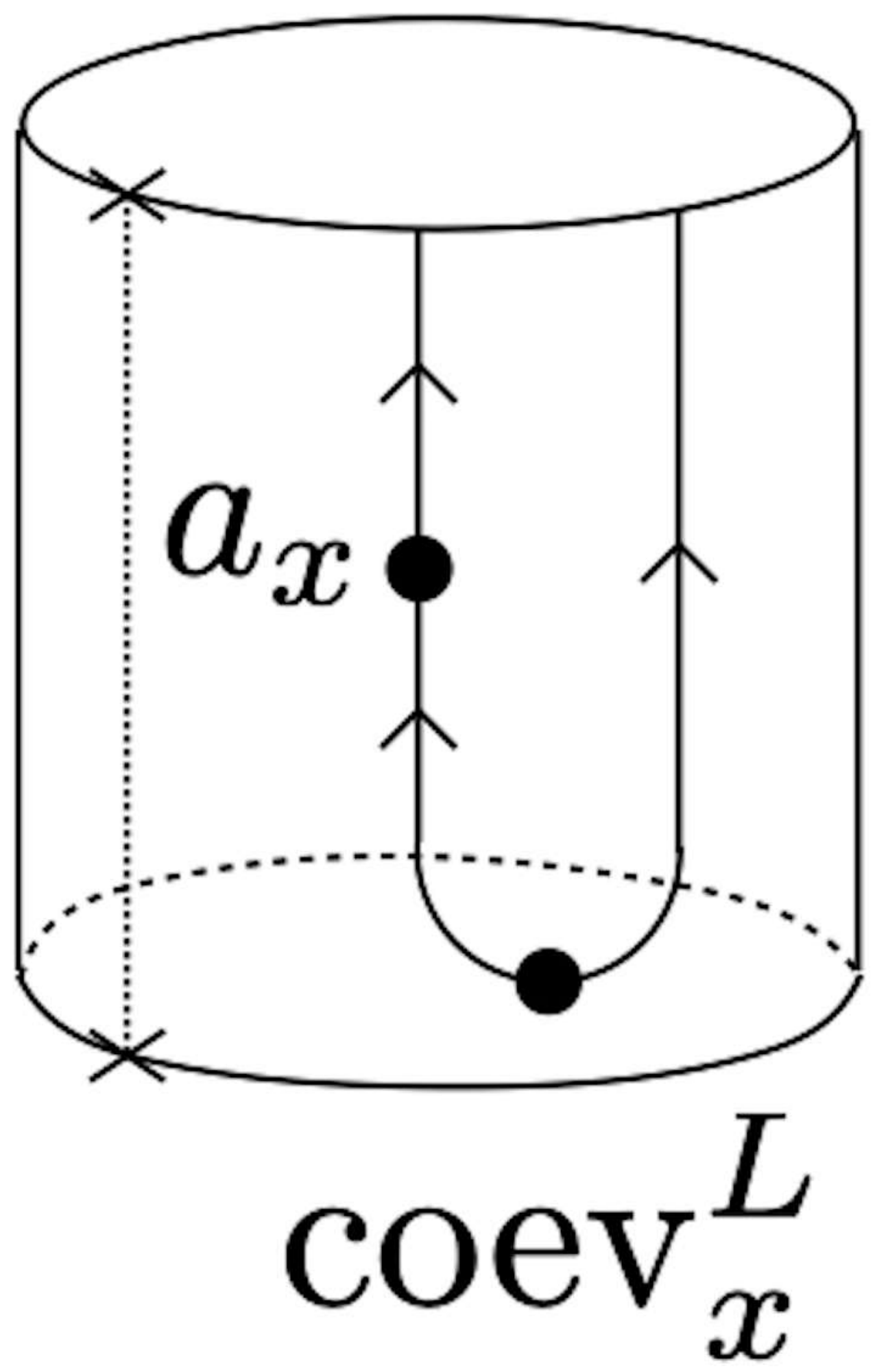} ~ = ~ 
\adjincludegraphics[valign = c, width = 1.5cm]{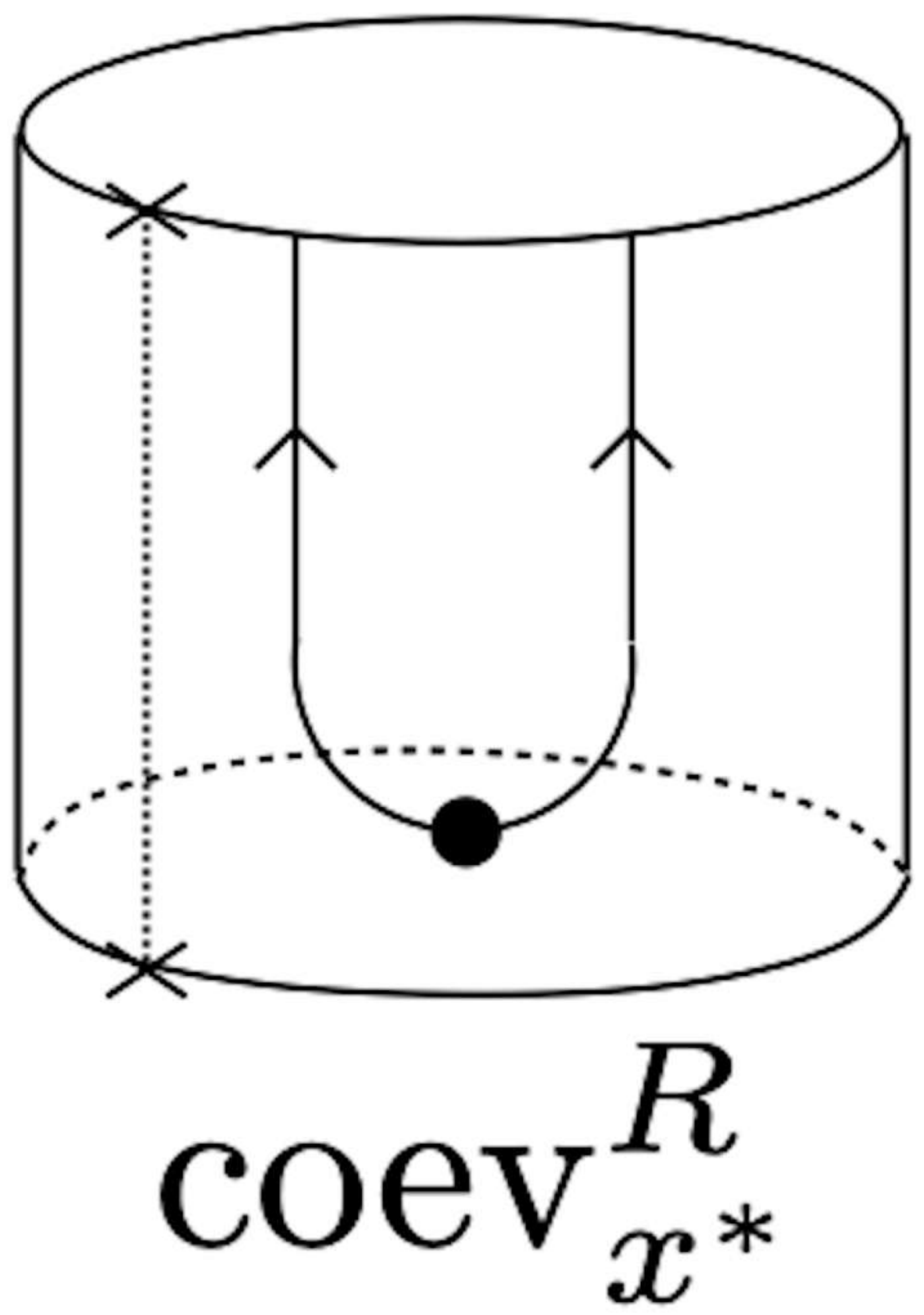},
\label{eq: U5 LHS}
\end{equation}
where the second equality follows from the relation (\ref{eq: canonical}) between the left and right coevaluation morphisms.
On the other hand, the right-hand side can be computed as
\begin{equation}
\mathrm{(RHS)} = \adjincludegraphics[valign = c, width = 1.5cm]{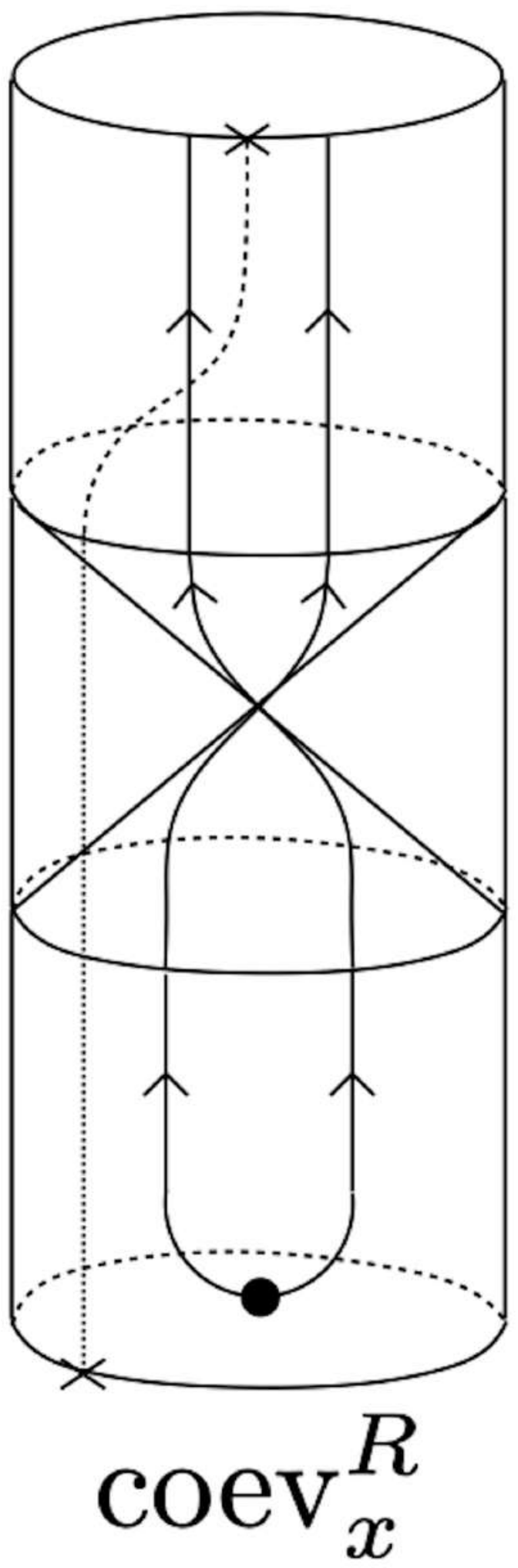} ~ = ~ 
\adjincludegraphics[valign = c, width = 1.5cm]{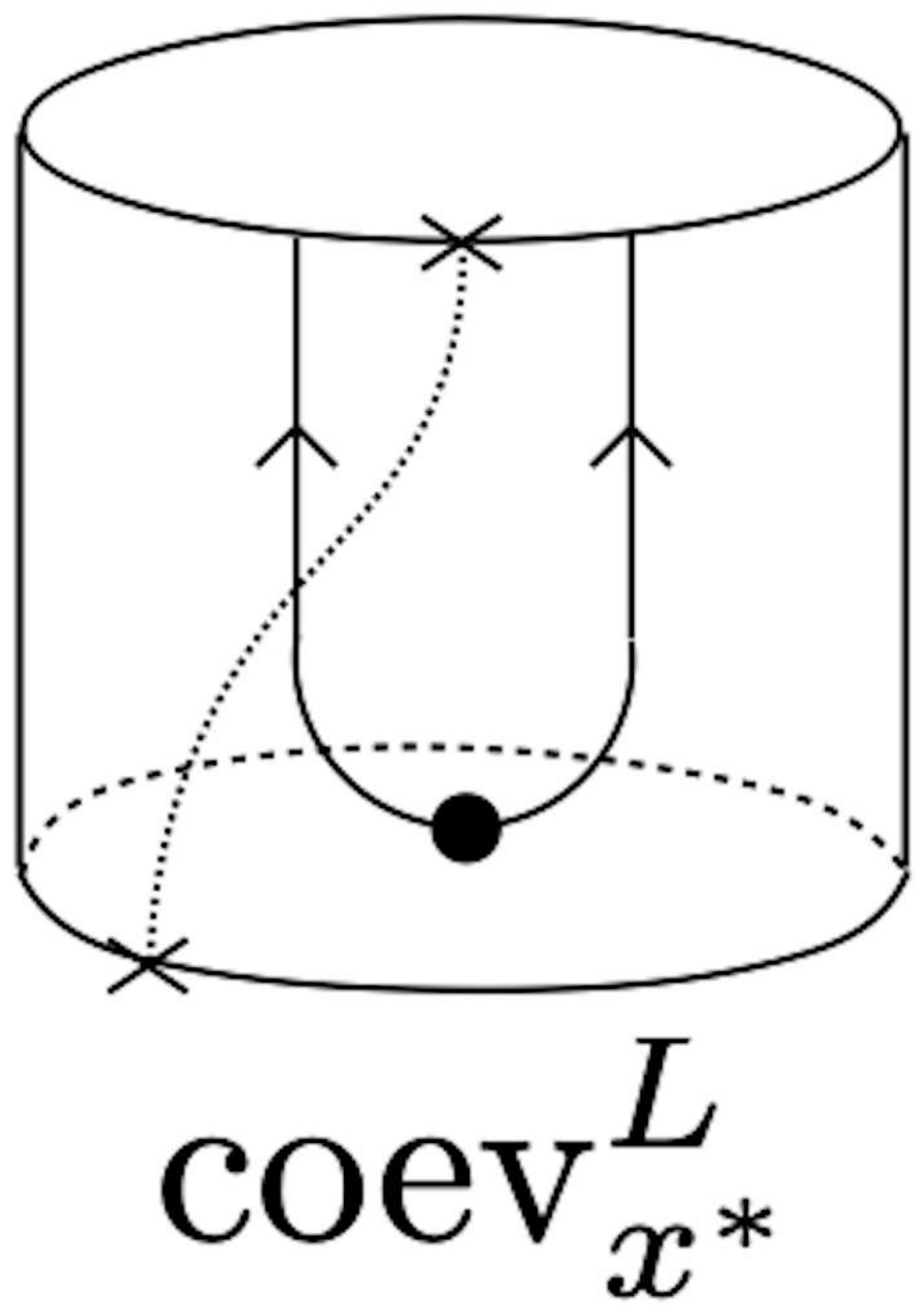} ~ = ~ 
\adjincludegraphics[valign = c, width = 1.5cm]{U5-4.pdf} = \mathrm{(LHS)}.
\label{eq: U5 RHS}
\end{equation}
The second equality follows from the fact that the orientation reversal of the right coevaluation morphism $\mathrm{coev}^R_{x}$ is given by $\overline{\mathrm{coev}^R_{x}} = \mathrm{coev}^L_{x^*}$, which can be seen from the unitarity of $\phi_x$ and the first equation of (\ref{eq: orientation reversal of f}).
Indeed, when $\phi_x$ is unitary, the orientation reversal of the coevaluation morphism defined by the first equation of (\ref{eq: orientation reversal of f}) satisfies
\begin{equation}
\adjincludegraphics[valign = c, width = 1.35cm]{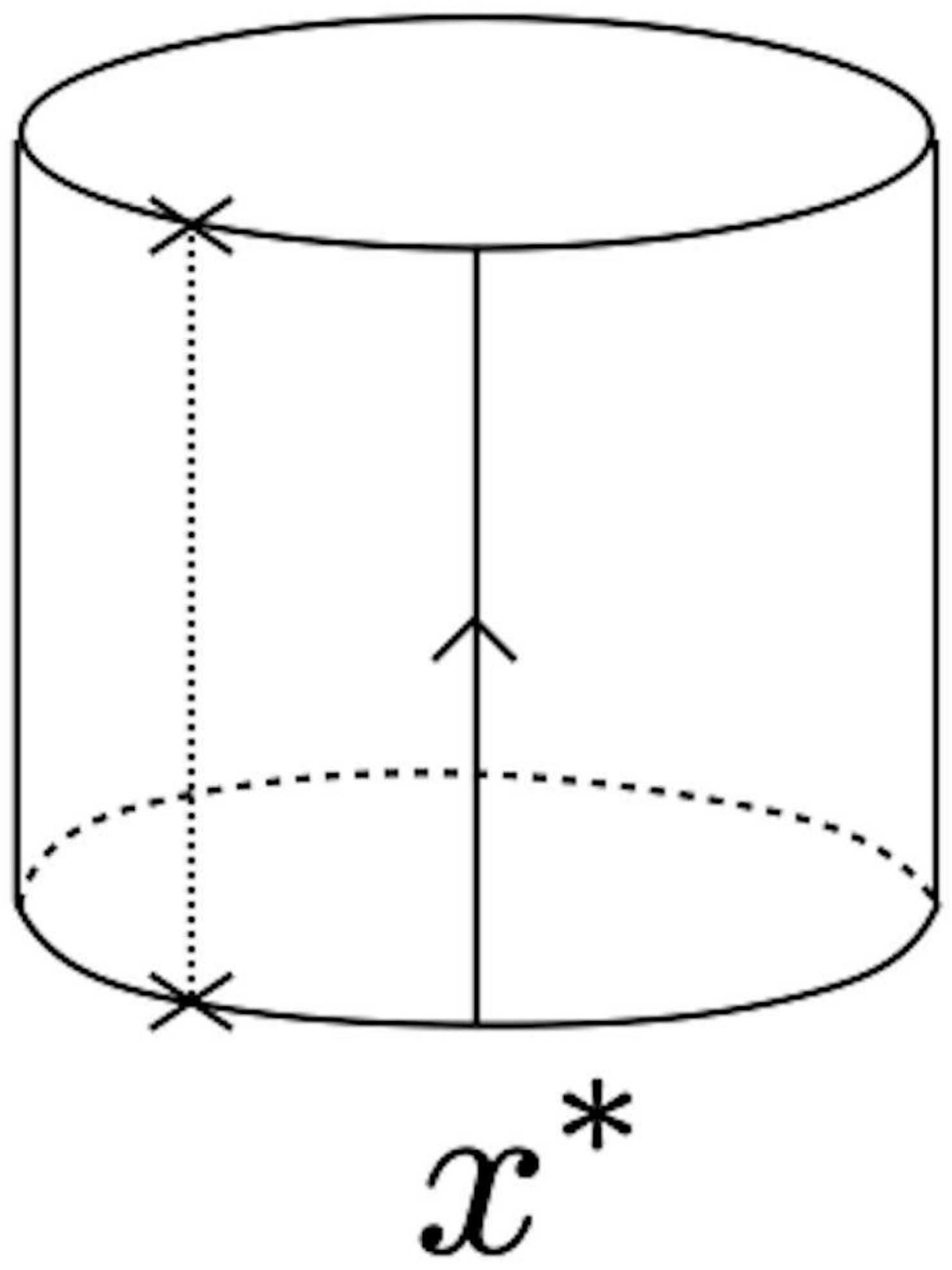} ~ = ~ \adjincludegraphics[valign = c, width = 2.25cm]{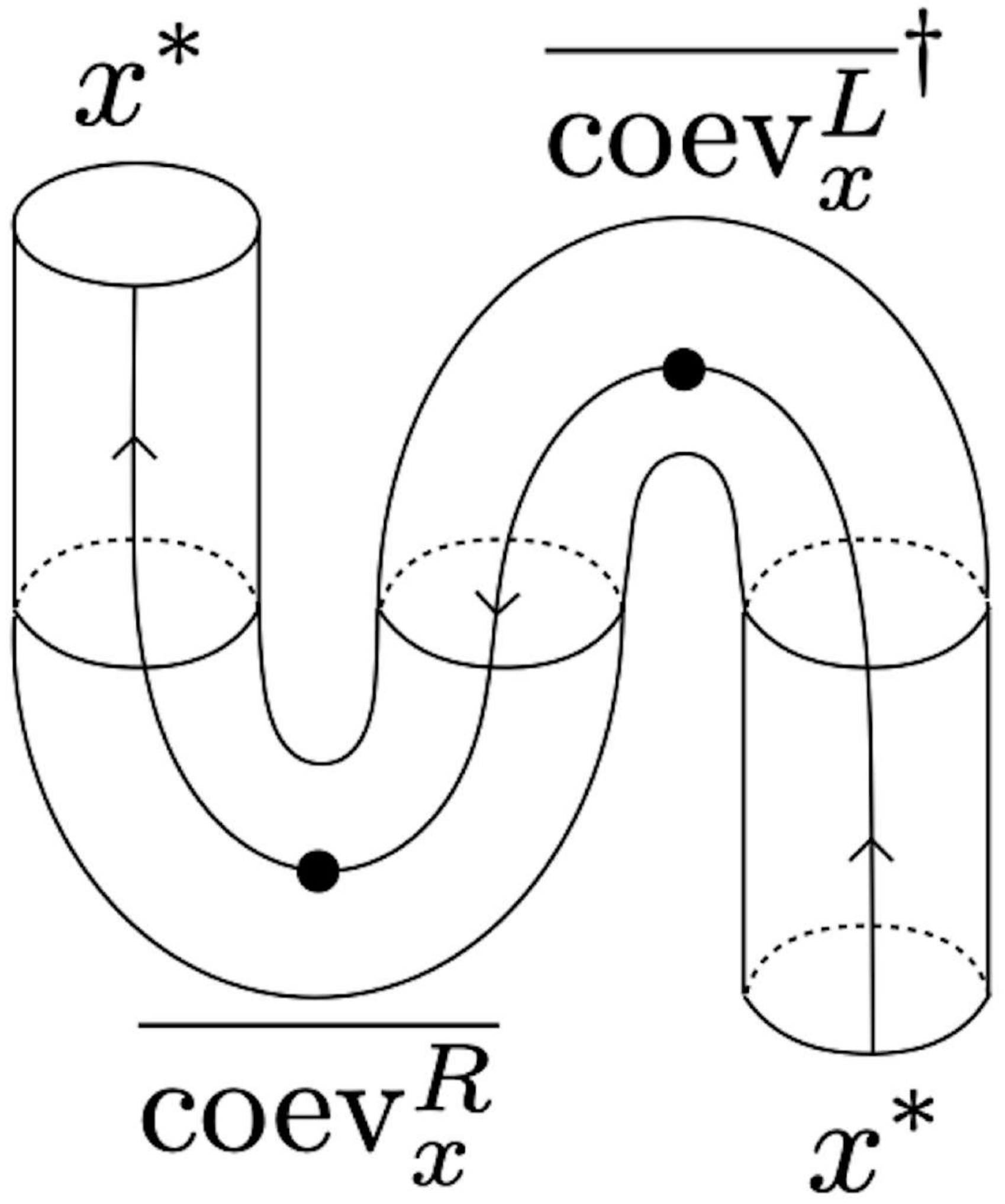},
\end{equation}
which implies $\overline{\mathrm{coev}^R_{x}} = \mathrm{coev}^L_{x^*}$ and $\overline{\mathrm{coev}^L_{x}} = \mathrm{coev}^R_{x^*}$.
We note that this shows the first equation of (\ref{eq: reconnection}). 
Finally, we consider the second equation of (\ref{eq: reconnection}).
The left-hand side and the right-hand side can be written as
\begin{equation}
\mathrm{(LHS)} = \adjincludegraphics[valign = c, width = 2.25cm]{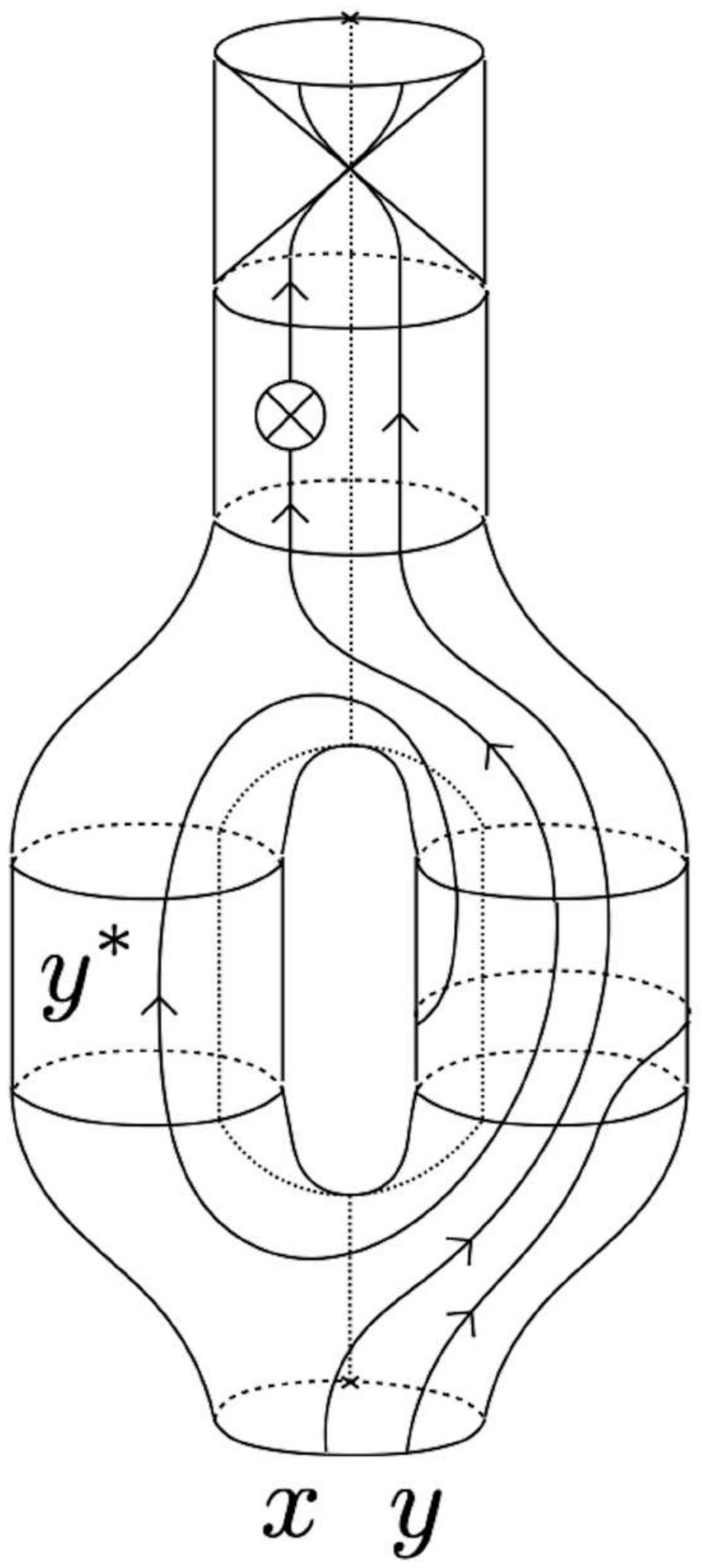}, \quad \quad
\mathrm{(RHS)}  = \adjincludegraphics[valign = c, width = 1.35cm]{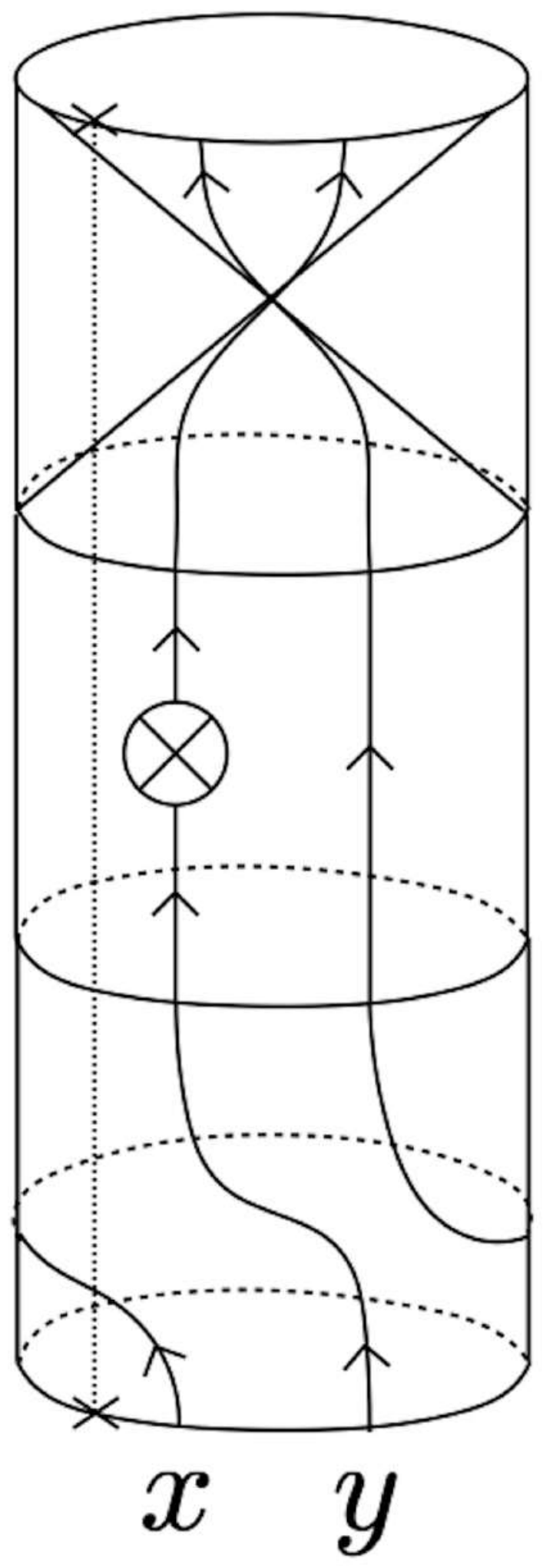}.
\end{equation}
We notice that both the left-hand side and the right-hand side have the orientation-reversing isomorphism and the cross-cap amplitude in common, which can be eliminated from the equality $(\mathrm{LHS}) = (\mathrm{RHS})$ because they are isomorphisms.
Hence, this equation reduces to the equality of the transition amplitudes for oriented surfaces.
By considering the commutativity of diagrams that consist only of $M$, $c$, $Z(\alpha)$, and $X$ as we discussed in section \ref{sec: Bosonic fusion category SPT phases without time-reversal symmetry}, we find that the second equation of (\ref{eq: reconnection}) is equivalent to the following commutative diagram:
\begin{equation}
\begin{tikzcd}[column sep = large, row sep = 12pt]
V_{x \otimes y} \arrow[rr, "X_{x, y}"] \arrow[dd, "Z(\mathrm{coev}_y^R)"'] \arrow[dr, "Z(\mathrm{coev}_y^R)"'] & & V_{y \otimes x}\\
& V_{x \otimes y \otimes y^{*} \otimes y} \arrow[dr, "X_{x, y \otimes y^* \otimes y}"] & \\
V_{y^* \otimes y \otimes x \otimes y} \arrow[rr, "X_{y^* \otimes y \otimes x, y}"'] \arrow[ur, "X_{y^* \otimes y, x \otimes y}"] & & V_{y \otimes y^* \otimes y \otimes x} \arrow[uu, "Z(\mathrm{ev}_y^R)"']
\end{tikzcd}
\end{equation}
The transition amplitudes for the surfaces flipped horizontally satisfy a similar equation.

Thus, the solution of the consistency conditions (\ref{eq: composition on a cylinder})--(\ref{eq: consistency on the torus}) and (\ref{eq: invariant unit})--(\ref{eq: Klein}) for an SPT phase with symmetry $\mathbb{Z}_2^T \times \mathcal{C}$ is given by a quintuple $(Z, M, i, \phi, \theta)$ such that the triple $(Z, M, i)$ is a fiber functor of $\mathcal{C}$, the isomorphism $\theta_{x, y}: V_{x \otimes y} \rightarrow V_{x^* \otimes y}$ is given by $\theta_{x, y} = M_{x^*, y} \circ (\theta_{x, 1} \otimes \mathrm{id}_{V_y}) \circ M_{x, y}^{-1}$ where $\theta_{x, 1}$ satisfies $\theta_{x^*, 1} \circ \theta_{x, 1} = Z(a_x)$, and the isomorphism $\phi_{x}: V_{x} \rightarrow V_{x^*}$ is given by $\phi_{x \otimes y} = \theta_{y, x^*} \circ X_{x^*, y} \circ \theta_{x, y}$ where $X_{x, y}$ is defined by eq. (\ref{eq: X = McM^{-1}}).
The quintuple $(Z, M, i, \phi, \theta)$ that satisfies the above conditions is equivalent to the following algebraic data $(Z, M, i, s, \phi)$:
\begin{itemize}
\item a fiber functor $(Z, M, i): \mathcal{C} \rightarrow \mathrm{Vec}$,
\item a sign $s \in \mathbb{Z}_2 = \{ \pm 1\}$,
\item an isomorphism $\phi_x: V_x \rightarrow V_{x^*}$ that is involutive up to pivotal structure $a_x$ and commutes with the multiplication $M$ up to isomorphism $X$ defined by eq. (\ref{eq: X = McM^{-1}}): 
\begin{equation}
\phi_{x^*} \circ \phi_x = Z(a_x), \quad
X_{x^*, y^*} \circ M_{x^*, y^*} \circ (\phi_x \otimes \phi_y) = \phi_{x \otimes y} \circ M_{x, y}.
\label{eq: involutive equivariance}
\end{equation}
\end{itemize}
When we have a quintuple $(Z, M, i, \phi, \theta)$, the sign $s \in \mathbb{Z}_2$ and the isomorphism that satisfies eq. (\ref{eq: involutive equivariance}) are given by $\theta_{1, 1}$ and $\phi$ respectively.
Conversely, when we have a quintuple $(Z, M, i, s, \phi)$, we can define the cross-cap amplitude as $\theta_{x, y} := M_{x^*, y} \circ (\theta_{x, 1} \otimes \mathrm{id}_{V_y}) \circ M^{-1}_{x, y}$ where $\theta_{x, 1} := s \phi_x$.

The quintuples $(Z, M, i, s, \phi)$ and $(Z^{\prime}, M^{\prime}, i^{\prime}, s^{\prime}, \phi^{\prime})$ correspond to the same SPT phase if they are related by a change of the bases of the vector spaces. The change of the bases is described by a tensor natural isomorphism $\eta: Z \rightarrow Z^{\prime}$, which guarantees that the transition amplitudes for oriented surfaces transform appropriately under the change of the bases.
For the classification of SPT phases with time-reversal symmetry, we need to impose an additional condition on the tensor natural isomorphism $\eta$ so that the transition amplitudes for unoriented surfaces also transform appropriately.
Specifically, we require that the following diagram commutes:
\begin{equation}
\begin{tikzcd}
Z(x) \arrow[r, "\phi_x"] \arrow[d, "\eta_x"'] & Z(x^{*}) \arrow[d, "\eta_{x^*}"]\\
Z^{\prime}(x) \arrow[r, "\phi^{\prime}_x"'] & Z^{\prime}(x^*)
\end{tikzcd}
\end{equation}
The quintuples $(Z, M, i, s, \phi)$ and $(Z^{\prime}, M^{\prime}, i^{\prime}, s^{\prime}, \phi^{\prime})$ are said to be equivalent when they are related by a tensor natural isomorphism $\eta$ that satisfies the above commutative diagram.
Therefore, bosonic fusion category SPT phases with time-reversal symmetry are classified by equivalence classes of quintuples $(Z, M, i, s, \phi)$.

\subsection{Examples}
\label{sec: Examples (unoriented)}

\subsubsection{Finite group symmetry}
1+1d bosonic SPT phases with finite group symmetry $\mathbb{Z}_2^T \times G$ are classified by the group cohomology $H^2(\mathbb{Z}_2^T \times G, \mathrm{U}_T(1))$ where $\mathrm{U}_T(1)$ is $\mathrm{U}(1)$ on which time-reversal symmetry $\mathbb{Z}_2^T$ acts as the complex conjugation \cite{CGLW2013, KT2017}.
By using a K\"{u}nneth formula, we can factorize the group cohomology $H^2(\mathbb{Z}_2^T \times G, \mathrm{U}_T(1))$ as
\begin{equation}
H^2(\mathbb{Z}_2^T \times G, \mathrm{U}_T(1)) = \mathbb{Z}_2 \times H^1(G, \mathrm{U}(1))/2H^1(G, \mathrm{U}(1)) \times {}_2H^2(G, \mathrm{U}(1)),
\label{eq: Kunneth}
\end{equation}
where $2H^1(G, \mathrm{U}(1))$ is the group of even elements of $H^1(G, \mathrm{U}(1))$ and ${}_2H^2(G, \mathrm{U}(1))$ is the 2-torsion subgroup of $H^2(G, \mathrm{U}(1))$, which consists of elements $x \in H^2(G, \mathrm{U}(1))$ such that $2x = 0$.
We note that the first factor $\mathbb{Z}_2$ captures the classification of 1+1d bosonic SPT phases only with time-reversal symmetry.

In the following, we will reproduce this classification as a special case of the classification of fusion category SPT phases.
We begin with noticing that the sign $s \in \mathbb{Z}_2$ gives the first term on the right-hand side of eq. (\ref{eq: Kunneth}).
Furthermore, for a finite group symmetry $G$,\footnote{As a fusion category symmetry, a finite group symmetry $G$ is described by the category of $G$-graded vector spaces $\mathrm{Vec}_G$.} fiber functors $(Z, M, i)$ are classified by the group cohomology $H^2(G, \mathrm{U}(1))$.
However, some elements of $H^2(G, \mathrm{U}(1))$ are not allowed when the time-reversal symmetry is taken into account.
We will see shortly that $H^2(G, \mathrm{U}(1))$ reduces to its 2-torsion subgroup ${}_2H^2(G, \mathrm{U}(1))$ in the presence of time-reversal symmetry.
The remaining term $H^1(G, \mathrm{U}(1)) / 2H^1(G, \mathrm{U}(1))$ comes from the classification of isomorphisms $\phi$ that satisfy eq. (\ref{eq: involutive equivariance}).

Once we fix a basis $b_g$ of each vector space $V_g$, the isomorphism $\phi_g: V_g \rightarrow V_{g^{-1}}$ can be represented by a phase factor $\phi(g)$ as $\phi_g(b_g) = \phi(g) b_{g^{-1}}$.
For this choice of the basis, equation (\ref{eq: involutive equivariance}) can be written as
\begin{equation}
\phi(g^{-1}) \phi(g) = 1, \quad \phi(g) \phi(h) = \frac{\nu(g, h)}{\nu(h^{-1}, g^{-1})} \phi(gh),
\label{eq: phi equivariance}
\end{equation}
where a 2-cocycle $\nu \in Z^2(G, \mathrm{U}(1))$ represents the multiplication $M_{g, h} (b_g \otimes b_h) = \nu(g, h) b_{gh}$ with respect to the basis $\{b_g\}$.
If we choose the basis so that $\nu(g, h) = \nu(h^{-1}, g^{-1})^{-1}$, the second equation indicates that the cohomology class of $\nu^2$ vanishes.
This means that $\nu$ takes values in ${}_2H^2(G, \mathrm{U}(1))$, which gives the last term on the right-hand side of eq. (\ref{eq: Kunneth}).

Given a 2-cocycle $\nu$, the solutions of eq. (\ref{eq: phi equivariance}) are in one-to-one correspondence with the first group cohomology $H^1(G, \mathrm{U}(1))$.
To see this, we suppose that $\phi_1$ and $\phi_2$ are solutions of eq. (\ref{eq: phi equivariance}).
The ratio $\Phi := \phi_1 / \phi_2$ satisfies
\begin{equation}
\Phi(g) \Phi(h) = \Phi(gh),
\end{equation}
which indicates that $\Phi$ is a homomorphism from $G$ to $\mathrm{U}(1)$.
Therefore, if eq. (\ref{eq: phi equivariance}) has a solution, the solutions of eq. (\ref{eq: phi equivariance}) are in one-to-one correspondence with the set of homomorphisms from $G$ to $\mathrm{U}(1)$, which is the first group cohomology $H^1(G, \mathrm{U}(1))$.
The existence of a solution can be seen by explicitly constructing a solution in the following way \cite{KT2017}.
Let $\widetilde{\nu} \in Z^2(\mathbb{Z}_2^T \times G, \mathrm{U}_T(1))$ be a group 2-cocycle that agrees with $\nu \in Z^2(G, \mathrm{U}(1))$ on the orientation-preserving subgroup $G$, i.e. $\widetilde{\nu}(g, h) = \nu(g, h)$ for all $g, h \in G$.
If we define
\begin{equation}
\phi_0(g) := \frac{\widetilde{\nu}(T, g^{-1}) \widetilde{\nu}(Tg^{-1}, T) \widetilde{\nu}(g, g^{-1})}{\widetilde{\nu}(T, T)}
\end{equation}
where $T$ is the generator of time-reversal symmetry $\mathbb{Z}_2^T$, we find that $\phi_0$ satisfies (\ref{eq: phi equivariance}).
Thus, the set of the orientation-reversing isomorphisms are in one-to-one correspondence with $H^1(G, \mathrm{U}(1))$.

The pairs $(\nu, \phi)$ and $(\nu^{\prime}, \phi^{\prime})$ are equivalent if they are related by a change of the basis $b_g \rightarrow b^{\prime}_g := \mu(g) b_g$ where $\mu$ is a $\mathrm{U}(1)$-valued function on $G$.
Under this change of the basis, $\nu$ and $\phi$ transform as
\begin{equation}
\nu(g, h) \rightarrow \nu^{\prime}(g, h) = \frac{\mu(g)\mu(h)}{\mu(gh)} \nu(g, h), \quad
\phi(g) \rightarrow \phi^{\prime}(g) = \frac{\mu(g)}{\mu(g^{-1})} \phi(g).
\end{equation}
To classify the solutions $\phi$ of eq. (\ref{eq: phi equivariance}) for a fixed $\nu$, we demand that the multiplication $\nu$ is invariant under the change of the basis.
This means that $\mu$ is a group homomorphism from $G$ to $\mathrm{U}(1)$.
Thus, equivalent solutions $\phi$ and $\phi^{\prime}$ are related by the multiplication of a squared group homomorphism $\mu(g)/\mu(g^{-1}) = \mu(g)^2$, which forms an abelian group $2H^1(G, \mathrm{U}(1))$.

Therefore, orientation-reversing isomorphisms are classified up to change of the basis by the quotient group $H^1(G, \mathrm{U}(1)) / 2H^1(G, \mathrm{U}(1))$, which gives the second term on the right-hand side of (\ref{eq: Kunneth}).
This completes the classification of 1+1d bosonic SPT phases with symmetry $\mathbb{Z}_2^T \times G$.

\subsubsection{Duality symmetry}
\label{sec: Duality symmetry (unoriented)}
The invariance of a theory under gauging a finite abelian group symmetry $A$ is called duality symmetry.
A theory with duality symmetry is said to be self-dual.
The duality symmetry is described by a Tambara-Yamagami category $\mathrm{TY}(A, \chi, \epsilon)$ where $\chi: A \times A \rightarrow \mathrm{U}(1)$ is a non-degenerate symmetric bicharacter of $A$ and $\epsilon \in \{ \pm 1\}$ is a sign \cite{TY1998}.
The set of simple objects of $\mathrm{TY}(A, \chi, \epsilon)$ consists of group-like objects $a \in A$ and the duality object $m$, whose fusion rules are given by
\begin{equation}
a \otimes b = ab, \quad m \otimes a = a \otimes m = m, \quad m \otimes m = \bigoplus_{a \in A} a.
\label{eq: TY fusion rule}
\end{equation}
From the above fusion rules, we find that the quantum dimensions of a group-like object $a$ and the duality object $m$ are given by $\mathop{\mathop{\mathrm{dim}}} a = 1$ and $\mathop{\mathop{\mathrm{dim}}} m = \sqrt{|A|}$ respectively.
The nontrivial associators of $\mathrm{TY}(A, \chi, \epsilon)$ are summarized as follows:
\begin{equation}
\alpha_{amb} = \chi(a, b) \mathrm{id}_m,\quad
\alpha_{mam} = \bigoplus_{b \in A} \chi(a, b) \mathrm{id}_b,\quad
(\alpha_{mmm})_{ab} = \frac{\epsilon}{\sqrt{|A|}} \chi(a, b)^{-1} \mathrm{id}_m.
\label{eq: TY associators}
\end{equation}
The last equation means that the associator $\alpha_{mmm}$ consists of the morphisms $\epsilon \chi(a, b)^{-1} \mathrm{id}_m / \sqrt{|A|}$ from the $a$th component of the source object $\bigoplus m$ to the $b$th component of the target object $\bigoplus m$.

In the following, we classify bosonic self-dual SPT phases with time-reversal symmetry by group-theoretical data.
We first review the classification of fiber functors of the Tambara-Yamagami category $\mathrm{TY}(A, \chi, \epsilon)$ \cite{Tam2000}.
To simplify the notation, we denote the isomorphisms $M_{xy}: V_x \otimes V_y \rightarrow V_{x \otimes y}$ for simple objects $x$ and $y$ as
\begin{align}
v_a v_b & := M_{ab}(v_a \otimes v_b), \label{eq: TY multiplication1}\\
v_a \cdot w & := M_{am}(v_a \otimes w), \label{eq: TY multiplication2}\\
w \cdot v_a & := M_{ma}(w \otimes v_a), \label{eq: TY multiplication3}\\
[w, w^{\prime}] & := M_{mm}(w \otimes w^{\prime}), \label{eq: TY multiplication4}
\end{align}
where $v_a \in V_a$, $v_b \in V_b$, and $w, w^{\prime} \in V_m$.
The first equation (\ref{eq: TY multiplication1}) gives an associative multiplication on $V = \bigoplus V_a$.
In general, this multiplication is twisted by a group 2-cocycle $\xi \in Z^2(A, \mathrm{U}(1))$
\begin{equation}
u_a u_b = \xi(a, b) u_{ab},
\end{equation}
where $u_a$ is a basis of $V_a$.
The second equation (\ref{eq: TY multiplication2}) and the third equation (\ref{eq: TY multiplication3}) give the left and right actions of $V$ on $V_m$ respectively.
The left and right actions are related by an involutive anti-automorphism $f: V \rightarrow V$ as follows:
\begin{equation}
w \cdot v_a = f(v_a) \cdot w.
\end{equation}
The involutive anti-automorphism $f$ on $V$ induces an involutive automorphism $\sigma$ on $A$ as
\begin{equation}
f(u_a) = \nu(a) u_{\sigma(a)},
\label{eq: def of nu}
\end{equation}
where $\nu(a)$ is a complex number with absolute value $1$.

We can show that a fiber functor of the Tambara-Yamagami category $\mathrm{TY}(A, \chi, \epsilon)$ is characterized by a triple $(\sigma, \xi, \nu)$ such that
\begin{align}
\chi(a, b) & = \xi(a, \sigma(b)) / \xi(\sigma(b), a),\label{eq: triple1}\\
\nu(a) \nu(b) / \nu(ab) & = \xi(a, b) / \xi(\sigma(b), \sigma(a)), \label{eq: triple2}\\
\nu(a) \nu(\sigma(a)) & = 1, \label{eq: triple3}\\
\sum_{a \in A \text{ s.t. } \sigma(a) = a} \nu(a) & = \epsilon \sqrt{|A|}. \label{eq: triple4}
\end{align}
We note that the cohomology class of $\xi$ must be nontrivial due to the non-degeneracy of $\chi$.

If we change the basis of $V_a$ from $u_a$ to $u^{\prime}_a := \psi(a) u_a$, the triple $(\sigma, \xi, \nu)$ changes to
\begin{equation}
\sigma^{\prime} = \sigma, \quad \xi^{\prime}(a, b) = \frac{\psi(a) \psi(b)}{\psi(ab)} \xi(a, b), \quad \nu^{\prime}(a) = \frac{\psi(a)}{\psi(\sigma(a))} \nu(a).
\label{eq: change of basis}
\end{equation}
Thus, the fiber functors characterized by the triples $(\sigma, \xi, \nu)$ and $(\sigma^{\prime}, \xi^{\prime}, \nu^{\prime})$ are naturally isomorphic if they are related by eq. (\ref{eq: change of basis}).
Such triples are said to be equivalent to each other.
Therefore, 1+1d bosonic SPT phases with duality symmetry are classified by equivalence classes of triples $(\sigma, \xi, \nu)$ that satisfy eqs. (\ref{eq: triple1})--(\ref{eq: triple4}).

To classify self-dual SPT phases with time-reversal symmetry, we also need to consider the orientation-reversing isomorphisms $\phi_x: V_x \rightarrow V_{x^*}$ that satisfy eq. (\ref{eq: involutive equivariance}).
Considering that the pivotal structure of the duality object $m$ is given by $a_m = \epsilon \mathop{\mathrm{id}}_m$ (see e.g. \cite{Shi2011}), we find that eq. (\ref{eq: involutive equivariance}) for the Tambara-Yamagami category $\mathrm{TY}(A, \chi, \epsilon)$ reduces to
\begin{align}
\phi_{a^{-1}} (\phi_a(v_a)) & = v_a, \label{eq: phi-a involution}\\
\phi_m(\phi_m(w)) & = \epsilon w, \label{eq: phi-m involution}\\
\phi_b(v_b) \phi_a(v_a) & = \phi_{ab}(v_a v_b), \label{eq: phi-b phi-a}\\
\phi_m(w) \cdot \phi_a(v_a) & = \phi_m(v_a \cdot w), \label{eq: phi-m phi-a}\\
\phi_a(v_a) \cdot \phi_m(w) & = \phi_m(w \cdot v_a), \label{eq: phi-a phi-m}\\
[\phi_m(w^{\prime}), \phi_m(w)]_{a} & = \phi_{a^{-1}} ([w, w^{\prime}]_{a^{-1}}). \label{eq: phi-m phi-m}
\end{align}
The subscript $a$ on the left-hand side of the last equation indicates the $a$-twisted sector $V_a$ in $V_m \otimes V_m = \bigoplus V_a$.
By choosing a basis $u_a \in V_a$, we can represent $\phi_a$ by a phase factor $\phi(a)$ as $\phi_a(u_a) = \phi(a) u_{a^{-1}}$.
Similarly, by choosing a basis of $V_m$, we can represent $\phi_m$ by a $\sqrt{|A|} \times \sqrt{|A|}$ unitary matrix $\Phi$ as $\phi_m(w) = \Phi w$.
In terms of $\phi(a)$ and $\Phi$, eqs. (\ref{eq: phi-a involution}) and (\ref{eq: phi-m involution}) can be written as
\begin{align}
\phi(a) \phi(a^{-1}) & = 1, \label{eq: phi(a) involution}\\
\Phi^2 & = \epsilon. \label{eq: Phi involution}
\end{align}
If we recall that the multiplication on $V = \bigoplus V_a$ is twisted by a 2-cocycle $\xi$, we can write eq. (\ref{eq: phi-b phi-a}) as
\begin{equation}
\phi(a) \phi(b) \xi(b^{-1}, a^{-1}) = \phi(ab) \xi(a, b),
\label{eq: quintuple1}
\end{equation}
which includes eq. (\ref{eq: phi(a) involution}) as a special case.
Equation (\ref{eq: phi-m phi-a}) is equivalent to $f(\phi_a(u_a)) \cdot \phi_m(w) = \phi_m(u_a \cdot w)$, whose matrix representation is given by
\begin{equation}
\phi(a) \nu(a^{-1}) U_{\sigma(a^{-1})} \Phi = \Phi U_a
\label{eq: phi-m phi-a matrix rep}
\end{equation}
where $U_a$ is the matrix representation of the left action of $u_a$.
We note that $U_a$ obeys the same algebra as $u_a$, i.e. $U_a U_b = \xi(a, b) U_{ab}$.
Equation (\ref{eq: phi-m phi-a matrix rep}) means that the unitary transformation $U_a \mapsto \Phi U_a \Phi^{-1}$ induced by the orientation-reversing isomorphism $\Phi$ on $V_m$ agrees with $f \circ \phi: U_a \mapsto \phi(a) \nu(a^{-1}) U_{\sigma(a^{-1})}$.
Put differently, the linear map $f \circ \phi$ is represented by the unitary matrix $\Phi$, which gives the orientation-reversing isomorphism on $V_m$.
Similarly, eq. (\ref{eq: phi-a phi-m}) gives rise to the same equation (\ref{eq: phi-m phi-a matrix rep}).
Furthermore, as we will see in appendix \ref{sec: Matrix representation of eq. (phi-m phi-m)}, eq. (\ref{eq: phi-m phi-m}) reduces to
\begin{equation}
c(\sigma(a)) \nu(\sigma(a)) = c(a)
\label{eq: symmetry of Gamma Phi}
\end{equation}
where $c(a)$ is the $U_a$-component of $\Phi$, i.e. $\Phi = \sum_{a \in A} c(a) U_a$.
This expansion is always possible because $\{U_a\}$ spans the vector space of all $\sqrt{|A|} \times \sqrt{|A|}$ matrices, which will also be shown in appendix \ref{sec: Matrix representation of eq. (phi-m phi-m)}.

Therefore, a self-dual SPT phase with time-reversal symmetry is given by the algebraic data $(\sigma, \xi, \nu, s, \phi, \Phi)$ where $s \in \mathbb{Z}_2$ is a sign and $(\sigma, \xi, \nu, \phi, \Phi)$ satisfies (\ref{eq: triple1})--(\ref{eq: triple4}) and (\ref{eq: Phi involution})--(\ref{eq: symmetry of Gamma Phi}).
If we change the bases of $V_a$ and $V_m$ by a phase factor $\psi(a)$ and a unitary matrix $U$ respectively, $(\sigma, \xi, \nu)$ transforms as eq. (\ref{eq: change of basis}) and $(s, \phi, \Phi)$ transforms as
\begin{equation}
s^{\prime} = s, \quad \phi^{\prime}(a) = \frac{\psi(a)}{\psi(a^{-1})} \phi(a), \quad \Phi^{\prime} = U \Phi U^{-1}.
\label{eq: change of basis (unoriented)}
\end{equation}
The sextuples $(\sigma, \xi, \nu, s, \phi, \Phi)$ and $(\sigma^{\prime}, \xi^{\prime}, \nu^{\prime}, s^{\prime}, \phi^{\prime}, \Phi^{\prime})$ are equivalent if they are related by the change of the bases (\ref{eq: change of basis}) and (\ref{eq: change of basis (unoriented)}).
Thus, time-reversal invariant bosonic SPT phases with duality symmetry $\mathrm{TY}(A, \chi, \epsilon)$ are classified by equivalence classes of $(\sigma, \xi, \nu, s, \phi, \Phi)$.

As a concrete example, we explicitly compute the classification of self-dual SPT phases when $A = \mathbb{Z}_2 \times \mathbb{Z}_2 = \{(0, 0), (1, 0), (0, 1), (1, 1)\}$.
In this case, we have four different Tambara-Yamagami categories $\mathrm{TY}(\mathbb{Z}_2 \times \mathbb{Z}_2, \chi_{\pm}, \pm 1)$ where the non-degenerate symmetric bicharacters $\chi_{\pm}$ are given by \cite{TY1998}
\begin{align}
\chi_+((1, 0), (1, 0)) & = \chi_+((0, 1), (0, 1)) = +1, \quad \chi_+((1, 0), (0, 1)) = -1,\\
\chi_-((1, 0), (1, 0)) & = \chi_-((0, 1), (0, 1)) = -1, \quad \chi_-((1, 0), (0, 1)) = +1.
\end{align}
The involutive automorphism $\sigma$ is uniquely determined by eq. (\ref{eq: triple1}) where $\xi$ is a representative of the nontrivial cohomology class in $H^2(\mathbb{Z}_2 \times \mathbb{Z}_2, \mathrm{U}(1)) = \mathbb{Z}_2$.
Specifically, $\sigma$ is equal to the identity map when $\chi = \chi_+$, while $\sigma$ is a nontrivial involution that maps $(1, 0)$ to $(0, 1)$ when $\chi = \chi_-$.
Here, we choose a group 2-cocycle $\xi$ as $\xi(a, b) = (-1)^{a_1 b_2}$ where $a = (a_1, a_2)$ and $b = (b_1, b_2)$.
For this choice of $\xi$, the two-dimensional projective representation of $\mathbb{Z}_2 \times \mathbb{Z}_2$ is given by using the Pauli matrices as follows:
\begin{equation}
U_{(0, 0)} = 1, \quad U_{(1, 0)} = \sigma_x, \quad U_{(0, 1)} = \sigma_z, \quad U_{(1, 1)} = i \sigma_y.
\end{equation}
Furthermore, the solutions $\phi$ of eq. (\ref{eq: quintuple1}) are classified by $H^1(\mathbb{Z}_2 \times \mathbb{Z}_2, \mathrm{U}(1)) / 2H^1(\mathbb{Z}_2 \times \mathbb{Z}_2, \mathrm{U}(1)) = \mathbb{Z}_2 \times \mathbb{Z}_2$ as we discussed in the previous subsection.
If we denote $\phi$ as $\phi = (\phi(0, 0), \phi(1, 0), \phi(0, 1), \phi(1, 1))$, the solutions of (\ref{eq: quintuple1}) are explicitly written as
\begin{equation}
\phi_1 = (1, -1, 1, 1), \quad \phi_2 = (1, 1, -1, 1), \quad \phi_3 = (1, 1, 1, -1), \quad \phi_4 = (1, -1, -1, -1).
\end{equation}
These solutions do not necessarily give rise to time-reversal invariant self-dual SPT phases due to eqs. (\ref{eq: phi-m phi-a matrix rep}) and (\ref{eq: symmetry of Gamma Phi}).
In the following, we classify SPT phases $(\sigma, \xi, \nu, s, \phi, \Phi)$ for each $\mathbb{Z}_2 \times \mathbb{Z}_2$ Tambara-Yamagami category.
Since the classification of $(\sigma, \xi, \nu)$ is already known \cite{Tam2000, TW2019}, it suffices to consider the classification of $(s, \phi, \Phi)$.
In particular, $\mathrm{TY}(\mathbb{Z}_2 \times \mathbb{Z}_2, \chi_-, -1)$ does not admit SPT phases because eqs. (\ref{eq: triple1})--(\ref{eq: triple4}) do not have any solution.

For $\mathrm{TY}(\mathbb{Z}_2 \times \mathbb{Z}_2, \chi_+, +1)$, we have three inequivalent $\nu$, which we denote by $\nu_1$, $\nu_2$, and $\nu_3$:
\begin{align}
\nu_1(1, 0) & = -1, \quad \nu_1(0, 0) = \nu_1(0, 1) = \nu_1(1, 1) = 1, \\
\nu_2(0, 1) & = -1, \quad \nu_2(0, 0) = \nu_2(1, 0) = \nu_2(1, 1) = 1, \\
\nu_3(1, 1) & = -1, \quad \nu_3(0, 0) = \nu_3(1, 0) = \nu_3(0, 1) = 1.
\end{align}
We find that there are four equivalence classes of the solutions of (\ref{eq: Phi involution})--(\ref{eq: symmetry of Gamma Phi}) for each $\nu$:\footnote{Any solution of eqs. (\ref{eq: Phi involution})--(\ref{eq: symmetry of Gamma Phi}) is equivalent to one of the solutions listed in eqs. (\ref{eq: solution1})--(\ref{eq: solution3}). For example, we have a solution $(\nu, \phi, \Phi) = (\nu_1, \phi_2, -\sigma_y)$, which is equivalent to $(\nu_1, \phi_2, \sigma_y)$.}
\begin{align}
\nu = \nu_1 \Rightarrow (\phi, \Phi) & = (\phi_1, +1), (\phi_1, -1), (\phi_2, \sigma_y), (\phi_3, \sigma_z), \label{eq: solution1}\\
\nu = \nu_2 \Rightarrow (\phi, \Phi) & = (\phi_1, \sigma_y), (\phi_2, +1), (\phi_2, -1), (\phi_3, \sigma_x), \label{eq: solution2}\\
\nu = \nu_3 \Rightarrow (\phi, \Phi) & = (\phi_1, \sigma_z), (\phi_2, \sigma_x), (\phi_3, +1), (\phi_3, -1). \label{eq: solution3}
\end{align}
If we take into account the choice of a sign $s \in \mathbb{Z}_2$, we obtain 24 different SPT phases in total.

For $\mathrm{TY}(\mathbb{Z}_2 \times \mathbb{Z}_2, \chi_+, -1)$, we have a single $\nu$: $\nu(0, 0) = 1, \nu(1, 0) = \nu(0, 1) = \nu(1, 1) = -1$.
The solutions of (\ref{eq: Phi involution})--(\ref{eq: symmetry of Gamma Phi}) are given by $(\phi, \Phi) = (\phi_4, +i), (\phi_4, -i)$, which are not equivalent to each other.
Since the classification of SPT phases also depends on the choice of a sign $s \in \mathbb{Z}_2$, the bosonic SPT phase with duality symmetry $\mathrm{TY}(\mathbb{Z}_2 \times \mathbb{Z}_2, \chi_+, -1)$ splits into four different SPT phases in the presence of time-reversal symmetry.

For $\mathrm{TY}(\mathbb{Z}_2 \times \mathbb{Z}_2, \chi_-, +1)$, we have one equivalence class of $\nu$, whose representative is given by $\nu(a) = 1$ for all $a \in \mathbb{Z}_2 \times \mathbb{Z}_2$.
The solutions of (\ref{eq: Phi involution})--(\ref{eq: symmetry of Gamma Phi}) are in the equivalence class of $(\phi, \Phi) = (\phi_3, (\sigma_x + \sigma_z)/\sqrt{2})$.
Therefore, the bosonic SPT phase with duality symmetry $\mathrm{TY}(\mathbb{Z}_2 \times \mathbb{Z}_2, \chi_-, +1)$ splits into two different time-reversal invariant SPT phases, which are distinguished by the choice of a sign $s \in \mathbb{Z}_2$.
The number of bosonic SPT phases for each duality symmetry $\mathrm{TY}(\mathbb{Z}_2 \times \mathbb{Z}_2, \chi, \epsilon)$ is summarized in tables \ref{tab: duality SPT phases1} and \ref{tab: duality SPT phases2}.
\begin{table}
\begin{center}
\begin{minipage}{0.475 \hsize}
\begin{center}
\begin{tabular}{c|cc}
& $\epsilon = +1$ & $\epsilon = -1$\\ \hline
$\chi = \chi_+$ & 3 & 1\\
$\chi = \chi_-$ & 1 & 0
\end{tabular}
\caption{The number of $\mathbb{Z}_2 \times \mathbb{Z}_2$ self-dual SPT phases without time-reversal symmetry.}
\label{tab: duality SPT phases1}
\end{center}
\end{minipage}%
\begin{minipage}{0.05 \hsize}
\begin{center}
\end{center}
\end{minipage}%
\begin{minipage}{0.475 \hsize}
\begin{center}
\begin{tabular}{c|cc}
& $\epsilon = +1$ & $\epsilon = -1$\\ \hline
$\chi = \chi_+$ & 24 & 4\\
$\chi = \chi_-$ & 2 & 0
\end{tabular}
\caption{The number of $\mathbb{Z}_2 \times \mathbb{Z}_2$ self-dual SPT phases with time-reversal symmetry.}
\label{tab: duality SPT phases2}
\end{center}
\end{minipage}
\end{center}
\end{table}

Finally, we notice that some duality symmetries do not admit time-reversal invariant SPT phases even if they admit SPT phases without time-reversal symmetry.
For example, a $\mathbb{Z}_{2n+1} \times \mathbb{Z}_{2n+1}$ self-duality $\mathrm{TY}(\mathbb{Z}_{2n+1} \times \mathbb{Z}_{2n+1}, \chi, +1)$ admits an SPT phase in the absence of time-reversal symmetry due to Proposition 4.2 of \cite{Tam2000}.
Here, the non-degenerate symmetric bicharacter $\chi$ is given by
\begin{equation}
\chi(a, a) = \chi(b, b) = 1, \quad \chi(a, b) = e^{2 \pi i/ (2n+1)},
\end{equation}
where $a$ and $b$ are the generators of $\mathbb{Z}_{2n+1} \times \mathbb{Z}_{2n+1}$.
However, this self-duality does not admit time-reversal invariant SPT phases because the 2-torsion subgroup of $H^2(\mathbb{Z}_{2n+1} \times \mathbb{Z}_{2n+1}, \mathrm{U}(1)) = \mathbb{Z}_{2n+1}$ is trivial and hence there is no nontrivial cocycle $\xi$ that satisfies eqs. (\ref{eq: triple1}) and (\ref{eq: quintuple1}) simultaneously.
This may be regarded as a mixed anomaly between time-reversal symmetry and the duality symmetry $\mathrm{TY}(\mathbb{Z}_{2n + 1} \times \mathbb{Z}_{2n + 1}, \chi, +1)$.

\section{Conclusion}
In this paper, we discussed the classification of 1+1d bosonic fusion category SPT phases with and without time-reversal symmetry.
We showed that bosonic fusion category SPT phases without time-reversal symmetry are classified by isomorphism classes of fiber functors, which agrees with the previous result \cite{TW2019, KORS2020}.
We obtained this result by explicitly solving the consistency conditions of oriented TQFTs with fusion category symmetry formulated in \cite{BT2018}.
Our derivation revealed that the data of a fiber functor naturally appear in the low energy limit of SPT phases.
We also classified bosonic fusion category SPT phases with time-reversal symmetry when the total symmetry splits into time-reversal symmetry and fusion category symmetry.
To accomplish the classification, we axiomatized unoriented TQFTs with fusion category symmetry by generalizing the oriented case.
We found that unoriented TQFTs for SPT phases are in one-to-one correspondence with equivalence classes of quintuples $(Z, M, i, s, \phi)$ where $(Z, M, i)$ is a fiber functor, $s$ is a sign, and $\phi$ is a collection of isomorphisms that satisfy eq. (\ref{eq: involutive equivariance}).
As an application, we specified the group-theoretical data that classify bosonic self-dual SPT phases with time-reversal symmetry and found that some duality symmetries do not have SPT phases only when time-reversal symmetry is imposed.

There are several future directions.
One is to generalize our result to the case where the total symmetry does not split into time-reversal symmetry and fusion category symmetry.
For a finite group symmetry that is not necessarily a direct product of time-reversal symmetry and internal symmetry, the axioms of unoriented equivariant TQFTs are given in \cite{KT2017}.
In particular, invertible unoriented equivariant TQFTs are shown to be classified by twisted group cohomology.
It would be interesting to extend this result to more general fusion category symmetries.

Another future direction is the classification of fermionic fusion category SPT phases.
For fermionic theories, fusion category symmetry can further be generalized to superfusion categories, which incorporate the information of Majorana fermions that reside on topological defect lines \cite{KORS2020, NR2020, GK2016, BGK2017, BE2017, Ush2018, BGHNPRW2017, ALW2019, LSCH2020}.
Moreover, fermionic topological field theories, which describe the low energy limit of fermionic SPT phases, depend on a choice of a variant of spin structure.
It would be interesting to formulate fermionic topological field theories with superfusion category symmetry and solve them to classify fermionic fusion category SPT phases.

\section*{Acknowledgments}
The author thanks Ryohei Kobayashi and Masaki Oshikawa for comments on the manuscript.
The author is supported by FoPM, WINGS Program, the University of Tokyo.

\appendix

\section{Unoriented equivariant TQFTs}
\label{sec: Unoriented equivariant TQFTs}
In this appendix, we show that the consistency conditions (\ref{eq: composition on a cylinder})--(\ref{eq: consistency on the torus}) and (\ref{eq: invariant unit})--(\ref{eq: Klein}) of unoriented TQFTs with fusion category symmetry reduce to the consistency conditions of unoriented equivariant TQFTs when the symmetry is given by a finite group $\mathbb{Z}_2^T \times G$.
We begin with recalling the algebraic data of oriented equivariant TQFTs.
We first assign a vector space $V_g$ to a circle with a topological defect $g \in G$.
The symmetry action $\alpha: G \rightarrow \mathrm{Aut}(V)$ on the vector space $V = \bigoplus_{g \in G} V_g$ is a homomorphism where $\alpha_h$ is a linear map from $V_g$ to $V_{hgh^{-1}}$.
The vector space $V$ has an associative multiplication $M_{g, h}: V_g \otimes V_h \rightarrow V_{gh}$, which is simply denoted by $M_{g, h} (\psi_g \otimes \psi_h) = \psi_g \psi_h$ for all $\psi_g \in V_g$ and $\psi_h \in V_h$.
Furthermore, we have a $G$-invariant linear map $\epsilon: V_1 \rightarrow \mathbb{C}$ such that the pairing $\epsilon \circ M_{g^{-1}, g}$ is non-degenerate.
The adjoint of $\epsilon$ with respect to this non-degenerate pairing gives the unit of the multiplication.
Due to this non-degenerate pairing, the vector space $V_{g^{-1}}$ is regarded as the dual vector space of $V_{g}$.
The dual bases of $V_g$ and $V_{g^{-1}}$ are denoted by $\{\xi_i^g\}$ and $\{\xi_g^i\}$ respectively.
For oriented equivariant TQFTs, the consistency conditions on these algebraic data are summarized as follows: \cite{MS2006}
\begin{align}
\left. \alpha_{g} \right|_{V_g} & = \mathrm{id}_{V_g}, \label{eq: OE1}\\
\alpha_h(\psi_g) \psi_h & = \psi_h \psi_g, \quad \forall \psi_g \in V_g, \forall \psi_h \in V_h, \label{eq: OE2}\\
\sum_i \alpha_h (\xi_i^g) \xi_g^i & = \sum_i \xi_i^h \alpha_g (\xi_h^i). \label{eq: OE3}
\end{align}

For unoriented equivariant TQFTs, we also have the following additional data: the cross-cap state $\Theta_g \in V_{g^2}$ and an involutive anti-automorphism $\alpha_T: V \rightarrow V$ that maps $V_g$ to $V_{g^{-1}}$.
The involutive anti-automorphism $\alpha_T$ represents the action of orientation-reversing symmetry, which preserves the unit and is compatible with internal symmetry $\alpha_{T} \circ \alpha_g = \alpha_g \circ \alpha_T$.
The consistency conditions on these algebraic data are given by \cite{KT2017}
\begin{align}
\alpha_h (\Theta_g) & = \Theta_{hgh^{-1}}, \label{eq: UE1}\\
\alpha_T (\Theta_g) & = \Theta_{g^{-1}}, \label{eq: UE2}\\
\Theta_g \psi_h & = \alpha_{g} (\alpha_T (\psi_h)) \Theta_{gh}, \quad \forall \psi_h \in V_h, \label{eq: UE3}\\
\sum_i \alpha_g (\alpha_T (\xi_{gh}^i)) \xi_{i}^{gh} & = \Theta_g \Theta_h. \label{eq: UE4}
\end{align}

Now, let us reproduce the above consistency conditions from (\ref{eq: composition on a cylinder})--(\ref{eq: consistency on the torus}) and (\ref{eq: invariant unit})--(\ref{eq: Klein}).
First of all, the vector space $V = \bigoplus_{g \in G}V_g$ is a unital associative algebra with a non-degenerate pairing due to (\ref{eq: non-degenerate}), (\ref{eq: unit constraint}), and (\ref{eq: associativity constraint}).
The symmetry action on this vector space is given by the change of the base point 
\begin{equation}
X_{g, h} = \left. \alpha_{h} \right|_{V_{gh}}
\end{equation}
because moving the base point from the left of $g$ to the left of $h$ is equivalent to winding the topological defect $h$ around a cylinder.
Under this identification of the symmetry action and the change of the base point, we find that $\alpha: G \rightarrow \mathrm{Aut}(V)$ becomes a homomorphism as a consequence of (\ref{eq: C2}), (\ref{eq: coherence}), and (\ref{eq: uniqueness of the multiplication}).
Furthermore, the consistency condition (\ref{eq: C2}) reduces to eq. (\ref{eq: OE1}).
Equations (\ref{eq: OE2}) and (\ref{eq: OE3}) follows from the twisted commutativity (\ref{eq: twisted commutativity}) and the consistency on the torus (\ref{eq: consistency on the torus}) respectively.
This completes the consistency conditions of oriented equivariant TQFTs.
We note that we used all the consistency conditions (\ref{eq: composition on a cylinder})--(\ref{eq: consistency on the torus}) except for the conditions on topological point operators.

We can also check the consistency conditions of unoriented equivariant TQFTs in a similar way.
The orientation-reversing isomorphism $\phi$ is identified with the action $\alpha_T$ of orientation-reversing symmetry, which is an involutive anti-automorphism due to (\ref{eq: involution}) and (\ref{eq: equivariance}).
Moreover, the consistency conditions (\ref{eq: invariant unit}) and (\ref{eq: commutativity of X and phi}) indicate that $\alpha_T$ preserves the unit and is compatible with internal symmetry.
The cross-cap state $\Theta_g \in V_{g^2}$ is defined as
\begin{equation}
\Theta_g := \adjincludegraphics[valign = c, width = 1.5cm]{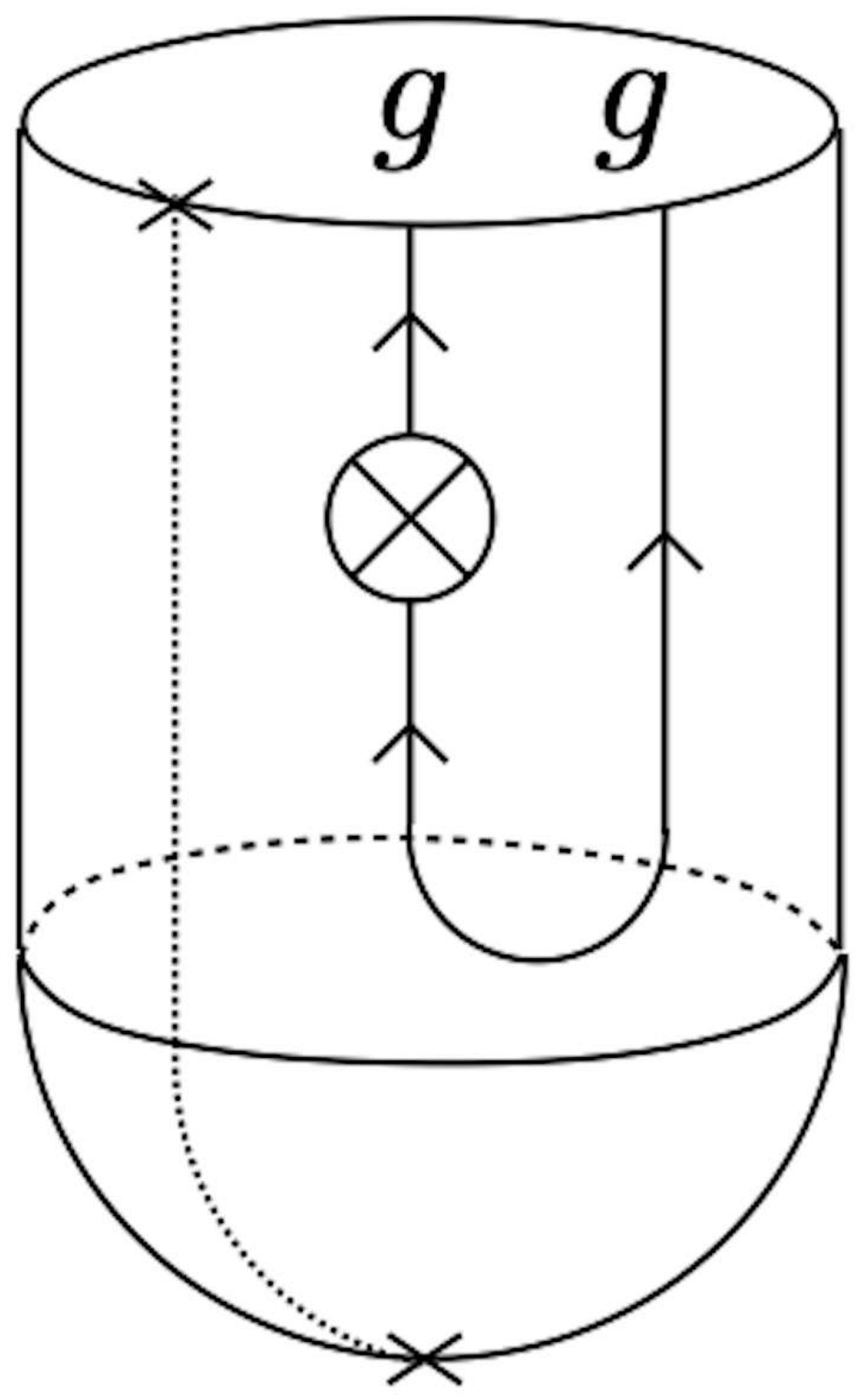},
\end{equation}
which depends only on the topology of the diagram on the right-hand side due to (\ref{eq: compatibility of left-theta and right-theta}) and (\ref{eq: compatibility of theta and M}).
The consistency condition (\ref{eq: reconnection}) implies that the action $\alpha_h$ of internal symmetry $h \in G$ maps the cross-cap state $\Theta_g$ to another cross-cap state $\Theta_{hgh^{-1}}$, which shows eq. (\ref{eq: UE1}).
Similarly, the consistency condition (\ref{eq: compatibility of theta and phi}) indicates that the action $\alpha_{T}$ of orientation-reversing symmetry maps $\Theta_g$ to $\Theta_{g^{-1}}$, which shows eq. (\ref{eq: UE2}).
The remaining equations (\ref{eq: UE3}) and (\ref{eq: UE4}) follows from the M\"{o}bius identity (\ref{eq: Mobius}) and the Klein identity (\ref{eq: Klein}) respectively.
Thus, the consistency conditions of unoriented TQFTs with fusion category symmetry reduce to those of unoriented equivariant TQFTs when the symmetry is given by a finite group $\mathbb{Z}_2^T \times G$.
We again notice that we used all the consistency conditions (\ref{eq: invariant unit})--(\ref{eq: Klein}) except for (\ref{eq: orientation reversal of f}), which is a condition on topological point operators.

\section{Derivation of equation (\ref{eq: symmetry of Gamma Phi})}
\label{sec: Matrix representation of eq. (phi-m phi-m)}
In this appendix, we derive eq. (\ref{eq: symmetry of Gamma Phi}).
For this purpose, it is convenient to represent the multiplication of $V_m$ by a non-degenerate pairing $\gamma$ as \cite{Tam2000}
\begin{equation}
[w, w^{\prime}]_1 = \gamma(w, w^{\prime}) u_1,
\end{equation}
where the basis $u_1 \in V_1$ is the multiplicative unit of $V = \bigoplus V_a$.
The other components of $[w, w^{\prime}] \in V$ are given by $[w, w^{\prime}]_a = \gamma(u_{a^{-1}} \cdot w, w^{\prime}) u_{a^{-1}}^{-1}$.
The non-degenerate pairing $\gamma$ is symmetric or anti-symmetric $\gamma(w, w^{\prime}) = \epsilon \gamma(w^{\prime}, w)$ depending on whether $\epsilon = +1$ or $-1$.
By choosing a basis of $V_m$, $\gamma$ can be represented by a symmetric or anti-symmetric non-degenerate matrix $\Gamma$ as 
\begin{equation}
\gamma(w, w^{\prime}) = w^{T} \Gamma w^{\prime}.
\end{equation}
The matrix representation of $f(u_a)$ is also given in terms of $\Gamma$ as $\Gamma^{-1} U_a^{T} \Gamma$.
In particular, eq. (\ref{eq: def of nu}) can be written as
\begin{equation}
\Gamma^{-1} U_a^{T} \Gamma = \nu(a) U_{\sigma(a)}.
\label{eq: matrix rep of nu}
\end{equation}
For this matrix representation, eq. (\ref{eq: phi-m phi-m}) is expressed as
\begin{equation}
\Phi^{T} U_{a^{-1}}^{T} \Gamma \Phi = \phi(a^{-1}) \Gamma^{T} U_a.
\label{eq: phi-m phi-m matrix rep}
\end{equation}
When $a$ is the identity element, this equation indicates that $\Gamma \Phi$ is a symmetric or anti-symmetric matrix
\begin{equation}
(\Gamma \Phi)^{T} = \epsilon \Gamma \Phi.
\label{eq: symmetric Gamma Phi}
\end{equation}
Equation (\ref{eq: phi-m phi-m matrix rep}) for general $a \in A$ follows from eqs. (\ref{eq: Phi involution}), (\ref{eq: phi-m phi-a matrix rep}), (\ref{eq: matrix rep of nu}), and (\ref{eq: symmetric Gamma Phi}).
Therefore, it suffices to consider eq. (\ref{eq: symmetric Gamma Phi}).
If we expand the unitary matrix $\Phi$ as $\Phi = \sum_{a \in A} c(a) U_a$, which can always be done without ambiguity because $\{U_a\}$ is a basis of $\sqrt{|A|} \times \sqrt{|A|}$ matrices as we will see below, eq. (\ref{eq: symmetric Gamma Phi}) reduces to
\begin{equation}
c(\sigma(a)) \nu(\sigma(a)) = c(a).
\end{equation}
This completes the derivation of eq. (\ref{eq: symmetry of Gamma Phi}).

Finally, we show that $\{U_a\}$ is a basis of $\sqrt{|A|} \times \sqrt{|A|}$ matrices as we mentioned above. If we assume that $\{U_a\}$ is not linearly independent, there exists $a \in A$ such that $U_a$ is expanded by the other matrices $\{U_b \mid b \neq a\}$.
By multiplying $U_a^{-1}$ to both sides of this expansion, we find that the identity matrix $U_1 = I$ is not linearly independent from the other matrices.
If we denote the linearly independent subset of $\{U_a \mid a \neq 1\}$ as $\{ U_a \mid a \in \widetilde{A}\}$, we can uniquely express the identity matrix $I$ as a linear combination
\begin{equation}
I = \sum_{a \in \widetilde{A}} c_I(a) U_a.
\end{equation}
By multiplying $U_b$ from the left, we obtain $U_b = \sum c_I(a) \xi(b, a) U_{ab}$ where the summation is taken over $a \in \widetilde{A}$.
On the other hand, if we multiply $U_b$ from the right, we have $U_b = \sum c_I(a) \xi(a, b) U_{ab}$.
These equations indicate that $\xi(a, b) = \xi(b, a)$ for all $b \in A$ because $\{U_{ab} \mid a \in \widetilde{A}\}$ is linearly independent.
This implies that $a = 1$ due to the non-degeneracy of $\xi(a, b) / \xi(b, a) = \chi(a, \sigma(b))$, see eq. (\ref{eq: triple1}).
However, the set $\widetilde{A}$ does not contain the identity element $1$ by definition.
This is a contradiction.
Therefore, we find that $\{U_a \mid a \in A\}$ is linearly independent.
Furthermore, since the number of elements in $\{U_a \mid a \in A\}$ is $|A| = \sqrt{|A|} \times \sqrt{|A|}$, the set $\{U_a \mid a \in A\}$ spans the whole vector space of $\sqrt{|A|} \times \sqrt{|A|}$ matrices.

\bibliographystyle{JHEP}
\bibliography{bibliography}

\end{document}